\documentclass[aps,prb,twocolumn,superscriptaddress,showpacs]{revtex4-1}
\usepackage{amsmath,amssymb,graphicx}

\usepackage[T1]{fontenc}
\usepackage{chancery}

\usepackage{xcolor}
\usepackage{physics}
\usepackage[bottom]{footmisc}

\IfFileExists{newtxtext.sty}   {\usepackage{newtxtext,newtxmath}}
{\IfFileExists{stix.sty}
	{\usepackage{stix}}
	{\IfFileExists{mathptmx.sty}
		{\usepackage{mathptmx}}{} } }

\usepackage{textcomp}

\usepackage{bm}
\usepackage{mathbbol}

\IfFileExists{siunitx.sty}{\usepackage{booktabs,siunitx}}{}

\usepackage{color}
\definecolor{LinkColor}{rgb}{0.256,0.439,0.588}
\usepackage{hyperref}
\hypersetup{
	pdfauthor={Tsung-Cheng Lu},
	colorlinks=true,
	citecolor=LinkColor,
	linkcolor=LinkColor,
	urlcolor=LinkColor
}

\usepackage{pifont}
\usepackage{epstopdf}
\usepackage{caption}
\usepackage{subcaption}
\usepackage{simplewick}
\renewcommand{\raggedright}{\leftskip=0pt \rightskip=0pt plus 0cm}
\let\vec\mathbf
\newcommand{\beq}{\begin{eqnarray}}
\newcommand{\eeq}{\end{eqnarray}}

\begin{document}

    \title{Entanglement transitions as a probe of quasiparticles and quantum thermalization}

	\author{Tsung-Cheng Lu} 
\affiliation{Department of Physics, University of California at San
	Diego, La Jolla, California 92093, USA}
\affiliation{Kavli Institute for Theoretical Physics, University of California at Santa Barbara, California 93106, USA}
	\author{Tarun Grover}
\affiliation{Department of Physics, University of California at San
		Diego, La Jolla, California 92093, USA}
	
	\begin{abstract}

We introduce a  diagnostic for quantum thermalization based on mixed-state entanglement. Specifically, given a pure state on a tripartite system $ABC$, we study the scaling of entanglement negativity between $A$ and $B$.  For representative states of self-thermalizing systems, either eigenstates or states obtained by a long-time evolution of product states, negativity shows a sharp transition from an area-law scaling to a volume-law scaling when the subsystem volume fraction is tuned across a finite critical value. In contrast, for a system with quasiparticles, it exhibits a volume-law scaling  irrespective of the subsystem fraction. For many-body localized systems, the same quantity shows an area-law scaling for eigenstates, and volume-law scaling for long-time evolved product states, irrespective of the subsystem fraction. We provide a combination of numerical observations and analytical arguments in support of our conjecture. Along the way, we prove and utilize a `continuity bound' for negativity: we bound the difference in negativity for two density matrices in terms of the Hilbert-Schmidt norm of their difference. 
	\end{abstract}

	\maketitle

\section{Introduction} \label{sec:intro}

\begin{figure}
	\centering
	\begin{subfigure}[b]{0.12\textwidth}
		\includegraphics[width=\textwidth]{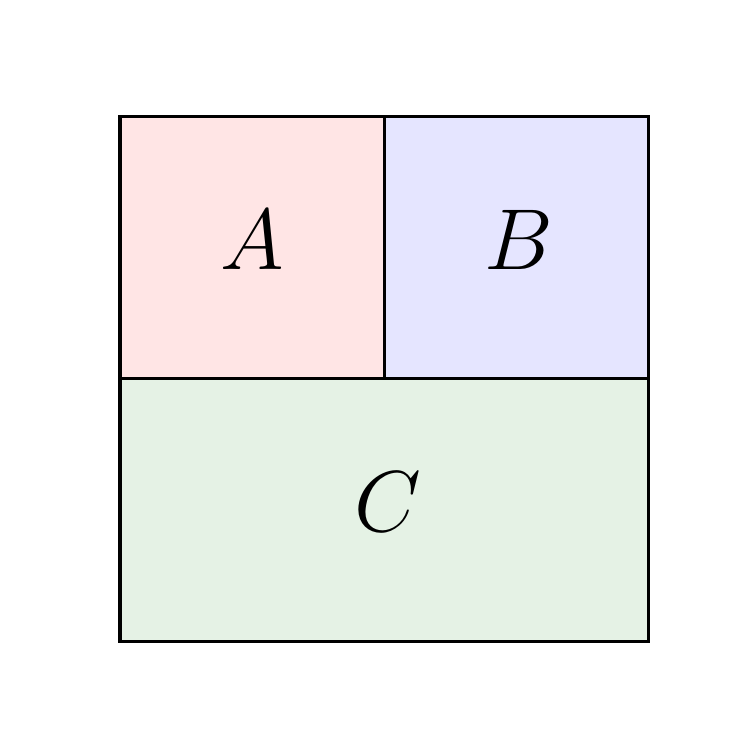}
	\end{subfigure}
	\begin{subfigure}[b]{0.5\textwidth}
		\includegraphics[width=\textwidth]{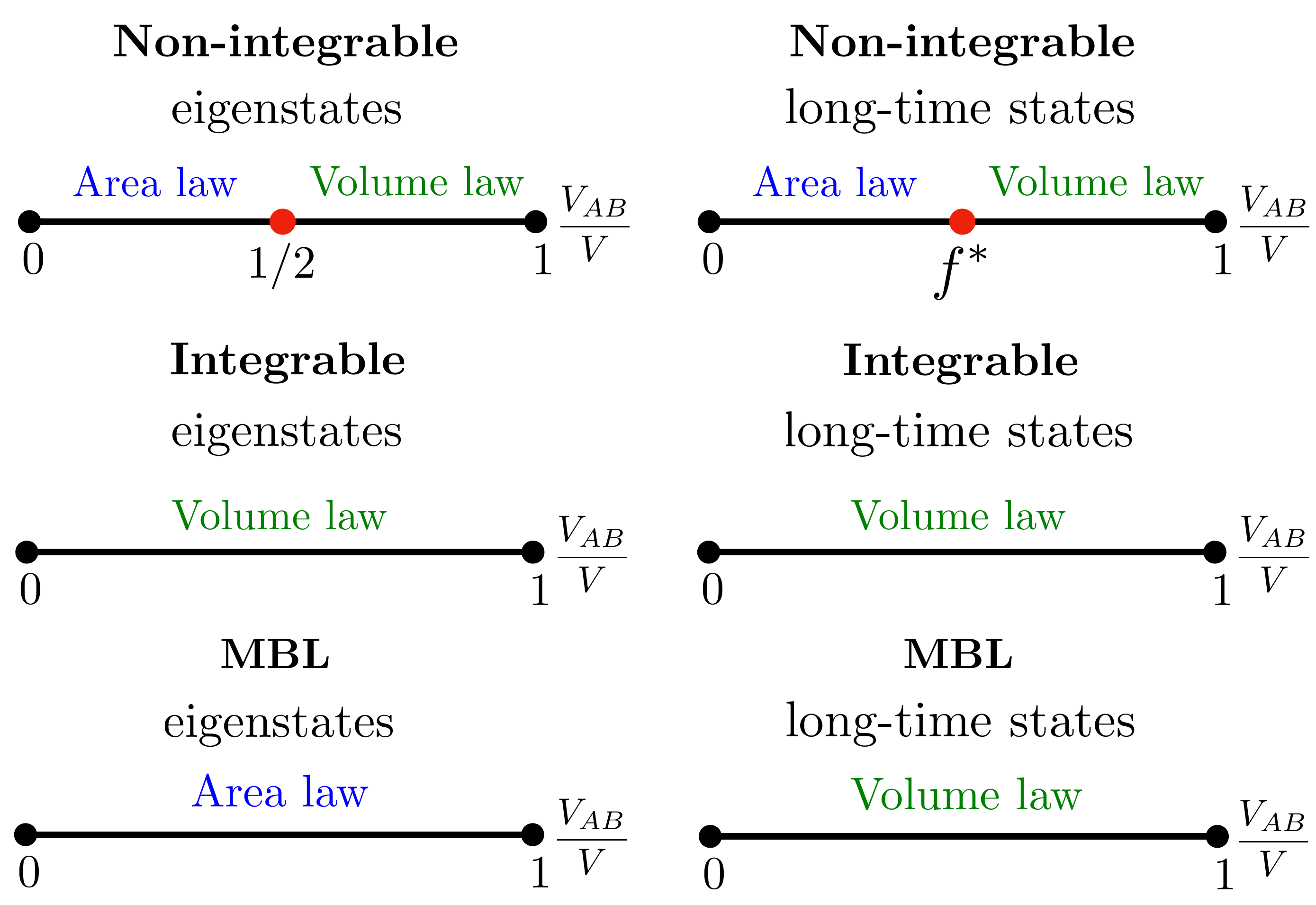}
	\end{subfigure}
	\caption{ Given a pure state in a tripartite system, we study the entanglement negativity $E_N$ between two subsystems $A$ and $B$. In non-integrable systems, given finite-energy density eigenstates or time-evolved states at long time from simple product states, subsystem negativity $E_N$ exhibits a a transition from area-law phase to volume-law phase by tuning the subsystem volume fraction $V_{AB}/V$. The transition for the former is at $1/2$ while the later is at $f^*= O(1) \in ( 0,1/2  ]$, where the exact value of $f^*$ may depend on initial states. In integrable systems studied here, which can be interacting (such as the Heisenberg spin chain) or non-interacting (such as free fermions), given finite-energy density eigenstates or time-evolved states at long time from simple product states, $E_N$ exhibits a volume law for any $V_{AB}/V$. In many-body localized (MBL) systems, $E_N$ exhibits an area law in eigenstates and a volume law in time-evolved states at long time from simple product states for any $V_{AB}/V$.}  \label{fig:intro}
\end{figure}

Consider a system where eigenstate thermalization hypothesis (ETH) \cite{deutsch1991,srednicki1994chaos,srednicki1998, rigol2008, rigol_review} holds true. For a finite-energy density pure state of such a system, the reduced density matrix of a subsystem is thermal when the ratio $f$ of a subsystem to the total system approaches zero. However, this is no longer true when $f$ is $O(1)$, e.g., Renyi entropies do not match their thermal counterpart \cite{Lu_renyi_2019, murthy2019structure, dong2019holographic}. This effect is most dramatic when $f > 1/2$, a regime where entanglement entropy \textit{decreases} with increasing subsystem size, indicating that the rest of the system is acting as a poor `thermal bath' for the subsystem. Monogamy of entanglement suggests that if one were to divide the subsystem further into two parts, these parts would be highly entangled with each other in this regime. Equivalently, one expects that when $f > 1/2$, the reduced density matrix of the subsystem would have a large bipartite mixed-state entanglement. Does there exist a sharp transition as function of $f$ in the mixed-state entanglement of the subsystem? How does this behavior change when one considers product states that have been evolved for a long time with an integrable or a many-body localized (MBL) Hamiltonian\cite{Huse_2007_mbl,Huse_mbl_2010,huse_lbits,huse2015mbl,ros_lbits,altman2015mbl,imbrie2016,alet2018mbl,abanin2019review}?

Motivated by above questions, in this work we discuss a new kind of entanglement transition which occurs within a single quantum state without tuning any parameters in the Hamiltonian. Our setup is as follows: we divide a system described by a pure state into three regions labelled by $A$, $B$, $C$, and study the entanglement between $A$ and $B$, see Fig.\ref{fig:intro}.  Since $A\bigcup B$ ($\equiv AB$) is not a closed system, one requires a mixed state entanglement measure to characterize the entanglement between $A$ and $B$, which we chose as the entanglement negativity \cite{eisert99, vidal2002, plenio2005logarithmic}. This setup allows for a transition where the scaling of negativity changes as the ratio of the region $AB$ to the total system ($\equiv V_{AB}/V$) is tuned.

In fact, this kind of entanglement transition has been noticed in the study of random pure states\cite{aubrun2012,aubrun2012_Ye,bhosale2012entanglement}. When $V_{AB}/V<1/2$, negativity $E_{N}$ between $A, B$ is zero in thermodynamic limit, while for $V_{AB}/V > 1/2$, $E_{N}$ scales with the number of spins in $AB$, i.e. exhibiting a volume entanglement. Below, we first review and provide an intuitive understanding for this transition using entanglement monogamy, and  then show that Renyi negativity, a proxy of entanglement negativity, can also detect this transition.

Going beyond random pure states, here we first consider finite-energy density eigenstates of Hamiltonians that are believed to satisfy ETH and find evidence of a similar transition (as a shorthand notation, we will  denote these states as  `chaotic eigenstates'). We perform three different calculations in support of this transition. Firstly, using exact diagonalization (ED) on finite size systems, we numerically find signatures of such a transition for relatively small systems even though the transition is defined only in the thermodynamic limit. Secondly, by applying ETH and using a slight generalization of the bound for bipartite negativity in a Gibbs thermal state\cite{sherman2016}, we analytically prove the area law for subsystem negativity when $V_{AB}/V<1/2$. Finally, motivated from earlier work, we consider a random tripartite ansatz for chaotic eigenstate, and calculate third Renyi negativity, and find it also exhibits a transition from area law to volume law at $V_{AB}/V= 1/2$.

In sharp contrast, for integrable systems, either interacting or non-interacting, we find that the negativity $E_N$ between $A$ and $B$ for a finite-energy density eigenstate follows the volume-law scaling for any $V_{AB}/V$. We focus on two different systems: a one dimensional spin-1/2 Heisenberg model  (interacting integrable), and free fermion Hamiltonians  (non-interacting integrable). For free fermion Hamiltonians, we analytically derive the volume-law coefficient for subsystem negativity, averaged over all eigenstates, when $V_{AB}/V\ll 1$. Using entanglement monotonicity of negativity\cite{vidal_monotones}, this implies that the volume-law coefficient is non-zero for any $V_{AB}/V$ . For the Heisenberg spin-chain, we perform ED on system sizes up to 18 sites, and find signatures of transition in negativity from a volume-law scaling to an area-law scaling when introducing an integrability-breaking term for $V_{AB}/V < 1/2$, in line with our aforementioned expectation.

In addition to subsystem negativity for eigenstates, we also study the same quantity for pure states obtained from a global quench. These states are more physical compared to the eigenstates in the sense that they can be prepared in an experimental set-up (see, e.g., Refs. \cite{gring2012relaxation, schreiber2015observation, Kaufman794, zhang2017observation, bernien2017probing, weld2019}). Specifically, given an initial product state, we study the subsystem negativity at long time when the subsystem reduced density matrix has reached a steady state. We find that the aforementioned scaling behaviors for eigenstates apply to the steady-state behavior of negativity as well, i.e. for non-integrable Hamiltonians, subsystem negativity has area-law to volume-law transition at a finite critical $V_{AB}/V$ while for integrable models, negativity satisfies volume-law scaling for arbitrary $V_{AB}/V$. Our argument for the integrable models relies only on the assumption that the  quasiparticle picture for entanglement \cite{cardy_quench_2005} holds true.

Finally we discuss the long-time negativity under quantum quench for a disordered Hamiltonian that hosts transition from a many-body localized phase to a chaotic phase. We find that the long-time negativity in the MBL phase exhibits a volume-law scaling in negativity for arbitrary $V_{AB}/V$, similar to the aforementioned integrable models. This is consistent with the emergent integrability in the MBL phase, and it is a consequence that a product state evolved with an MBL Hamiltonian does not look thermal locally despite possessing a volume-law bipartite entanglement. Therefore, as disorder increases, the negativity for $V_{AB}/V <\frac{1}{2}$ undergoes a transition from an area law (chaotic phase) to a volume law (MBL phase).

The paper is organized as follows: In Sec.\ref{sec:random} we demonstrate the phase transition in subsystem negativity as a function of $V_{AB}/V$ for random Haar states. In Sec.\ref{sec:comparison} we first numerically study subsystem negativity in local spin-chain models, and find that chaotic systems show an area to volume-law transition in subsystem negativity, while integrable models always have a volume-law scaling. We provide analytical understanding of these results using eigenstate thermalization hypothesis, and an analysis of free fermions using correlation matrix technique. In Sec.\ref{sec:quench} we discuss negativity of time evolved product states, and show that the distinction between integrable and non-integrable systems is similar to that for their corresponding eigenstates. We derive and utilize a continuity bound of negativity to understand the results for non-integrable models, and a quasiparticle-based argument to understand integrable models. In Sec.\ref{sec:mbl} we study states time evolved with a disordered Hamiltonian. We find that in the ergodic phase, the subsystem negativity is area-law as expected from previous sections, while in the MBL regime, it is volume-law. Finally, in Sec. \ref{sec:mutual}, we compare our protocol with the one based on mutual information, and discuss examples where mutual information and negativity qualitatively behave differently. We conclude with a summary and dicussion of our results in Sec.\ref{sec:summary}.

\section{Negativity transition in a random state} \label{sec:random}

Let us briefly introduce entanglement negativity \cite{eisert99, vidal2002, plenio2005logarithmic}. Unlike most of the entanglement measures for mixed states, negativity can be computed without requiring an optimization of a function over an infinitely large set of states. Therefore, it has been widely applied to various many-body systems, including free bosonic and fermionic systems\cite{audenaert2002entanglement,Eisler_2015,Tonni_negativity_2015,Bianchini_2016_free_boson,Eisler_2016_free_lattice,Shapourian2017,Shapourian2018}, one dimensional conformal field theory\cite{calabrese2012_negativity,tonni_quench_cft,negativity_large_c_2014_kulaxizi,calabrese2015_negativity,Tonni_negativity_cft_2015}, spin chains\cite{Bose_2009_spin_chains,Calabrese_2013_critical_ising,Calabrese_random_spin_chain_2016,gray2018fast,heyl_2018_transverse_field_Ising,turkeshi2019negativity}, and topologically ordered phases\cite{vidal2013,Castelnovo2013,Ryu_chern_simons_2016,ryu_2016_edge_theory,castelnovo2018,lu2019_topo_nega}. To define negativity, consider a density matrix $\rho_{AB}$ on the bipartite Hilbert space $\mathcal{H}=\mathcal{H}_A\otimes  \mathcal{H}_B$: $\rho_{AB}  =\sum_{a,b;a',b'} \rho_{a,b;a',b'} \ket{a,b} \bra{a',b'}$, taking its partial transpose on $B$ gives $\rho^{T_B}_{AB}  =\sum_{a,b;a',b'} \rho_{a,b;a',b'} \ket{a,b'} \bra{a',b}  $. Entanglement negativity is defined as $E_N =\log\left(    \norm{\rho_{AB}^{T_B}}_1   \right)$. 

In this section we consider a random pure state over a tripartite system $ABC$, and study the negativity between $A$ and $B$. We first review a result in Ref.\cite{bhosale2012entanglement}, which shows that this quantity undergoes a transition from zero to a volume-law scaling as the ratio  of the subsystem $AB$ to $C$ is tuned. We will provide an intuitive understanding for the transition, and then show that Renyi negativity, a proxy of entanglement negativity, exhibits such a transition as well.

To be concrete, consider $V$ spin-1/2 degrees of freedom in a random pure state $\ket{\psi}$.  We select $V_A$ spins for the subsystem $A$, $V_B$ spins for the subsystem $B$, and the rest $V_C=V-V_A-V_B$ spins for the subsystem $C$. For simplicity, we set $V_A=V_B=V_{AB}/2$. It was proved that the spectrum of $\rho^{T_B}_{AB}$, the reduced density matrix on $AB$ acted by partial transpose on $B$, follows a semi-circle law\cite{aubrun2012,aubrun2012_Ye}. Based on this result, Ref.\cite{bhosale2012entanglement} calculated the negativity $E_N$ between $A$ and $B$. In the limit $V\to \infty$, one finds,
\begin{equation}\label{maineq:random_nega} 
E_N =\begin{cases} 
0 \quad \text{for} ~ \frac{V_{AB}}{V} < \frac{1}{2}\\
\frac{1}{2} \left(V_{AB}-V_C\right)\log 2  +O(1)  \quad   \text{for} ~ \frac{V_{AB}}{V} >\frac{1}{2}  
\end{cases}
\end{equation}
i.e. $E_N$ exhibits a transition from zero to a volume-law scaling at $V_{AB}/V=1/2$. This transition is consistent with the following heuristic argument based on the notion of `entanglement monogamy' \cite{coffman2000, terhal2001family}. For $V_{AB}/V< 1/2$, entanglement entropy between $AB$ and $C$ is $S_{AB, C}=V_{AB}\log 2$\cite{lubkin1978,page1993average}. Intuitively, this implies every degree of freedom in $AB$ is maximally entangled with $C$. The principle of entanglement monogamy then suggests no entanglement can exist between $A$ and $B$, hence resulting in the vanishing negativity between $A$ and $B$. A different perspective is provided by considering the mutual information between $A$ and $B$: $I(A:B) =S_A+S_B  -S_{AB}=0 $, indicating no correlation exists between $A$ and $B$. This can also be observed from the reduced density matrix $\rho_{AB}$ on $AB$. The maximal entanglement between $AB$ and $C$ implies that $\rho_{AB}$ is a normalized identity matrix $\frac{\mathbb{1}_{AB}}{d_{AB}}= \frac{\mathbb{1}_{A}}{d_{A}}  \otimes \frac{\mathbb{1}_{B}}{d_{B}}$, where both classical and quantum correlations are absent. In other words, the complement of $AB$ can be regarded as an infinite temperature heat bath to destroy any correlations in $AB$. 

On the other hand, for $V_{AB}/V>1/2$, $S_{AB, C}=V_C\log 2$, implying every degree of freedom in $C$ is maximally entangled with $AB$. Since $V_{AB}>V_C$, there will be some degrees of freedom in $AB$ who are not entangled with $C$, and thus can participate in the entanglement between $A$ and $B$. The number of those degrees of freedom is $V_{AB}-V_C$, which suggests the entanglement between $A$ and $B$ of equal size will be $\frac{1}{2}\left( V_{AB}-V_C  \right) \log 2$, which exactly matches the volume-law component of negativity.

As a generalization of the aforementioned result on negativity (Eq.\ref{maineq:random_nega}), one can also consider Renyi negativity $R_n$, a variant of entanglement negativity which has been studied in various contexts\cite{calabrese2012_negativity,Chiamin:2014repqmc,lu2019_topo_nega,wu2019entanglement,Pollmann_2020_renyi_nega}. $R_n$ is defined as

\begin{equation}\label{eq:renyi}
R_n= b_n\log \left\{  \frac{ \tr\left[ \left(\rho^{T_B}_{AB}  \right)^{n}  \right] }{  \tr \rho^{n}_{AB}} \right\}, 
\end{equation}
where $\rho_{AB}= \tr_C{ \ket{\psi } \bra{\psi} }$ is the reduced density matrix on $AB$, and $b_n=\frac{1}{1-n}, \frac{1}{2-n}$ for odd $n$ and even $n$ respectively. Note that $b_n$ is chosen such that when $\rho_{AB}$ is pure, $R_n=S_{n},S_{n/2}$ for odd $n$ and even $n$ respectively, where $S_n $ denotes the n-th Renyi entanglement entropy between $A$ and $B$. Note that entanglement negativity $E_N$ can be obtained from Renyi negativity $R_n$ of even integer $n$ using an analytic continuation: $\lim_{\text{even} ~n \to 1} R_n =E_N$. In the context of random pure states, Ref.\cite{bhosale2012entanglement} also calculated the quantity $\tr \left(\rho^{T_B}_{AB}  \right)^{3} $ although the quantity $R_3$ was not considered. Here we will consider general $n$.

To calculate $R_n$ for a random pure state $\ket{\psi}$, we decompose the state as  $\ket{\psi}  =\sum_{a,b,c}\psi(a,b,c) \ket{a,b,c}$, where  $a$, $b$, and $c$ label bases in $A$, $B$, and $C$ respectively, and the wave function $\psi(a,b,c)$ is a random complex number. It follows that 
\begin{equation}
 \tr \rho^{n}_{AB}=\sum_{     \{a_i,b_i,c_i| i=1,\cdots, n  \}}     \prod_{i=1}^n\left[ \psi(a_i,b_i,c_i)  \psi^*(a_{i+1},b_{i+1},c_i)  \right],
 \end{equation}
and 
\begin{equation}
 \tr\left[ \left(\rho^{T_B}_{AB}  \right)^{n}  \right]=  \sum_{     \{a_i,b_i,c_i| i=1,\cdots, n  \}}   \prod_{i=1}^n  \left[       \psi(a_i,b_i,c_i)   \psi^*(a_{i+1},b_{i-1},c_i)\right],
\end{equation}
where $i+n\equiv i $. By taking the ensemble average over random states $\psi(a,b,c)$ for $   \prod_{i=1}^n\left[ \psi(a_i,b_i,c_i)  \psi^*(a_{i+1},b_{i+1},c_i)  \right]  $ and $\prod_{i=1}^n  \left[       \psi(a_i,b_i,c_i)   \psi^*(a_{i+1},b_{i-1},c_i)\right]$, we calculate $\overline{ \tr \rho^{n}_{AB}}$ and $\overline{ \tr\left[ \left(\rho^{T_B}_{AB}  \right)^{n}  \right]}$ in the thermodynamic limit $V\to \infty$ (see Appendix.\ref{appendix:random_renyi}). Note that in this limit, taking average before or after the logarithm gives the same result, i.e. $\log \overline{\tr \rho_{AB}^n} =  \overline{  \log \tr \rho_{AB}^n    }  $, and $ \log \left\{  \overline{\tr\left[ \left(\rho^{T_B}_{AB}  \right)^{n}  \right]   }  \right\} =  \overline{   \log \left\{  \tr\left[ \left(\rho^{T_B}_{AB}  \right)^{n}  \right] \right\}    }  $ as proved in Appendix.\ref{appendix:prove_random_eq} using an approach presented in Ref.\cite{Lu_renyi_2019}. Finally, one finds that in the thermodynamic limit, the volume law coefficient of the averaged Renyi negativity exactly equals that of the entanglement negativity (Eq.\ref{maineq:random_nega}): 
\begin{equation}
\lim_{V\to \infty}\frac{\overline{R_n}}{V}= \lim_{V\to \infty} \frac{E_N}{V} \quad \text{for any integer} ~n>2.
\end{equation} 
Therefore, $\overline{R_n}$ also exhibits the aforementioned transition as the ratio of $AB$ to $C$ is tuned. 
The advantage of working with Renyi negativity is that for a fixed, small Renyi index $n$, (say $n =3$), it is typically much easier to calculate than the $E_N$ itself. Although for the case of a random pure state we are able to carry out the computation for any $n$, we will encounter a problem in Sec.\ref{sec:eth} where we will be limited to $n = 3$. The fact that $R_n$ qualitatively behaves similarly to $E_N$ for random pure states, as well as several other problems \cite{calabrese2012_negativity, wu2019entanglement} gives us some confidence that it is a useful object to study.

Although entanglement transitions in random  states are instructive, these states lack a notion of locality. Therefore, in the rest of the paper, we focus on the eigenstates as well as time-evolved states for local Hamiltonians.

\section{Negativity transitions in eigenstates: Integrable Vs Non-integrable systems} \label{sec:comparison}

We first consider a class of local spin-chain Hamiltonians, and numerically study negativity of their eigenstates using a protocol identical to that in the last section. We find that in non-integrable systems, there is an area-law to volume-law transition at $V_{AB}/V=1/2$, reminiscent of the random states studied in the previous section, while for integrable systems, subsystem negativity always exhibits a volume-law scaling for arbitrary $V_{AB}/V$. To further support our numerical result, using eigenstate thermalization hypothesis (ETH) in non-integrable systems, we analytically derive the area law in the subsystem negativity for $V_{AB}/V<1/2$. Furthermore, we propose an `ergodic tripartite states' ansatz to characterize the volume-law coefficient of chaotic eigenstates, and show that the third Renyi negativity $R_3$ computed from such ansatz exhibits an area-law to volume-law transition at $V_{AB}/V=1/2$, analogous to negativity. As for the integrable systems, we analytically calculate the subsystem negativity averaged over all eigenstates in free fermions for any spatial dimensions, and find a volume-law scaling for arbitrary $V_{AB}/V$.  

\subsection{Numerical Observations} \label{sec:numerics}

We consider a spin-1/2 chain of size $L$ with periodic boundary condition. The model Hamiltonian reads 
\begin{equation}\label{main_eq:heisenberg}
H=\sum_{i=1}^L  \left( J_1\vec{S}_i \cdot \vec{S}_{i+1}+ J_2 S^{z}_iS^{z}_{i+2}\right).
\end{equation}
We set $J_1=1$ and impose periodic boundary conditions. At $J_2=0$, this Hamiltonian is integrable \cite{Bethe}  while the term proportional to $J_2$ breaks integrability. In the former case, the energy spectrum exhibits Poissonian statistics, while in the latter case, it exhibits the Gaussian-orthogonal ensemble (GOE) level statistics. In any finite-size system,  instead of an abrupt transition at $J_2 = 0$, one would observe a crossover between these two regimes as a function of $J_2$, and we chose $J_2 = 0.8$ as a representative of the non-integrable regime, a point at which the level statistics is clearly GOE.

\begin{figure}
	\centering
	\begin{subfigure}[b]{0.45\textwidth}
		\includegraphics[width=\textwidth]{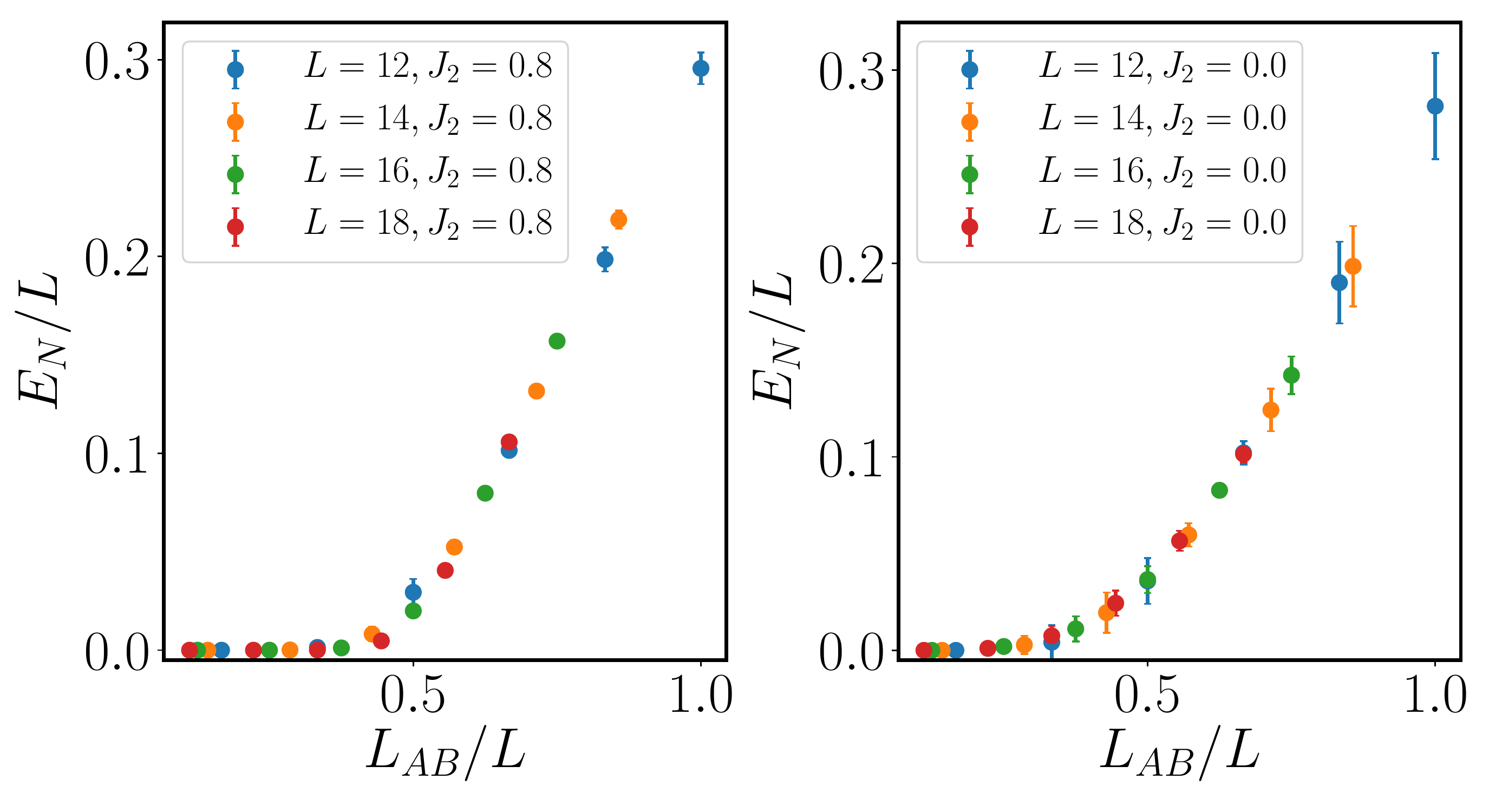}
	\end{subfigure}
	\begin{subfigure}[b]{0.45\textwidth}
		\includegraphics[width=\textwidth]{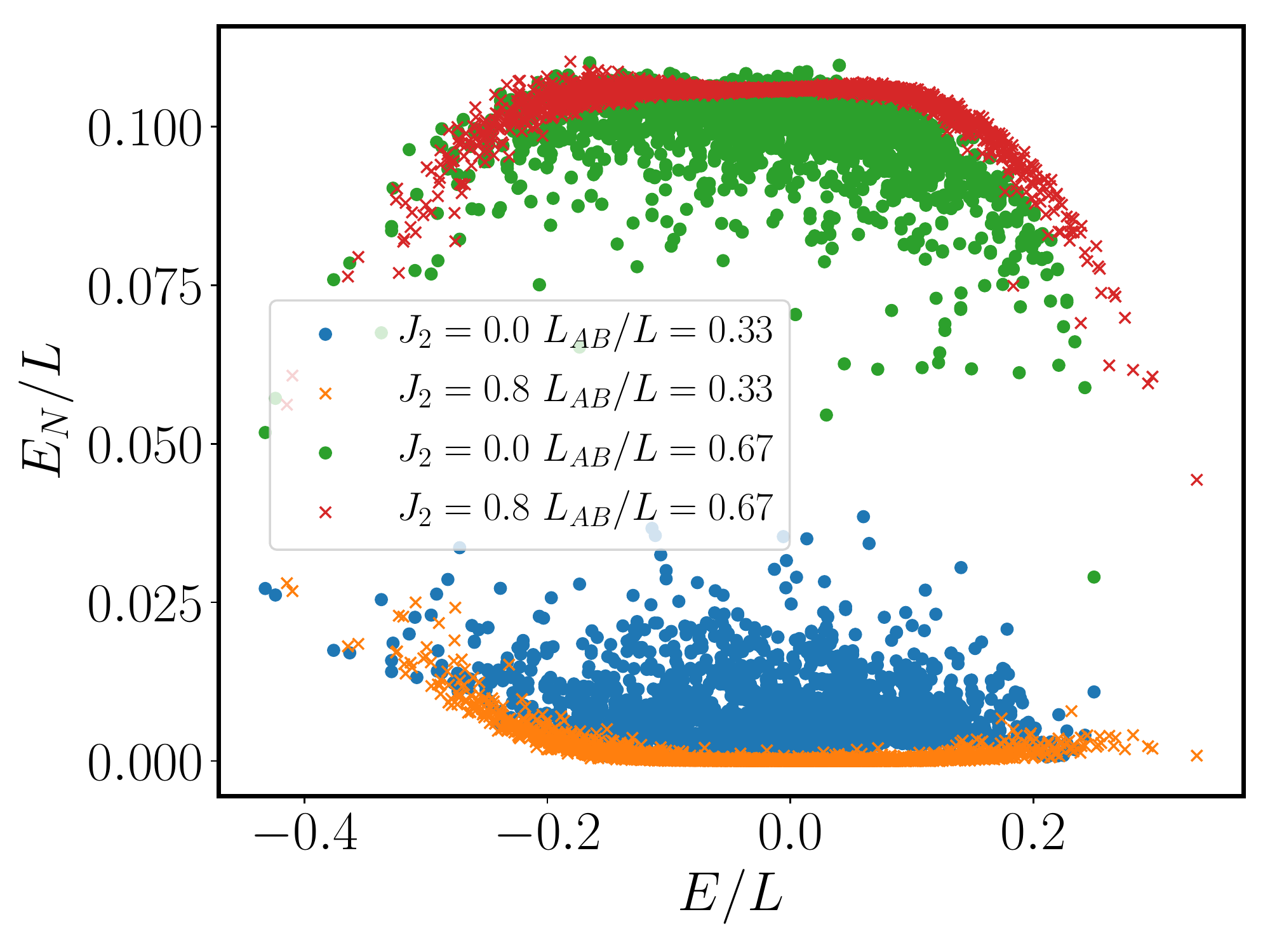}
	\end{subfigure}
	\caption{The subsystem negativity $E_N$, negativity between two subsystems $A$ and $B$, of eigenstates in $S_z=\sum_{i=1}^{L}S_i^z=0$ and momentum $k=0$ sector for the model defined in Eq.\ref{main_eq:heisenberg}. Upper left/right panel: in the non-integrable ($J_2=0.8$) / integrable  ($J_2=0.0$) system, $E_N$ divided by the total system size $L$ as a function of $L_{AB}/L$ averaged over all eigenstates in the energy window $E/L\in (-0.05,0)$ with error bars shown.  Lower panel: $E_N/L$ plotted with $E/L$ for integrable ($J_2=0.0$, marked with circles) and non-integrable ($J_2=0.8$, marked with crosses) of all eigenstates at $L=18$.}  \label{fig:eigen_heisenberg}
\end{figure}

First consider the non-integrable case, i.e., $J_2 = 0.8$  and perform an exact diagonalization using translation symmetry and $S_z=\sum_{i=1}^LS_i^z$ conservation. We divide the spin chain into three subregions $A$, $B$, and $C$ of size $L_{AB}/2$, $L_{AB}/2$, and $L-L_{AB}$ similar to the setup in Sec.\ref{sec:random} and calculate the negativity $E_N$ between $A$ and $B$ in each of the mixed states $\rho_{AB}$ corresponding to individual eigenstates. We then take an average of negativity over all eigenstates in the energy window $E/L\in (-0.05,0)$. In the upper left panel of Fig.\ref{fig:eigen_heisenberg}, we find $E_N/L\sim 0$ for $L_{AB}/L<1/2$ while $E_N/L$ deviates from zero and grows with $L_{AB}/L$ for $L_{AB}/L>1/2$, suggesting negativity between $A$ and $B$ exhibits an area (volume) law for $L_{AB}/L<1/2  \,\,(L_{AB}/L>1/2)$, similar to the behavior of a random pure state. Right at the critical point, i.e., $L_{AB}/L= 1/2$, one observes that $E_N/L$ decreases when increasing the system size $L$, suggesting it might vanish as $L\to \infty$ although it is hard to conclude this unequivocally due to limited system sizes in ED. The data shown here focus only on the eigenstates close to infinite temperature, but we find that eigenstates at finite temperatures exhibit the area-law to volume-law transition as well (see Appendix.\ref{appendix:finite_T_nega}).

Next, consider the integrable point $J_2=0$. We numerically find that negativity of finite-energy density eigenstates between $A$ and $B$ of equal size exhibits a volume law for \textit{any} $L_{AB}/L$, indicating the absence of entanglement transition (see Fig.\ref{fig:eigen_heisenberg} upper right panel). We also introduce an anisotropy in the spin chain to break the SU(2) symmetry down to U(1), and check that the the subsystem negativity is volume-law for any $L_{AB}/L$ as well (see Appendix.\ref{appendix:U(1)_nega}).

It's also instructive to plot subsystem negativity for all eigenstates with respect to their energy densities $E/L$ (lower panel in Fig.\ref{fig:eigen_heisenberg}). We find a distinct contrast between integrable systems ($J_2=0$) and non-integrable systems ($ J_2=0.8$). At a given fixed energy density,  $E_N/L$ has a much broader distribution at $J_2=0.0$ compared to $J_2=0.8$. This suggests that in non-integrable systems, subsystem negativity of finite-energy density eigenstates is possibly a universal (smooth) function of energy density, in a way similar to expectation values of local operators\cite{srednicki1998}, or even entanglement measures such as bipartite Renyi entropies \cite{Lu_renyi_2019,Murthy_2019_renyi}. Note that in both integrable and non-integrable models, although their low-energy eigenstates (i.e. those eigenstates with zero energy density above ground states) show a non-vanishing $E_N/L$ in the figure, we expect such result is due to a finite-size effect. Since these states do not possess an extensive bipartite entanglement, their subsystem negativity $E_N$ will naturally have a vanishing volume-law coefficient in the thermodynamic limit $L \to \infty$.

\subsection{Non-integrable systems: Bounds and scaling from Eigenstate Thermalization } \label{sec:eth}

One intuition for the scaling transition in negativity between $A$ and $B$ comes from their mutual information $I=S_A+S_B-S_{AB}$, akin to the case of random pure states studied in Sec.\ref{sec:random}. We recall that ETH implies a volume-law entanglement entropy between two complementary subsystems $R, \overline{R}$: $S_{R}\sim s_{th}\min(V_R,V_{\overline{R}})$, where $s_{th}$ is the thermal entropy density corresponding to the temperature of eigenstates\cite{garrison2015does}. This result then implies that the mutual information between $A$ and $B$ must exhibit a transition as well: $I\sim 0$ for $V_{AB}/V<1/2$ and $I\sim s_{th}(V_{AB}-V_C)$ for $V_{AB}/V>1/2$, consistent with our numerical observation in the area-law to volume-law transition of negativity. However, one drawback of this analysis is that negativity and mutual information do not necessarily exhibit the same scaling behavior for a general quantum state, as we will discuss further in \ref{sec:mutual}. Therefore, we now turn to a direct analysis of negativity to show that when the subsystem volume fraction of $AB$ is less the $1/2$, negativity between $A$ and $B$ obeys an area-law. The argument is valid in any spatial dimension, as long as the entire system is described by a chaotic eigenstate.

First consider the special case of vanishing volume fraction $V_{AB}/V \to 0$. In this limit, ETH implies that the reduced density matrix on $AB$ is essentially a thermal density matrix $\rho_{AB} \sim e^{-\beta H_{AB}}$, where $H_{AB}$ is the part of the Hamiltonian supported on $AB$. Since in such a thermal state, the negativity between any two complementary subsystems satisfies an area law as proved in Ref.\cite{sherman2016}, negativity between $A$ and $B$ for $V_{AB}/V\to 0$ follows an area law as well. 

For non-zero $V_{AB}/V$, we prove the area law assuming subsystem ETH\cite{dymarsky2016subsystem}, which states that, given a chaotic eigenstate $\ket{\psi}$ with energy $E$ for a local Hamiltonian $H= H_{R}+ H_{\bar{R}}  +   H_{R\bar{R}}  $, when $V_R< V_{\bar{R}}$, the reduced density matrix in $R$ takes the form 
\begin{equation}
\rho_{R}   =  \frac{1}{\mathcal{N}} \sum_i  e^{ S_{ \bar{R} } (E-E_i^R)   } \ket{i}\bra{i},
\end{equation}
where $\ket{i}$ is an eigenstate of $H_R$, and $e^{ S_{ \bar{R} } (E-E_i^R)   }  $ is the density of state of $H_{\bar{R}}$ at energy $E-E_i^R$. This equation indicates that the probability in $\ket{i}$ is proportional to the number of states in $\bar{R}$ consistent with the energy conservation, as if the entire system is described by a microcanonical ensemble. Here we outline the proof, and the detailed derivation can be found in Appendix.\ref{appendix:area_law_proof}. We first expand $ S_{ \bar{R} } (E-E_i^R) $ as
\begin{equation}
S_{ \bar{R} }(E-E_i^R)  =  \sum_{n=0}^{\infty}  \frac{  \left(-E^R_i  \right)^n   }{n!} \frac{\partial^n S_{\bar{R}}(E) }{  \partial E^n },
\end{equation}
then the reduced density matrix on $R$ can be written as an exponential of power series of $H_R$:
\begin{equation}
\rho_R  =\frac{1}{Z} e^{M}  , \quad  M=  \sum_{n=1}^{\infty    }    \frac{ s_{th}^{(n)} \left(-H_{R} \right)^n  }{n!  V_{\bar{R}}^{n-1} } 
\end{equation}
where $s_{th}^{(n)} $ is the $n$-th derivative of microcanonical entropy density at $E/V_{\bar{R}}$. Dividing $R$ into subsystems $A$ and $B$, one essentially needs to count the number of terms simultaneously acting on these two regions to bound the negativity between them\cite{sherman2016}. A detailed calculation gives the upper bound on negativity:
\begin{equation}\label{main:eigen_bound}
E_N   \leq  2J g(E/V_{\bar{R}}, JN_{AB}/V_{\bar{R}}  )\abs{\partial V_{AB}}.
\end{equation}
$J$ is the upper bound of each local term in the Hamiltonian $H$, $g$ is defined as $g(u, JN_{AB}/V_{\bar{R}})=  \sum_{n=0}^{\infty}   \frac{   \left(J N_{AB}/V_{\bar{R}}  \right)^{n}      }{(n)!  }     \abs{ \frac{ \partial^{n+1} s_{th}(u) }{  \partial u^{n+1}  } }    $, which is a function with $O(1)$ value, $N_{AB}$ is the number of terms in $H$ acting only on $A$ and $B$ excluding the terms across their shared boundary, and most importantly, $\abs{\partial V_{AB}}$ is the number of terms in $H$ acting on $A$ and $B$ simultaneously, which scales with the boundary area between $A$ and $B$.  This completes the proof of the area law in negativity for any bipartition of $AB$ when $V_{AB}/V<1/2$. Note that $g(u, JN_{AB}/V_{\bar{R}})$ is a function obtained by taking an absolute value for each term in the Taylor expansion of $s'_{th}(u+ J N_{AB}/ V_{\bar{R}}  )$ about $u(= E/V_{\bar{R}})$. As $V_{AB}/V \to 0$, $g$ reduces to inverse temperature $\beta$ of the eigenstate, hence giving the upper bound $2\beta J \abs{  \partial V_{AB}} $, which agrees with the bound given in Ref.\cite{sherman2016} for a Gibbs thermal state. 

The above argument demonstrates the area law of negativity for $V_{AB}/V<1/2$, but it does not provide any insight into the volume law for $V_{AB}/V > 1/2$. Furthermore, there is  a subtlety: although the reduced density matrix for a chaotic eigenstate is exponentially close to the one from subsystem ETH in their trace distance, it does not necessarily imply that their difference in non-local entanglement measures such as negativity will also be vanishing in the thermodynamic limit. Similar issue arises for the $n$-th Renyi entropy of chaotic eigenstates\cite{Lu_renyi_2019}, in which case for $n\geq 1$, a variety of arguments \cite{Lu_renyi_2019,Murthy_2019_renyi,dong2020_chaotic} provide a rather strong evidence that subsystem ETH indeed provides the correct answer.  Motivated by this, we now discuss an alternative approach for subsystem negativity of chaotic eigenstates, which is related to ETH, but allows one to study all fractions  $0 < V_{AB}/V < 1$. The basic idea is to generalize the `ergodic bipartition' ansatz for chaotic eigenstates discussed in Ref.\cite{Lu_renyi_2019}. We write the Hamiltonian as $H=H_{A}+H_{AB}+H_{B}+H_{BC}+H_C+H_{CA}$, where $H_A, H_B, H_C$ denote the part of $H$ supported only on the spatial region $A, B, C$, and $H_{AB}, H_{BC}, H_{CA}$ denote the interaction between $A$ and $B$, $B$ and $C$, $C$ and $A$. Introducing the chaotic eigenstates $\ket{E_a^A}$, $\ket{E_b^B}$,  $\ket{E_c^C}$ corresponding to the bulk Hamiltonians $H_A$, $H_B$, $H_C$ respectively, we propose the following `ergodic tripartite state' ansatz for a single chaotic eigenstate:

\begin{equation}\label{maineq:ansatz}
	\ket{E}=\sum_{E^A_a+E^B_b+E^C_c\in\left( E-\frac{1}{2}\Delta, E+\frac{1}{2} \Delta  \right)} \psi(a,b,c)\ket{E_a^A}\otimes\ket{E_b^B}\otimes\ket{E_c^C}, 
\end{equation}
where $\psi(a,b,c)$ are random complex numbers, and $\Delta$ is a small energy window. Following the calculation in Ref.\cite{Lu_renyi_2019}, one can immediately show that such an ansatz satisfies ETH for any operators of the form $O=O_AO_BO_C$ where $O_A$, $O_B$, $O_C$ are supported on $A$, $B$, $C$, and they are not close to the boundary between any two subsystems. Therefore, we expect this is a good ansatz for calculating any bulk quantity such as the volume-law coefficient of negativity between $A$ and $B$. 

\begin{figure}
	\centering
	\includegraphics[width=0.4\textwidth]{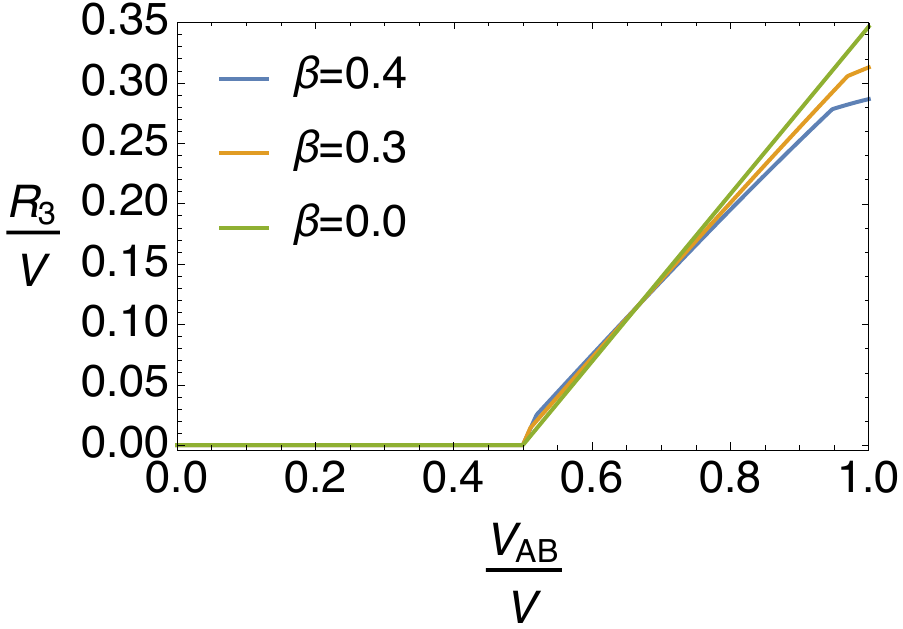}
	\caption{Third Renyi negativity $R_3$ of ergodic tripartite states defined in Eq.\ref{maineq:ansatz} as a function of $V_{AB}/V$ for various inverse temperatures assuming Gaussian density of states. At $\beta=0$, the volume-law component of $R_3$ exactly reproduces the prediction from a random pure state, exhibiting a transition from zero to a volume-law scaling at $V_{AB}/V=1/2$. At non-zero $\beta$, the ergodic tripartite states  exhibit such a transition at $V_{AB}/V=1/2$ as well, but there are two more singularities in the volume-law coefficient: one slightly above $V_{AB}/V=1/2$, and one slightly below $V_{AB}/V=1$.}
	\label{fig:main_text_R_3_ansatz}
\end{figure}

To make progress, we calculate the third Renyi negativity $R_3$ between $A$ and $B$ for the tripartite state $|E\rangle$, as detailed in Appendix.\ref{appendix:tripartite}.
To be concrete, we assume that the many-body density of states $D(u) \sim e^{ V s(u)   }$ is a Gaussian, i.e. the thermal entropy density is quadratic $s(u) = \log 2- \frac{1}{2}u^2$. We find that in the thermodynamic limit $V \to \infty$, $R_3/V$ is zero for $V_{AB}/V<1/2$ while nonzero for $V_{AB}/V>1/2$, faithfully capturing the area-law to volume-law transition (Fig.\ref{fig:main_text_R_3_ansatz}). In particular, $R_3$ for the ergodic tripartite state at infinite temperature $\beta =0$ exactly reproduces the prediction from the random pure states. For finite temperature $\beta \neq 0$, curiously, there are two extra singularities for the volume-law coefficient. It will be interesting to investigate in the future whether the same feature applies to negativity as well.

\subsection{Integrable systems: Volume-law scaling for free fermions} \label{sec:free}
In this section, we will discuss free fermions in one spatial dimension, and show that the subsystem negativity is volume-law for any subsystem volume fraction, as suggested by  the aforementioned ED study of integrable spin-chain. Although we will present detailed calculation only in one spatial dimension,  the same approach works in arbitrary dimensions, and the scaling of subsystem negativity also remains a volume law.

Consider a one dimensional lattice of $L$ sites with periodic boundary condition, the most general Hamiltonian for free fermions with translational symmetry and $U(1)$ charge conservation reads
     
     \begin{equation}\label{main_eq:free_fermion}
     H= - \sum_{x_1,x_2=1}^L \left(  t(x_1-x_2)  c_{x_1}^{\dagger}c_{x_2} + h.c.\right).
     \end{equation}

Dividing the system into three parts labeled by $A$ (sites from $x=1$ to $x=L_A$), $B$ (sites from $x=L_A+1$ to $x=L_A+L_B$), and $C$ (sites from $x=L_A+L_B+1$ to $x=L$), we are interested in the negativity between $A$ and $B$ for energy eigenstates.

Given a fermion eigenstate $\ket{\psi}$, which is a Gaussian state characterized by the correlation matrix $C_{0,xy}=\expval{c_x^{\dagger}c_y}$, we consider its reduced density matrix in $AB$: $\rho_{AB}=\tr_C\ket{\psi}\bra{\psi}$, where $\rho_{AB}$ is again a Gaussian state characterized by the correlation matrix $C$, a sub-block of $C_{0,xy}$ by restricting $x,y \in A B$.

As first shown in Ref.\cite{Shapourian2017}, a fermionic Gaussian state operated by the fermionic partial transpose remains a Gaussian, which allows for an efficient calculation of negativity using the correlation matrix technique. Specifically, let $\rho_{AB}^{T_B}$ be the partial transposed density matrix, one defines the normalized composite density matrix (remains a Gaussian) $\widetilde{\rho} =\rho_{AB}^{T_B} \left(  \rho_{AB}^{T_B}  \right)^{\dagger} /\widetilde{Z}$ , where $\widetilde{Z} = \tr \left[   \rho_{AB}^{T_B} \left(  \rho_{AB}^{T_B}  \right)^{\dagger}   \right]= \tr \rho_{AB}^2$. The negativity reads\cite{Shapourian2018}

\begin{equation}\label{main:fermion_nega}
\begin{split}
E_N= \log \left(   \tr \sqrt{\rho_{AB}^{T_B} \left(  \rho_{AB}^{T_B}  \right)^{\dagger}   }   \right) =  \log \left(   \tr   \widetilde{\rho }^{\frac{1}{2}}  \right)+\frac{1}{2} \log \left(  \tr \rho^2_{AB}  \right),
\end{split}
\end{equation}
where the above two terms can be calculated from the correlation matrices:

\begin{equation}\label{maineq:correlation}
\begin{split}
&\log \left(   \tr   \widetilde{\rho }^{\frac{1}{2}}  \right)      =     \tr{ \log\left[ \widetilde{C}^{\frac{1}{2}} + \left( 1-\widetilde{C} \right)^{\frac{1}{2}}   \right] } \\
&    \frac{1}{2} \log \left(  \tr \rho^2_{AB}  \right)     =      \frac{1}{2}  \tr{  \log \left[  C^2+\left(  1-C \right)^2    \right]     }.
\end{split}
\end{equation}
with $\widetilde{C}$ and $C$ being the correlation matrix of $\widetilde{  \rho  }$ and $\rho_{AB}$ respectively.

 The central idea of calculating the negativity averaged over all eigenstates is to perform an expansion for Eq.\ref{maineq:correlation} in powers of $\widetilde{\Gamma} (= \mathbb{I} -2  \widetilde{C}   )$ and $\Gamma (=  \mathbb{I} -2C  ) $ around $\widetilde{\Gamma}=0$ and $\Gamma=0$, analogous to the calculation in Ref.\cite{vidmar2017}, which studies the entanglement entropy averaged over all eigenstates of quadratic fermionic Hamiltonians. Therefore, negativity can be calculated from the moments of $\widetilde{ \Gamma  }$ and $\Gamma$.

\begin{figure}
     	\centering
     	\includegraphics[width=0.5\textwidth]{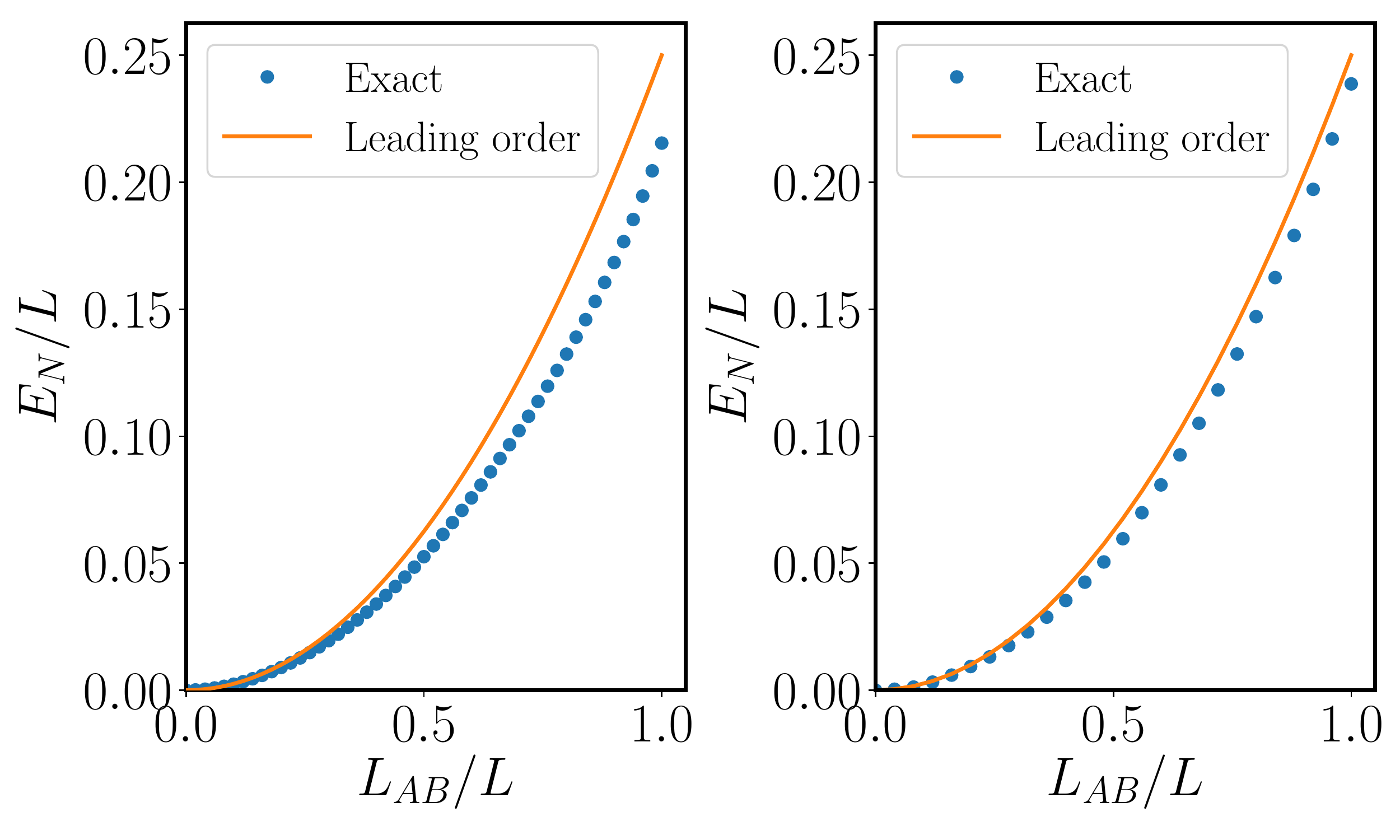}
     	\caption{Subsystem negativity $E_N$ as a function $L_{AB}/L$ in 1D free fermion model (defined in Eq.\ref{main_eq:free_fermion}) with nearest-neighboring hopping . Left: averaged $E_N$ over randomly chosen $10^5$ eigenstates at $L=200$. Right: Long-time $E_N$ of the state $\ket{\psi(t)}$ for large $t$ evolved from a product state at $t=0$ ($\ket{\psi_0}=\prod_{i=1,3,\cdots}^{L-1} c_i^{\dagger}\ket{0}$ where $\ket{0}$ is a vacuum state) at $L=100$. The data shown are the averaged $E_N$ in the time interval $[1000,1200]$. Leading order refers to $\overline{E_N}= \frac{1}{4} \frac{L_{AB}}{L}  L_{AB}$ (Eq.\ref{main_eq:fermion_leading}).
     	}
     	\label{fig:fermion}
     \end{figure}

In the limit $L_{AB}/L \ll 1$, we find the subsystem negativity averaged over all eigenstates follows a volume-law scaling (see Appendix.\ref{appendix:fermion} for details): 

\begin{equation}\label{main_eq:fermion_leading}
\boxed{
	\overline{E_N}  =  \alpha L_{AB}      =  \left[ \frac{L_{AB}}{4L}   \right]  L_{AB}}.
\end{equation}
For finite $L_{AB}/L$, the volume-law coefficient $\alpha$ is a power series of $ L_{AB}/L$: $\alpha=\sum_{n=1}^{\infty}  \alpha_n \left( \frac{L_{AB}}{L}    \right)^n$, similar to the bipartite entanglement entropy of free fermions discussed in Ref.\cite{vidmar2017}. By comparing the leading-order result (Eq.\ref{main_eq:fermion_leading}) with the exact numerical calculation of negativity, we find a good agreement when $L_{AB}/L \ll 1$ (see Fig.\ref{fig:fermion} left). Crucially, despite the fact that we are unable to calculate all moments of $\Gamma$ and $\widetilde{\Gamma}$ to obtain a closed-form expression for negativity, a positive volume-law coefficient when $L_{AB}/L \ll 1$ already ensures volume-law scaling for $E_N$ at \textit{any} $L_{AB}/L$. This is because being an entanglement monotone, negativity is non-increasing under a partial trace\cite{vidal_monotones}. It follows that negativity is non-decreasing when increasing the subsystem size fraction $L_{AB}/L$. Therefore, volume law in $L_{AB}/L  \ll 1 $ already implies volume law at any $L_{AB}/L$.

\section{Negativity transitions in a quantum quench} \label{sec:quench}
\begin{figure}
	\centering
	\begin{subfigure}[b]{0.45\textwidth}
		\includegraphics[width=\textwidth]{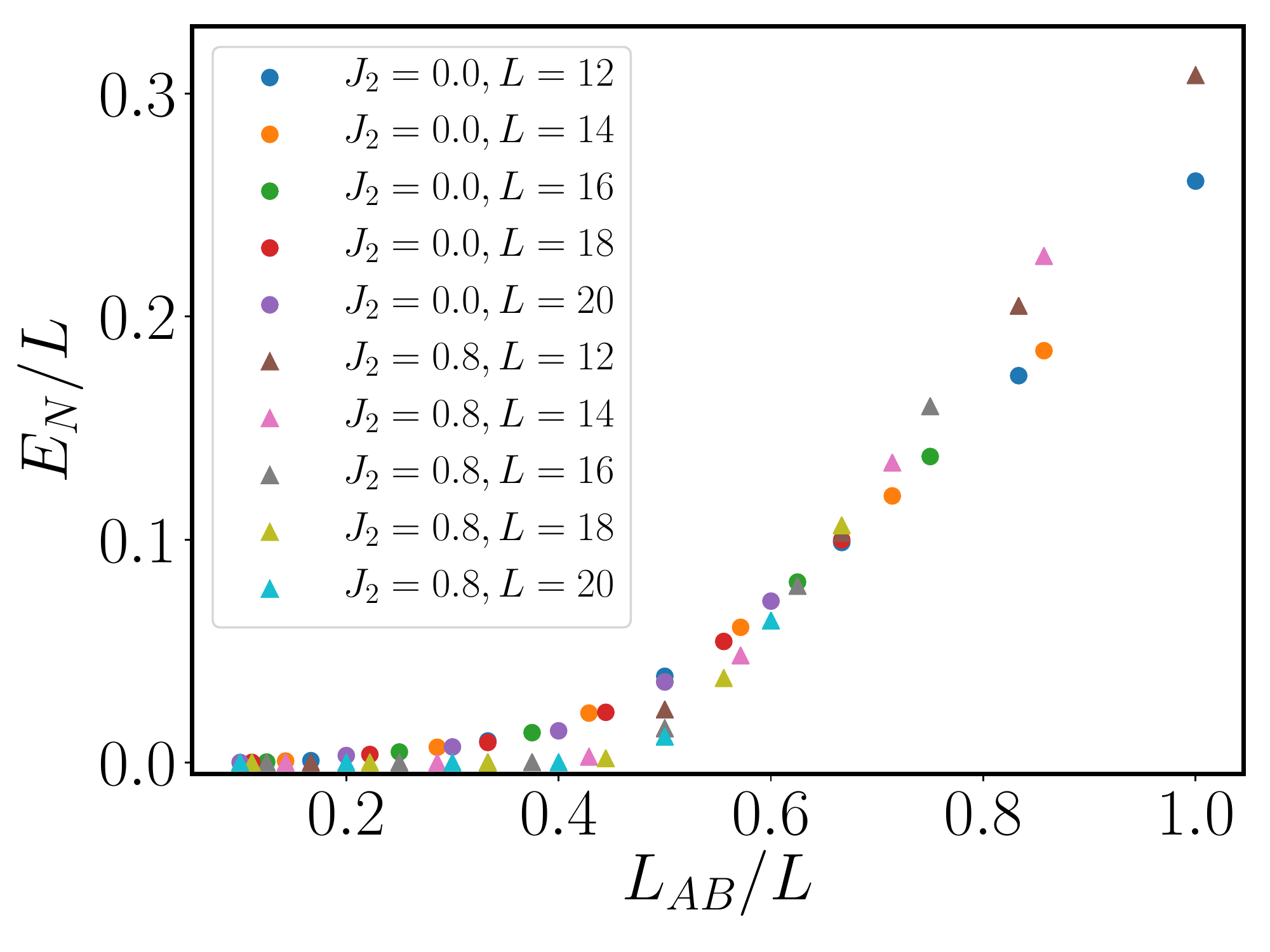}
	\end{subfigure}
	\caption{Comparison of a non-integrable Hamiltonian ($J_2=0.8$) and an interacting integrable Hamiltonian ($J_2=0.0$) defined in Eq.\ref{main_eq:heisenberg} for negativity $E_N$ between two subsystems $A$ and $B$ in a time-evolved state $\ket{\psi(t)}$ at large $t$. While the former exhibits an area-law to volume-law transition at a finite $L_{AB}/L\approx 1/2$, the latter exhibits a volume-law scaling for any $L_{AB}/L$. The data shown are the average of negativity over the time interval $[20,30]$.   }
	\label{fig:heisenberg_long_time}
\end{figure}

We now show that similar to its behavior in eigenstates, subsystem negativity of long-time evolved states also distinguishes an integrable system from a non-integrable system: the former exhibits a volume-law scaling for any $V_{AB}/V$ while the later exhibits an entanglement transition from area-law to volume-law at a certain finite $V_{AB}/V$. The numerical evidence for these statements can be seen in Fig.\ref{fig:heisenberg_long_time}, where we consider the spin chain Hamiltonian (Eq.\ref{main_eq:heisenberg}) with the initial state $\ket{\psi_0}$ as a N\'{e}el state, and study the subsystem negativity for its time-evolved state $\ket{\psi(t)}=e^{-iHt} \ket{\psi_0}$. We also study the long-time negativity for a initial product state evolved by a free fermion Hamiltonian, and find it exhibits a volume-law as well (see Fig.\ref{fig:fermion} right). In the following, we will provide analytical understanding for these numerical results.

\subsection{Non-integrable systems: a rigorous bound} \label{sec:quenchnonint}
Before presenting analytical understanding of subsystem negativity for quantum quench in non-integrable systems, we first present a continuity bound of negativity valid for arbitrary density matrices, which will be essential for our discussion later. \\\\
\noindent \underline{\textit{Continuity bound for negativity}}: Continuity bounds for various entanglement measures, such as the Fannes-Audenart inequality \cite{fannes1973, audenaert2007sharp, petz2007quantum} and the Fannes-Alicki inequality \cite{alicki2004continuity} have  found various applications in quantum information theory \cite{nielsen2002}. Here we derive a continuity bound for the entanglement negativity $E_N$.

Given arbitrary density matrices $\rho$ and $\omega$ acting on a $d$ dimensional bipartite Hilbert space $\mathcal{H}= \mathcal{H}_A\otimes \mathcal{H}_B$, we prove that

\begin{equation}\label{eq:bound}
\boxed{   \abs{  E_N(\rho)-  E_N(\omega)  } \leq\log\left(  1+ \sqrt{d} \norm{    \rho- \omega}_2  \right)}
\end{equation}
where $E_N(\rho) =  \log \left(  \norm{  \rho^{T_B}  }_1 \right), $ and $||\,\,||_2$ denotes the 2-norm (also known as the Hilbert-Schmidt norm).  To derive this bound, notice that $\abs{ ~\norm{      \rho^{T_B}    }_1 - \norm{   \omega^{T_B}  }_1    }  \leq \norm{    \rho^{T_B} - \omega^{T_B}   }_1 \leq  \sqrt{d}  \norm{  \rho^{T_B} -\omega^{T_B}  }_2  $, where we first utilize a reverse triangular inequality for the matrix 1-norm, and then utilize an inequality between the 1-norm and 2-norm \footnote{For any $n\cross n$ matrix $M$, $\norm{M}_1\leq\sqrt{n} \norm{M}_2 $.see e.g. Refs.\cite{popescu2006entanglement,winter_2009_equilibrium}}. Finally, using the fact that $\tr\left[ M^2\right]= \tr \left[(M^{T_B})^2 \right] $ for any matrix $M$, one finds $\abs{ ~\norm{      \rho^{T_B}}_1 - \norm{   \omega^{T_B}  }_1    } \leq   \sqrt{d} \norm{    \rho-\omega }_2 \equiv \Delta$. To proceed, we can assume $\norm{      \rho^{T_B} }_1 \geq  \norm{   \omega^{T_B}  }_1$ without any loss of generality. A simple manipulation shows that $\log\left( \norm{   \rho^{T_B} }_1 \right)  - \log\left( \norm{   \omega^{T_B}  }_1 \right) \leq \log \left( 1+ \Delta/ \norm{   \omega^{T_B} }_1\right) \leq  \log \left( 1+ \Delta \right)$, where the last inequality is due to $\norm{\omega^{T_B}}_1\geq 1$ for any density matrix $\omega$. This completes the proof of Eq.\ref{eq:bound}, and we will now employ this bound for proving the area-law subsystem negativity up to a finite critical $V_{AB}/V$. \\

\noindent \underline{\textit{Application to quantum quenches}}: 
For the quantum quench in non-integrable systems, we analytically show that the area-law for subsystem negativity persists up to a finite $V_{AB}/V$. To start, given a time-evolved state $\ket{\psi(t)}= e^{-i Ht }\ket{\psi_0}$, its reduced density matrix on $AB$ is

\begin{equation}
\rho_{AB}(t)  =\sum_{mn } c_m c_n^*  e^{- i  (E_m-E_n)t   }  \tr_C \left(\ket{m}\bra{n}  \right),
\end{equation}
where $c_m$ is the overlap between the eigenstates $\ket{m}$ and the initial state: $c_m= \bra{m}\ket{ \psi_0}$, and $E_m$ denotes the energy of $\ket{m}$. Define the diagonal ensemble $\omega$ by taking an infinite time average of $\rho(t)=\ket{\psi(t)} \bra{\psi(t)}$
\begin{equation}
\omega = \lim_{T\to \infty } \frac{1}{T}\int_0^{T}  dt~\rho(t)  =\overline{ \rho(t) }    =  \sum_m \abs{c_m}^2 \ket{m }\bra{m}, 
\end{equation}
and $\omega_{AB}=\tr_C\omega $ as the corresponding reduced density matrix on $AB$,  we utilize Eq.\ref{eq:bound} combined with the concavity of logarithm, and find

\begin{equation}\label{eq:bound_2_norm}
\overline{\abs{  E_N(\rho_{AB}(t))-  E_N(  \omega_{AB}  )  }} \leq  \log\left(  1+ \sqrt{d_{AB}} ~\overline{\norm{    \rho_{AB}-  \omega_{AB} }_2}  \right),
\end{equation}
where $d_{AB}= e^{V_{AB} \log 2  }$ is the Hilbert space dimension of $AB$. To further bound the time average of the 2-norm, we now employ a result derived in Ref.\cite{winter_2009_equilibrium}, which is valid for any Hamiltonian without degenerate energy spectrum (hence valid for the non-integrable Hamiltonians): $ \overline{   \norm{    \rho_{AB} - \omega_{AB} }_2   }     \leq  \sqrt{d_{AB}}  e^{-\frac{1}{2}  S_2(\omega)  }  $, where $S_2(\omega)$ is the second Renyi entropy of the diagonal ensemble $\omega$. Combining this result with Eq.\ref{eq:bound_2_norm}, we thus obtain the bound

\begin{equation}\label{eq:difference}
\overline{\abs{  E_N(\rho_{AB}(t))    -  E_N(  \omega_{AB}  )  }     } \leq  \log\left(  1+ d_{AB}  e^{-\frac{1}{2} S_2 (\omega)  }   \right),
\end{equation}

Since $S_2(\omega)$ in nonintegrable systems is extensive\cite{eisert_2019_equilibrium}: i.e. $S_2(\omega)= \alpha V$ with $0<\alpha\leq \log 2$, Eq.\ref{eq:difference} implies that in the regime $V_{AB}/V<  f^*=\alpha/(2 \log 2)$ for almost all times, the difference between $ E_N(\rho_{AB}(t))$ and $E_N(  \omega_{AB}  )$, are exponentially small in the total system volume. Therefore for almost all times $t$,

\begin{equation}\label{eq:equivalence}
\lim_{V\to \infty}  \left[ E_N(\rho_{AB}(t))    -  E_N(  \omega_{AB}  )       \right]  =0  \quad \text{for}~ \frac{V_{AB}}{V}<  f^*=\frac{\alpha}{2 \log 2}.
\end{equation}

Since all eigenstates satisfy area-law subsystem negativity for $V_{AB}/V\leq 1/2$ as argued in Eq.\ref{main:eigen_bound}, the subsystem negativity of reduced density matrix from the diagonal ensemble, i.e. $E_N(  \omega_{AB}  )   =\log \left(   \norm{    \omega_{AB}^{T_B}  }_1 \right)$, also follows an area law\cite{footnote:bound}, which hence indicates the area-law scaling of $E_N(t)$ for $V_{AB}/V<f^* \leq 1/2$ due to Eq.\ref{eq:equivalence}.

\subsection{Integrable systems: Volume-law scaling from quasiparticles} \label{sec:quenchint}
For quantum quenches in integrable systems, the quasiparticle picture, as first introduced in Ref.\cite{cardy_quench_2005}, has successfully described the growth of many-body entanglement\cite{Alba_qp_2017,Alba_qp_2018,Calabrese_qp_2018,Alba_qp_2019,Dutta_2020_mutual,alba2020open,Alba_revival2020}. In particular, Ref.\cite{Alba_qp_2019} showed that such picture allows for an exact prediction of time-evolved negativity under a quantum quench in a space-time scaling limit, whose validity is further supported by numerically studying negativity between two subsystems embedded in an infinite system of one-dimensional free bosons and free fermions. Here we instead consider finite subsystem size fraction $L_{AB}/L$, and adopt the quasiparticle picture to provide a heuristic argument for volume-law subsystem negativity for any $L_{AB}/L$ at long time. Although we specialize to one space dimension below, our argument applies to higher dimensions as well.

In the description of the quasiparticle picture, since an initial state typically has a finite-energy density with respect to the post-quench Hamiltonian, each point in space is a source of quasiparticle pairs, and the two particles in each pair are entangled while propagating with opposite momentum. Because a quasiparticle pair contributes to the entanglement between two spatial regions $A$ and $B$ only when one particle is in $A$ and its partner is in $B$, the total amount of entanglement between $A$ and $B$ can be obtained by counting the number of such quasiparticle pairs.

\begin{figure}
	\centering
	\begin{subfigure}[b]{0.3\textwidth}
		\includegraphics[width=\textwidth]{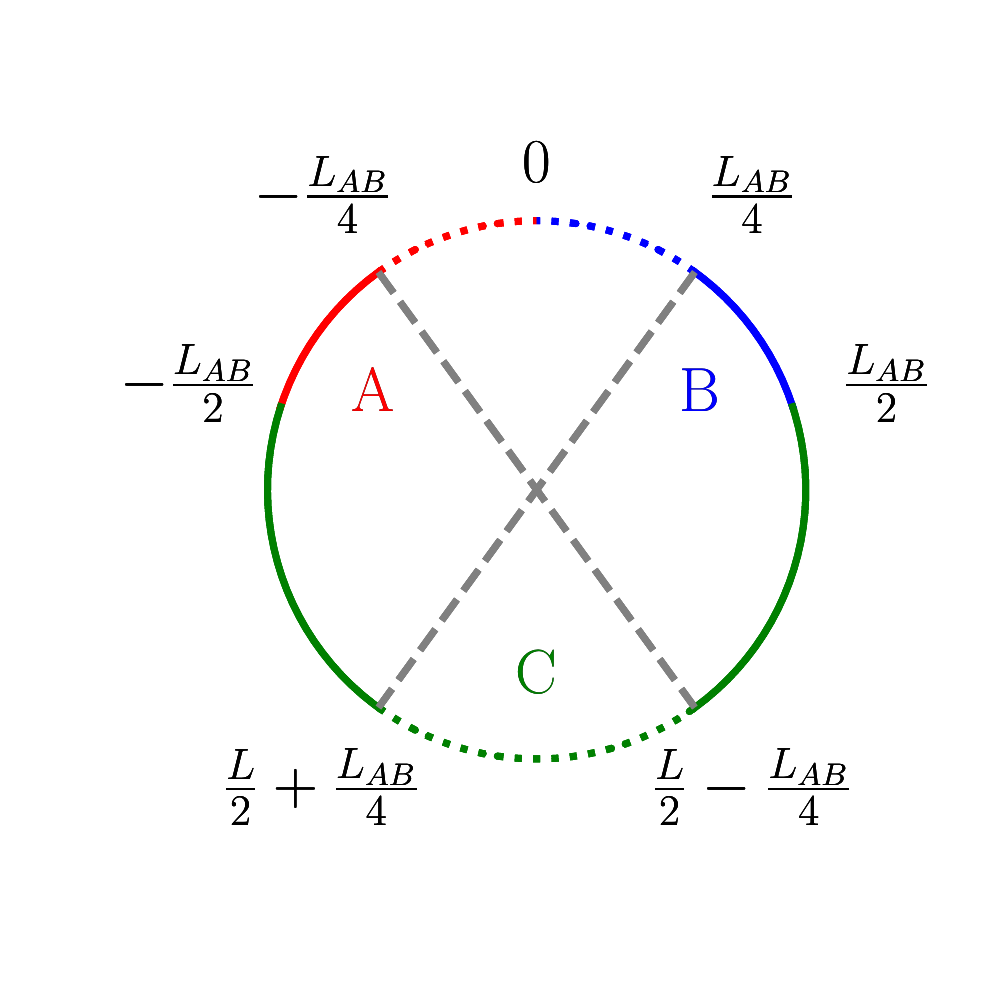}
	\end{subfigure}
	\caption{We divide a one dimensional ring into subregion $A$ (red), $B$ (blue), and $C$ (green), and study negativity between $A$ and $B$. Only those quasiparticle pairs generated in the dashed regions can be shared between $A$ and $B$ to contribute entanglement between these two regions.}
	\label{fig:tripartite_qp}
\end{figure}

Now we apply the quasiparticle picture to study the subsystem negativity. Given a 1D chain with periodic boundary condition ($x+L\equiv x$), let $A$ be the spatial interval $(-L_{AB}/2 ,0  )$, $B=(0, L_{AB}/2   )$, and $C$ be the rest of the chain (see Fig.\ref{fig:tripartite_qp}), at $t=0$, quasiparticle pairs with different momenta $k$ are generated uniformly in space. It is not hard to see that only when a pair is generated within the spatial interval $\mathcal{I}=(-L_{AB}/4 ,  L_{AB}/4  ) \bigcup (L/2-L_{AB}/4,L/2+L_{AB}/4)$ (marked by dashed lines), the two particles can reside in $A$ and $B$ simultaneously at some later times to entangle $A$ and $B$. Now we consider a pair of quasiparticles with velocities $v(k)$ and $v(-k)=-v(k)$, generated at $x$ in the interval $(-L_{AB}/4 ,  L_{AB}/4  ) $. These two particles initially both belong to either $A$ or $B$, and they begin to entangle $A$ and $B$ at $t_1=\abs{x/v}$ until one of the particle first moves into $C$ at $t_2=( L_{AB}/2-\abs{x}  )/\abs{v}$. Due to the periodic boundary condition, in a time period $T=L/\abs{v}$, the time duration for two particles simultaneously in $A$ and $B$ is $2(t_2-t_1)$ in a period. Thus the entanglement between $A$ and $B$ contributed from the quasiparticle pair averaged over the period $T$ is $s(k) 2(t_2-t_1)/T  $, where $s(k)$ is the amount of entanglement carried by the pair. Since all quasiparticle pairs emitted from  $\mathcal{I}=(-L_{AB}/4 ,  L_{AB}/4  ) \bigcup (L/2-L_{AB}/4,L/2+L_{AB}/4)$ with all possible momenta contribute to entanglement, one finds long-time averaged negativity between $A$ and $B$ within the quasiparticle picture reads
\begin{equation}\label{eq:qp_negativity}
\begin{split}
\overline{E}_{N,qp}   &= 2 \int \frac{dk}{2\pi}  \int_{-\frac{L_{AB}}{4}}^{-\frac{L_{AB}}{4}} dx ~ s(k)\left(  \frac{L_{AB}}{L}   -\frac{ 4\abs{x} }{L}  \right)\\
& =   2\int \frac{dk}{2\pi} s(k  )   \left( \frac{L_{AB}}{4L} \right) L_{AB}, 
\end{split}
\end{equation}     
which scales with the subsystem volume $L_{AB}$ with a volume-law coefficient $\sim L_{AB}/L$ for any subsystem volume fraction. In sum, quasiparticle picture allows to predict a volume-law scaling of negativity at long time in a quantum quench: $E_N \sim L_{AB}^2/L$. Such volume-law scaling results from the fact that the number of quasiparticle pairs that can entangle $A$ and $B$ scales with $L_{AB}$, and the fraction of time duration in which a pair entangles $A$ and $B$ in a period scales with $L_{AB}/L$. Note that Ref.\cite{Alba_qp_2019} also studied subsystem negativity for systems with quasiparticles and instead found it vanishes at long time. This is because they considered a different limit:   
$\lim_{L\to \infty } L_{AB}/L = 0 $.

\begin{figure}
	\centering
	\begin{subfigure}[b]{0.45\textwidth}
		\includegraphics[width=\textwidth]{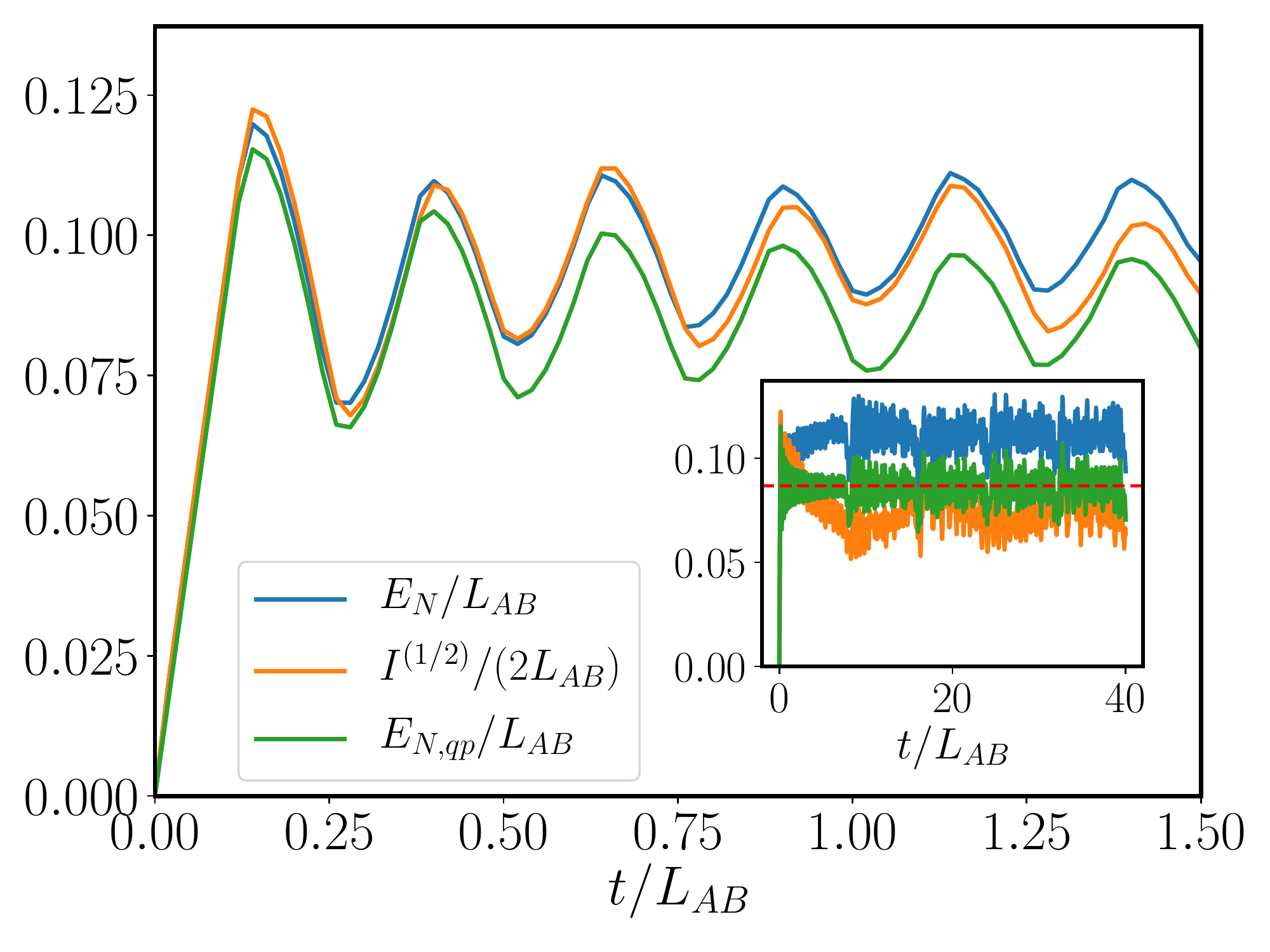}
	\end{subfigure}
	\caption{Negativity $E_N$ compared with Renyi mutual information $I^{(1/2)}$ and the prediction of negativity $E_{N,qp}$ from quasiparticle picture between two subsystems $A$ and $B$ both of size $L_{AB}/2$ as a function of time $t$ by evolving a product state $\ket{\psi_0} $ at $t=0$ with a nearest-neighboring hopping fermion model (Eq.\ref{main_eq:free_fermion}). The initial state is chosen as $\ket{\psi_0}=\prod_{i=1,3,\cdots}^{L-1} c_i^{\dagger}\ket{0}$ where $\ket{0}$ is a vacuum state. We choose the total system size $L=200$ and subsystem size $L_{AB}=100$. Inset: dynamics of $E_N$, $I^{(1/2)}$, and $E_{N,qp}$ up to an extremely long time $t\gg L_{AB}$. The dashed red line is given by Eq.\ref{eq:qp_negativity}, i.e. the infinite-time average of $E_{N,qp}$.}
	\label{fig:fermion_compare}
\end{figure}

In addition to predicting a volume-law subsystem negativity at long time, the quasiparticle picture also predicts the time evolution of negativity. As found in Ref.\cite{Alba_qp_2019}, $s(k)$, the entanglement negativity carried by a quasiparticle pair with momentum $k$, can be fixed by the entropy contribution of $k$-momentum mode in $S^{(1/2)}$, i.e. the Renyi entropy at index $1/2$, in the generalized Gibbs ensemble (GGE). Intuitively, this is because entanglement negativity between complementary systems in a pure state identically equals $S^{(1/2)}$. This implies that $E_N=I^{(1/2)}/2$ whenever quasiparticle picture holds\cite{Alba_qp_2019}, where $I^{(1/2)}(\equiv S^{(1/2)}_A+S^{(1/2)}_B-S^{(1/2)}_{AB})$ is the Renyi mutual information at index $1/2$. Here we test this claim in our setup for a quench in a one-dimensional free fermion model. We compare three different quantities: $E_N$, $I^{(1/2)}/2$, and $E_{N,qp}$ predicted from the quasiparticle picture (Fig.\ref{fig:fermion_compare}) . We find excellent agreement between these three quantities up to a time scale $t\sim L_{AB}$ while after that time, they start to deviate from each other as shown in the inset of Fig.\ref{fig:fermion_compare}. At extremely long time, i.e. $t\gg L_{AB}$, $E_{N,qp}$ typically oscillates between $I^{(1/2)}/2$ and $E_N$. Such deviation from the quasiparticle picture has also been observed in Ref.\cite{Alba_qp_2019}. Nonetheless, the quasiparticle picture provides a simple understanding of the volume-law subsystem negativity in integrable systems.

    \section{Distinguishing MBL from ETH phase using negativity transition} \label{sec:mbl}
\begin{figure}
	\centering
 \begin{subfigure}[b]{0.35\textwidth}
	\includegraphics[width=\textwidth]{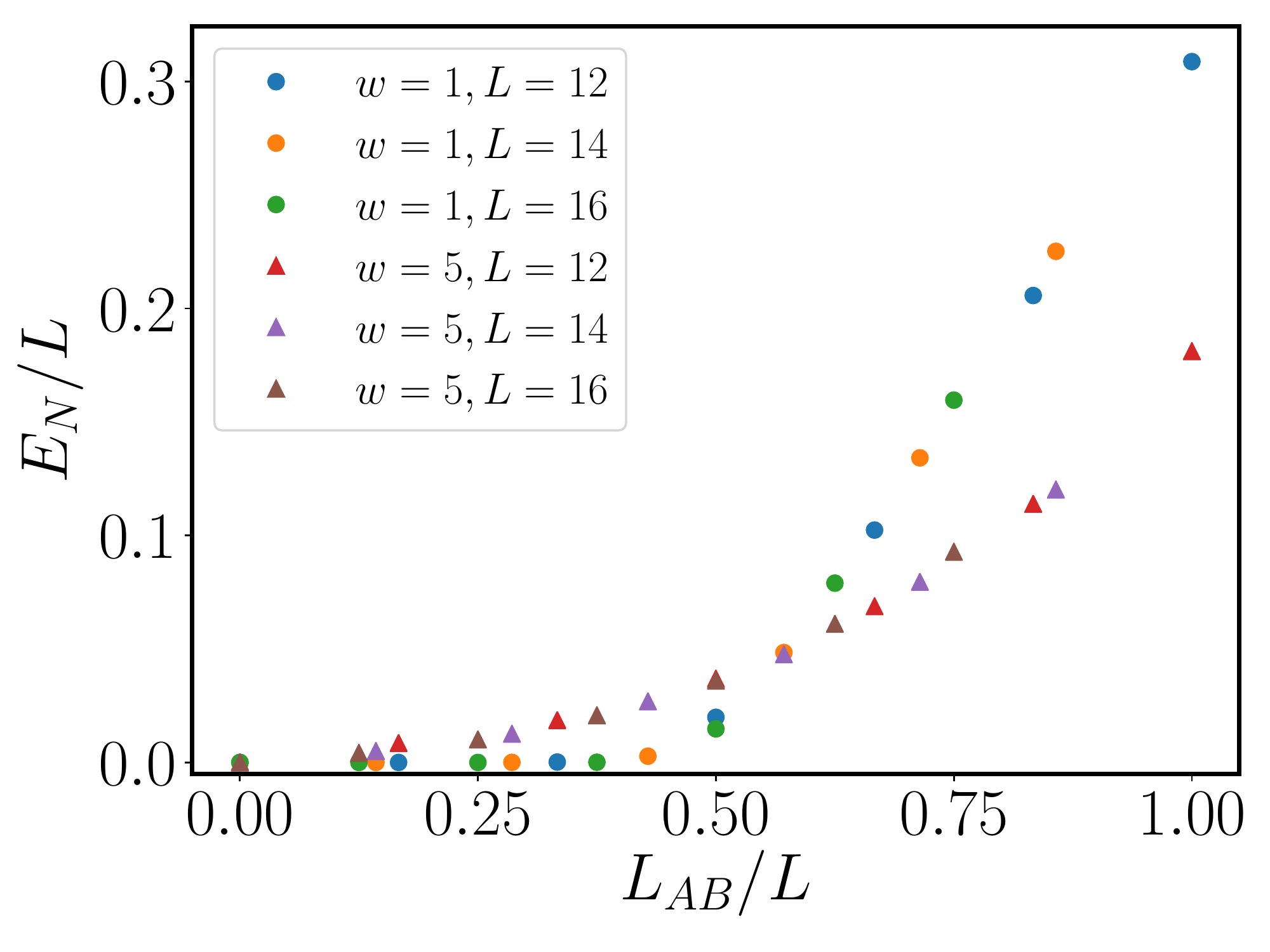}
\end{subfigure}

 \begin{subfigure}[b]{0.35\textwidth}
	\includegraphics[width=\textwidth]{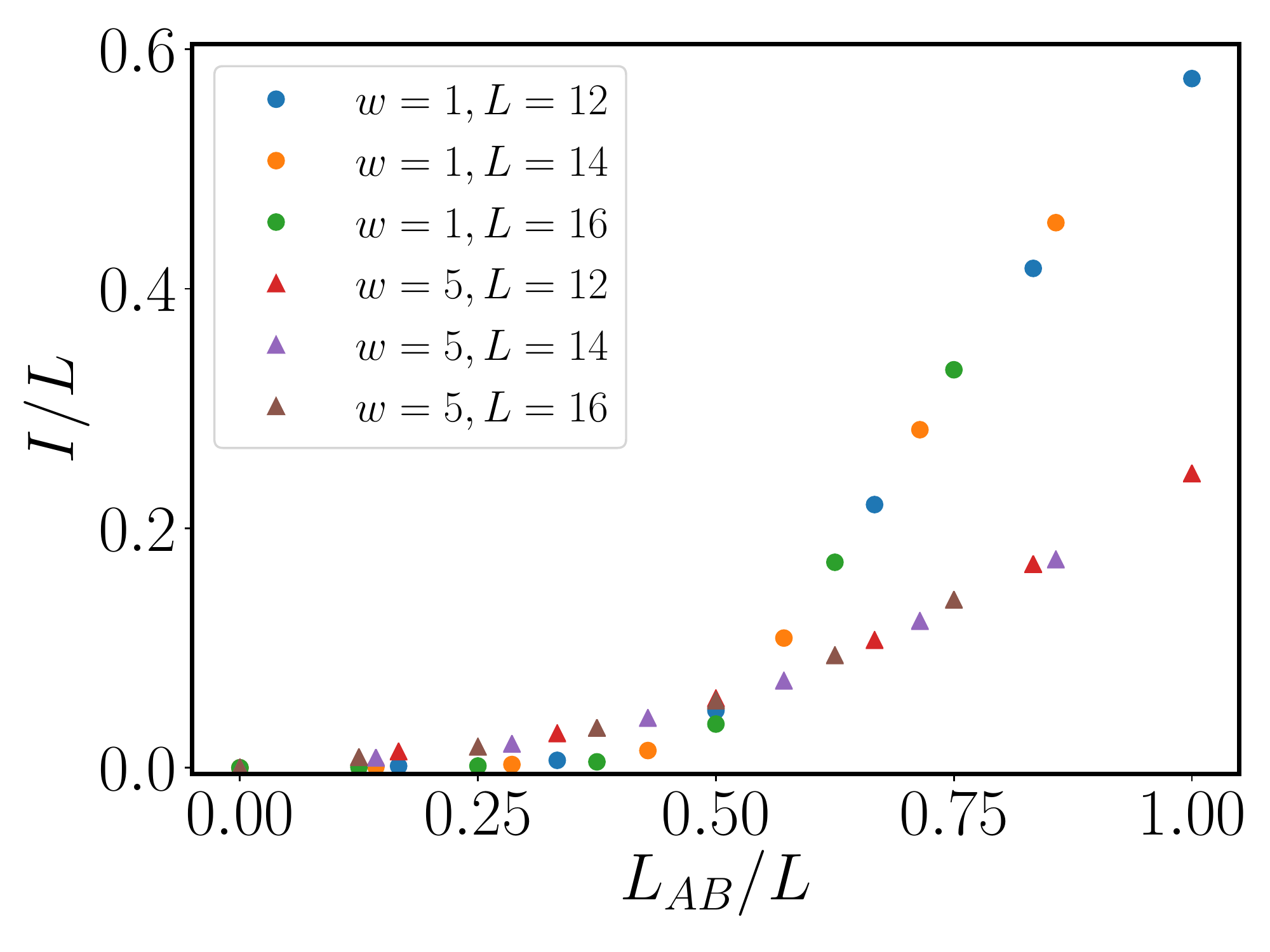}
\end{subfigure}
	\caption{Comparison of the ETH Hamiltonian ($w=1$) and the MBL Hamiltonian ($w=5$) in subsystem negativity $E_N$ (upper panel) and mutual information $I$ (lower panel) as a function $L_{AB}/L$ for a time-evolved state $\ket{\psi}$ at $t=1000$. The data for $L=12, 14, 16$ presented are averaged over $200, 100, 100$ random samples.}
	\label{fig:mbl}
\end{figure}

     Finally we discuss how signatures in subsystem negativity distinguish MBL phase from ETH phase. Deep in the MBL phase, all eigenstates are localized, exhibiting area-law scaling in the entanglement entropy between two complementary systems\cite{pal2010many, vznidarivc2008many, Bauer_2013_mbl,abanin_lbits}. Furthermore, eigenstates can be efficiently described by matrix product states of finite bond dimension\cite{eisert_mps}. Therefore, negativity between two subsystem $A$ and $B$ naturally follows an area-law scaling for any $L_{AB}/L$. Despite the presence of localized eigenstates, initial product states under time evolution in the MBL phase at long time exhibit volume-law scaling of bipartite entanglement entropy \cite{bardarson2012}, which can be understood as a dephasing mechanism given by an effective ``l-bits'' Hamiltonian\cite{ abanin_lbits,huse_lbits,ros_lbits}. Here we study the long-time evolved state, and find that the negativity between two subsystems exhibits a volume-law scaling as well. To obtain MBL phase, we introduce on-site random fields on spins in Eq.\ref{main_eq:heisenberg} to obtain the model Hamiltonian
     \begin{equation}\label{main_eq:mbl}
     	H=\sum_{i=1}^L  \left( J_1\vec{S}_i \cdot \vec{S}_{i+1}+ J_2 S^{z}_iS^{z}_{i+1}- h_i S^z_{i+1} \right),
     \end{equation}
where $h_i$ is randomly drawn from $[-w,w]$, and we set $J_1=1, J_2=0.8$. Choosing the initial state as a N\'{e}el state $\ket{\psi_0}$, we study the negativity of the state $\ket{\psi(t)}=e^{-iHt} \ket{\psi_0}$ at large $t$. We compare $w=1$ (ETH phase) and $w=5$ (MBL phase) in the long-time negativity $E_N$ between $A$ and $B$. We find a signature of volume-law scaling in $E_N$ for the MBL phase, similar to the cases of integrable systems discussed before, in contrast to the area-law to volume-law transition in the ETH phase, see Fig.\ref{fig:mbl}, upper panel.

To build intuition for the volume-law subsystem negativity in the MBL phase, we consider the mutual information between $A$ and $B$. As argued in Ref.\cite{maccormack2020}, the bipartite entanglement entropy for a single region of size $\ell$ scales as $S \sim \ell - \ell^2/L$ where $L$ is the total system size. This implies that the mutual information between $A$ and $B$ is a \textit{volume law} for any $L_{AB}/L$, unlike the ETH phase where there is an area-law to volume-law transition at a finite critical $L_{AB}/L$. We find evidence in support of this claim in our ED study, as shown in Fig.\ref{fig:mbl}, lower panel. Given that mutual information seems to follow the same scaling as subsystem negativity in all the other examples we have considered so far, this indicates that the subsystem negativity also satisfies a volume law for all $L_{AB}/L$.

We note that subsystem negativity for models that exhibit the MBL transition was also previously studied in Ref.\cite{gray2019_mbl}. However, the focus of Ref.\cite{gray2019_mbl} was different: they considered the scaling of subsystem negativity as a function of the separation between two disjoint blocks at the transition.

\section{Comparison with mutual information} \label{sec:mutual}

In all systems studied so far in this paper, the subsystem negativity and the mutual information both essentially have the same scaling form as a function of the subsystem volume fraction. Since mutual information is not a mixed state entanglement measure, it is  natural to ask whether there are physical situations related to quantum thermalization where these two quantities can qualitatively behave differently, and therefore necessitate a mixed-state entanglement (such as subsystem negativity) based protocol? We now motivate a few physical scenarios where that is indeed the case.

First,  consider separable states  $\rho=\sum_i p_i \rho_i^A\otimes \rho_i^B $, where $\sum_i p_i=1 $ with $p_i\geq 0$, and $\rho_i^A$, $\rho_i^B$ are density matrices on $A$, $B$. Such states manifestly have zero negativity, but  they allow for a volume-law mutual information between $A$ and $B$ as we show below. Such a construction relies on the intuition that mutual information measures the amount of information gained regarding one system by observing the other. Therefore one can imagine that when the index $i$ runs over a range that is exponentially large in the total system volume, observing a subsystem (say $A$) gives a great amount of knowledge for the other ($B$), which can result in a volume-law mutual information. A concrete example is given by the so-called thermo-mixed double state \cite{verlinde2020}, which has been proposed as a typical mixed state of a two-sided black hole: 

\begin{equation}
\rho_{\text{TMD}}=\sum_n \frac{e^{-\beta E_n} }{Z}  \left[  \ket{n }\bra{n}  \right]_A \otimes \left[\ket{n }\bra{n}\right]_B, 
\end{equation}
where $Z= \sum_n e^{-\beta E_n}$. It is not difficult to see that the mutual information $I(A,B)=S_A+S_B -  S_{AB}    =  S_{th} $, where $S_{th}$ is the extensive thermal entropy of a canonical ensemble for Hamiltonian $H= \sum_n E_n\ket{n}\bra{n}$ at inverse temperature $\beta$. Hence $\rho_{TMD}$ constitutes a class of states whose negativity and mutual information behave qualitatively differently. 

As an example motivated by condensed matter physics, consider eigenstates of a `quantum disentangled liquid' (QDL) \cite{grover2014quantum}. The Hilbert space of QDL consists of two kinds of particles, `heavy' and `light', with the property that a projective measurement of the heavy (light) particles results in a wavefunction of the light (heavy) particles that has an area-law (volume-law) bipartite entanglement. As an example, consider the following wavefunction where the sets $\{R\}$ and $\{r\}$ denote coordinates of the heavy and light particles respectively: $|\psi\rangle = \sum_{R}  \textrm{Det}\left( e^{i k_{i}.R_{j}}\right) \sqrt{p(\{R\})} |\phi_R\rangle |R\rangle$. Here  $\textrm{Det}\left( e^{i k_{i}.R_{j}}\right) $ denotes a slater determinant wavefunction with volume-law entanglement, state $|\phi_R\rangle$ is a state in the Hilbert space of light particles with area-law entanglement,  and $p(\{R\})$ is some probability distribution over the configurations of the heavy particles. As a specific example, let's now assume that the states $|\phi_R\rangle$ are all product states of the form $|\phi_R\rangle = |\phi_R\rangle_A  |\phi_R\rangle_B$ where $|\phi_R\rangle_A$ and $|\phi_{R'}\rangle_A$ are orthonormal whenever $\{R\}$ and $\{R'\}$ are distinct. Similarly, $|\phi_R\rangle_B$ and $|\phi_{R'}\rangle_B$ are also orthonormal. Then the density matrix for light particles is  given by $\rho = \sum_R p(R) |\phi_R\rangle \langle\phi_R|$, which is clearly separable. The mutual information, on the other hand, is given by $- \sum_R p(R) \log p(R)$, which is volume-law since the number of distinct states in the set $\{R\}$ scale exponentially with the system size. We note that an explicit demonstration of the area-law subsystem negativity for QDL-like states was provided in Ref.\cite{ben2020disentangling} in a 1D Hubbard model supplemented with a nearest-neighbor repulsive interaction \cite{garrison2017partial}.

As a final example, consider an initial state which does not have a sharply defined energy density with respect to a non-integrable Hamiltonian $H$. To be concrete, let's assume that the initial state has a support over two distinct energy densities which correspond to inverse temperatures $\beta_1$ and $\beta_2$. Unitary evolution of this state with $H$  for sufficiently long time will lead to a reduced density matrix of a region $AB$ (with $V_{AB}/V \ll 1$) that may be appoximated as: $\rho_{AB} \approx p \frac{e^{-\beta_1 H_{AB}}}{Z_{AB}(\beta_1)} + (1-p) \frac{e^{-\beta_2 H_{AB}}}{Z_{AB}(\beta_2)} $. Here $Z_{AB}$ denotes the partition function, and $0 < p < 1$. By a similar argument we utilized before\cite{footnote:bound}, one finds that the negativity of this state is area-law. However, the mutual information is generically expected to be volume-law. This can be seen by explicitly calculating the mutual second Renyi entropy between $A$ and $B$, or alternatively by noticing that the logarithm of $\rho_{AB}$ yields a highly non-local Hamiltonian.

\section{Discussion and summary} \label{sec:summary}

In this work, using analytical arguments and exact digonalization studies, we provided evidence that the subsystem negativity $E_N$ between two regions $A, B$ in a tripartite system is a useful quantity to distinguish three classes of systems: (a) systems that satisfy ETH and can therefore act as their own heat bath, (b) systems with well-defined quasiparticles, and (c) systems that many-body localize. For self-thermalizing eigenstates, $E_N$ exhibits the area-law scaling for $V_{AB}/V<1/2$ and the volume-law scaling for $V_{AB}/V>1/2$. In strong contrast, for eigenstates of an integrable system, either non-interacting or interacting, we find a volume-law scaling in $E_N$ for arbitrary $V_{AB}/V$. In support of our numerical evidence, we analytically calculated the volume-law coefficient of negativity, averaged over all free fermion eigenstates, and showed that it satisfies a volume-law scaling. We also provided evidence that similar distinction holds for long-time evolved states starting from a product state, and used the quasiparticle picture to understand the volume-law scaling in a system with quasiparticles. Finally, we provided evidence that for an MBL phase in one spatial dimension, $E_N$ of long-time evolved states shows a volume-law scaling for any $V_{AB}/V$, similar to the integrable models. We also calculated a Renyi version of subsystem negativity analytically for random Haar states and found that they show a transition from being zero to following a volume law as the `subsystem volume' (= logarithm of the subsystem Hilbert space dimension) across half of the total system volume. The eignstates of MBL of course show an area-law scaling for any $V_{AB}/V$. See Fig.\ref{fig:intro} for a summary.

We note that there are several other diagnostics that distinguish between integrable systems and non-integrable systems, including level statistics\cite{Berry_level_statistics_1977,Schmit_level_statistics_1984}, spectral form factor\cite{Hikami_ssf_1997,Shenker_ssf_2017,Liu_ssf_2018,prozen_sff_2018}, average entanglement entropy of eigenstates\cite{rigol_entanglement_entropy_2019}, growth of operator space entanglement entropy from simple local operators \cite{Zanardi_osee_2001,prozen_osee2007,Alba_osee_2019,Alba_osee_2020}, diagonal entropy in quantum quenches\cite{rigol_diag_entropy_2011,rigol_diag_entropy_2016}, mutual information in quantum quenches\cite{Alba_mutual_information_2019}, entanglement revival\cite{Alba_revival2020}, tripartite mutual information of local operators or unitary time-evolution operator\cite{Yoshida_chaos_2016,ryu_tripartite_2020},out-of-time-order correlator\cite{otoc_1969_larkin,Stanford_otoc_2015,Stanford_otoc_2016,yoshida_chaos_2017}. Our diagnostic is sensitive to presence/absence of quasiparticles but not scrambling, and requires time evolution only up to a time-scale that is polynomial in system size. The fact that it probes the presence of quasiparticles is most evident in our calculation for the subsystem negativity in integrable systems using the quasiparticle picture (Sec.\ref{sec:quench}). To see that the protocol is not sensitive to scrambling, consider discrete time evolution in a random Clifford circuit \cite{nahum2017quantum}. Here there are no well-defined quasiparticles but there is no scrambling either. In the steady state, the density matrix of a subregion $AB$ with $V_{AB}/V < 1/2$ is identity, and therefore, $\rho_{AB}$ is separable.  In this sense, our diagnostic is closer in spirit to operator space entanglement and mutual information, although as discussed in Sec.\ref{sec:mutual}, there are cases where mutual information is not a good measure, and the operator space entanglement is known to be not a mixed state entanglement measure either \cite{prozen_osee2007}.

Although for non-thermalizing systems we found that they always obey a volume law, and thus do not show a transition from area to volume law at $V_{AB}/V = 1/2$, there is still a possibility that they exhibit a weaker singularity in the coefficient of the volume law at $V_{AB}/V = 1/2$. An example of such a weaker singularity in an integrable system is provided by the bipartite entanglement in a one dimensional random quadratic fermion Hamiltonians studied in Ref.\cite{ydba2020eigenstate}, whose closed form expression was argued to be: $S = \left[1 - \frac{1+ f^{-1}(1-f) \log(1-f) }{\log(2)}\right] L_A \log(2)$ where $L_A$ is the subsystem size and $f$ is the subsystem fraction. Expanding this expression around $f = 1/2$, one notices that its $n$th derivatives for odd $n\geq 3$ are discontinuous at $f = 1/2$. Therefore, the mutual information between two regions $A$ and $B$ in one dimension would be singular at $L_{AB}/L = 1/2$ despite remaining a volume law for all $L_{AB}/L$. One may ask whether an analogous singularity exists in subsystem negativity.

A basic point that remains to be understood is the magnitude of the volume-law coefficient in both integrable (for all $V_{AB}/V$) and non-integrable (for $V_{AB}/V > 1/2$) systems. Relatedly, it will be of interest to extend our calculation for the third Renyi negativity in a tripartite ergodic state (Sec.\ref{sec:eth}) to arbitrary Renyi index so that it can be analytically continued to obtain an expression for the entanglement negativity.

Another question which needs further investigation is: for the long-time state evolved from a simple product state in non-integrable systems, what is the critical subsystem size fraction for the area-law to volume-law transition of subsystem negativity? As discussed in Sec.\ref{sec:quenchnonint}, utilizing the results from Refs.\cite{winter_2009_equilibrium,eisert_2019_equilibrium}, we only show the persistence of area-law scaling up to $V_{AB}/V=f^*= \alpha/(2\log 2)$, where $\alpha$ is the volume-law coefficient of the second Renyi entropy of the diagonal ensemble. It would be interesting to pinpoint the exact critical 
fraction $f^*$ in the future.

Another related  direction we did not address is:  to what extent can the features of subsystem negativity in quantum quench of integrable models be captured by Generalized Gibbs Ensemble (GGE) (see Ref.\cite{vidmar_2016_gge} for a review). It is known that GGE faithfully describes the long-time expectation values of local observables. It is then natural to ask whether GGE captures the nonlocal entanglement measured by negativity between two subsystem as well. It is natural to suspect that the extensive number of conserved quantities is responsible for the extensive negativity. The bound on negativity in Ref.\cite{sherman2016} is proportional to the number of terms in the entanglement Hamiltonian of $AB$ that cuts across the entanglement boundary between $A$ and $B$ (see, e.g., Eq.\ref{main:eigen_bound}), which may seem to suggest that GGE implies an extensive negativity. However, this approach only leads to an \textit{upper} bound, and is thus not directly helpful to show an extensive negativity.

Finally, it will be worthwhile to find experimental protocols to construct states with low mixed-state entanglement but large mutual information, perhaps along the lines discussed in Sec.\ref{sec:mutual}.

\acknowledgements{We are grateful to John McGreevy, Marcos Rigol, Hassan Shapourian, Xue-Yang Song, Ruben Verresen, Yi-Zhuang You for discussions, and especially Bowen Shi for pointing out a misconception on entanglement of stabilizer states. T. Grover is supported by an Alfred P. Sloan Research Fellowship and National Science Foundation under Grant No. DMR-1752417. T.-C. Lu acknowledges support from KITP Graduate Fellows Program and Graduate Student Research support from the University of California's Multicampus Research Programs and Initiatives (MRP-19-601445).}

\bibliography{v1bib}

\begin{thebibliography}{100}
\bibitem{footnote:bound}
Writing $\omega_{AB} = \sum_m |c_m|^2 \rho_{AB,m} $, where $\rho_{AB,m}$ is the
  reduced density matrix on $AB$ of the eigenstate $\ket{m}$, i.e.
  $\rho_{AB,m}= \tr_C \left( \ket{m}\bra{m} \right) $, one finds $\norm{
  \omega_{AB}^{T_B} }_1 = \norm{ \sum_m |c_m|^2 \rho^{T_B}_{AB,m} }_1 \leq
  \sum_m |c_m|^2 \norm{ \rho^{T_B}_{AB,m} }_1 \leq \text{Max}_{\ket{m}} \norm{
  \rho^{T_B}_{AB,m} }_1 $. Since all eigenstates satisfy area-law subsystem
  negativity for $V_{AB}/V<1/2$, $\log \norm{ \omega_{AB}^{T_B} }_1 $ obeys an
  area-law scaling as well.
\end{thebibliography}
\renewcommand\refname{Reference}
\bibliographystyle{unsrt}
\newpage

\appendix

\onecolumngrid

    \section{Renyi negativity of random pure states}
    \subsection{Calculation of volume-law coefficients}\label{appendix:random_renyi}
    We consider a system consisting of $V$ spins, and define a pure state $\ket{\psi} =  \sum_{i} \psi_i \ket{i} $, where $\ket{i}$ is an arbitrary orthonormal basis, and $\{\psi_i   \}$ is randomly sampled from the probability distribution: $ P( \{\psi_i \}  ) \sim  \delta( 1- \sum_i \abs{\psi}^2  ) $. Dividing the system into three parts labeled by $A$, $B$, and $C$ with $V_{AB}/2$, $V_{AB}/2$, and $V_C$ number of spin-1/2 particles, we here calculate the Renyi negativity $R_n$ between $A$ and $B$. $R_n$ with integer order $n$ ($n>2$) is defined as
    \begin{equation}
    	R_n= b_n\log \left\{  \frac{ \tr\left[ \left(\rho^{T_B}_{AB}  \right)^{n}  \right] }{  \tr \rho^{n}_{AB}} \right\}, 
    \end{equation}
    where $\rho_{AB}= \tr_C{ \ket{\psi } \bra{\psi} }$ is the reduced density matrix on $AB$, and $b_n=\frac{1}{1-n}, \frac{1}{2-n}$ for odd $n$ and even $n$ respectively.

\begin{figure}
    	\centering
    	\begin{subfigure}[b]{0.30\textwidth}
    		\includegraphics[width=\textwidth]{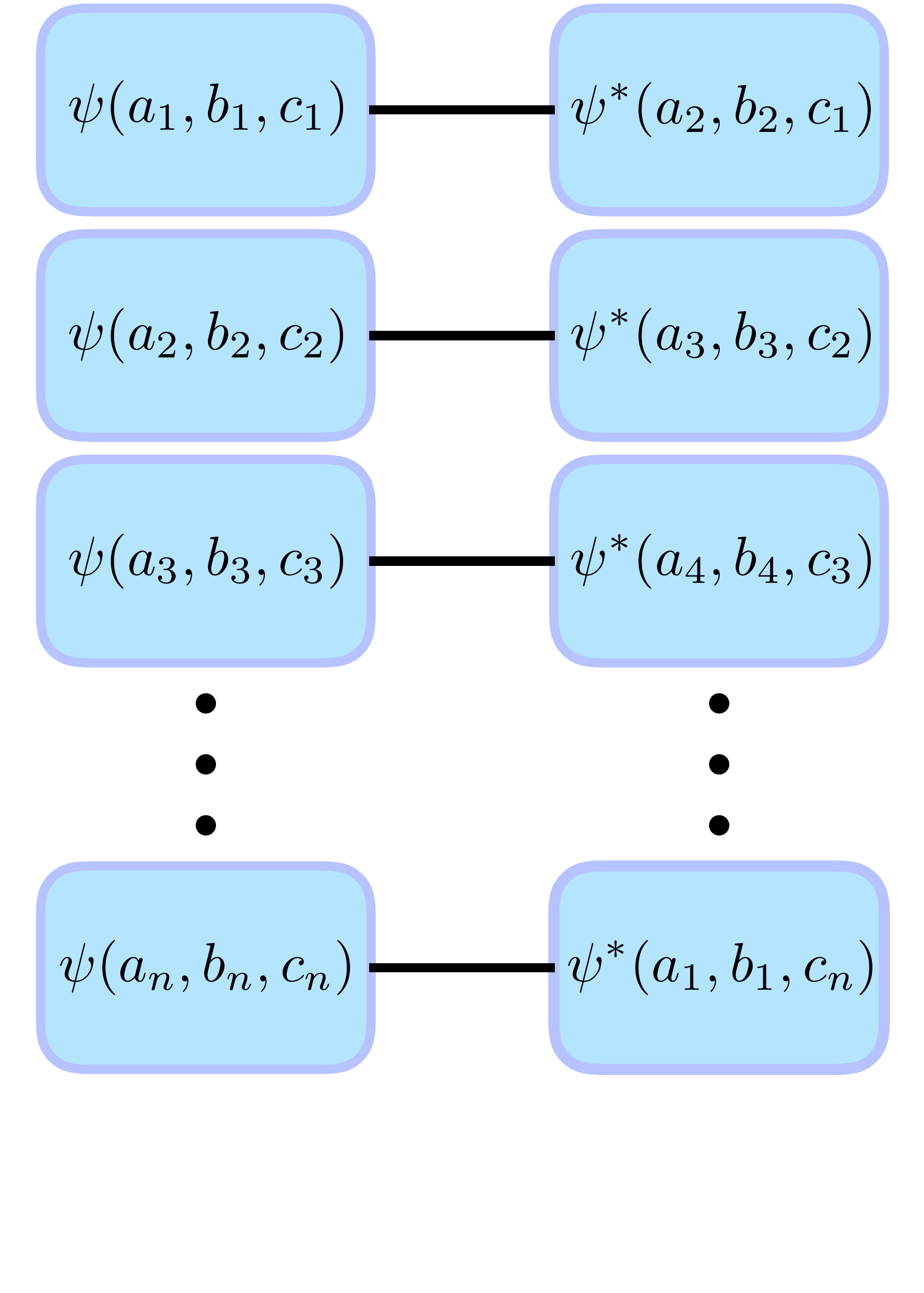}
    		\caption{$V_{AB}< V/2$    }
    		\label{fig:moment_less}
    	\end{subfigure}
    	\begin{subfigure}[b]{0.30\textwidth}
    		\includegraphics[width=\textwidth]{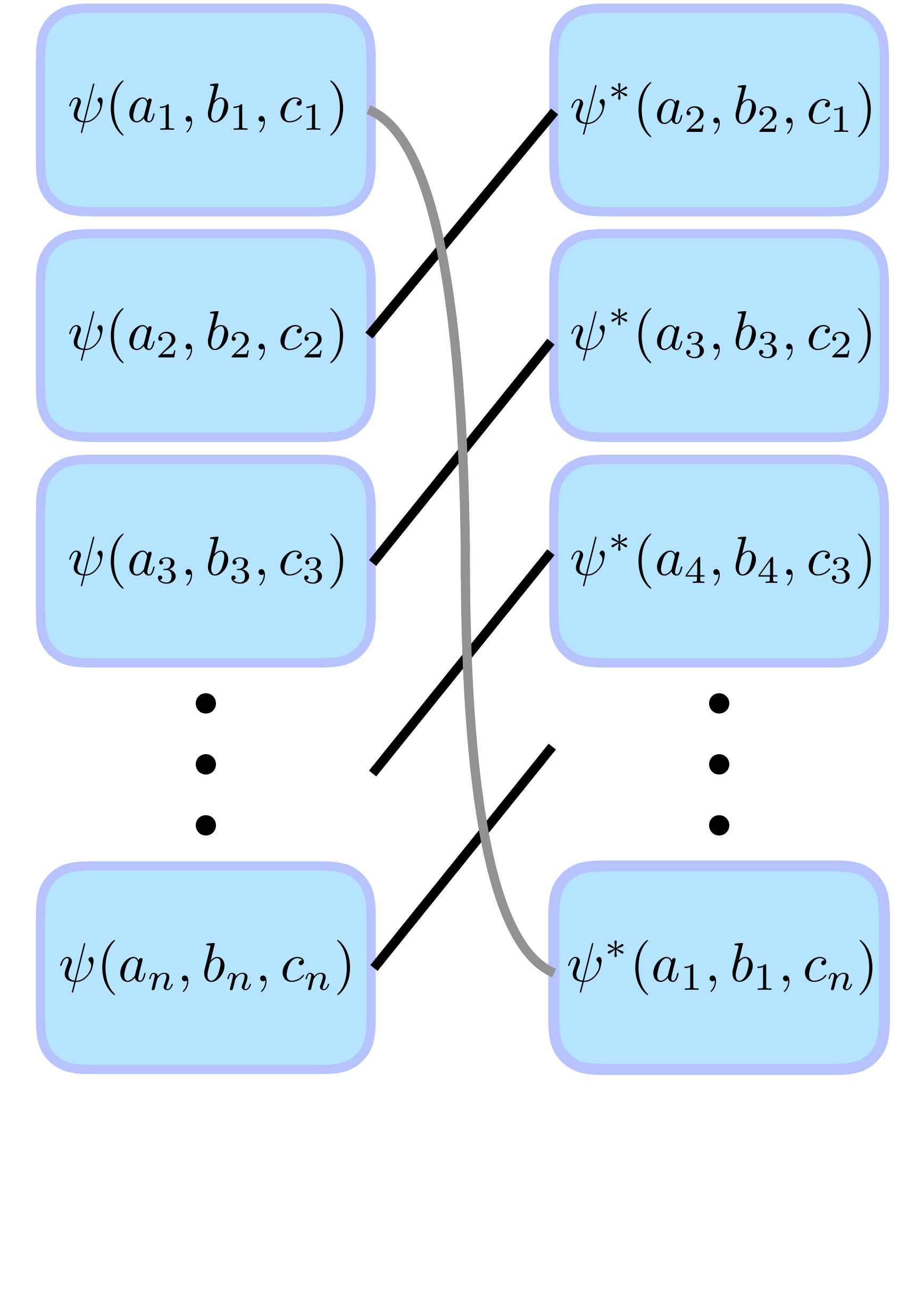}
    		\caption{$V_{AB}> V/2$    }
    		\label{fig:moment_greater}
    	\end{subfigure}
    	\caption{ Dominating terms in $ \overline{\tr \rho_{AB}^n }  $     }   
    \end{figure}

    Before proceeding to the calculation, we recall that given $ P( \{\psi_i \}  ) \sim  \delta( 1- \sum_i \abs{\psi}^2  )$, as the total Hilbert space dimension $d\to  \infty $, one has $\expval{ \psi_i^{*}\psi_j  }=  \frac{1}{d} \delta_{ij}$, $\expval{ \psi_i   \psi_j  }=0$, and any $2N$-point functions of finite $N$ follow the Wick's theorem: 
    
    \begin{equation}
    	\expval{  \prod_{n=1}^N\left( \psi_{i_n}  \psi^*_{j_n}    \right) }   = \sum_{\sigma} \prod_{n=1}^N \expval{ \psi_{i_n }  \psi^{*}_{j_{\sigma(n)}}   },
    \end{equation}
    where $N!$ possible permutations $\sigma$ are summed over. Using these results, we calculate the ensemble average $\overline{\tr \rho^{n}_{AB}}=\sum_{     \{a_i,b_i,c_i| i=1,\cdots, n  \}}     \overline{\prod_{i=1}^n\left[ \psi(a_i,b_i,c_i)  \psi^*(a_{i+1},b_{i+1},c_i)  \right]}$, where all possible Wick contracting terms contribute. As $V\to \infty$ at a fixed subsystem fraction, only one type of terms (may have degeneracy) dominates.  The leading-order contractions for $V_{AB}<\frac{1}{2} V$ and  $V_{AB}>\frac{1}{2} V$ are shown in Fig.\ref{fig:moment_less} and Fig.\ref{fig:moment_greater}, which give
    \begin{equation}
    	d^{n}  \tr \rho^{n}_{AB}= \begin{cases}
    		2^{V_A+V_B +nV_C} \quad \text{for }~ V_{AB}<\frac{1}{2} V\\  2^{n(V_A+V_B) +V_C} \quad  \text{for }~ V_{AB}>\frac{1}{2} V
    	\end{cases}
    \end{equation}

    \begin{figure}
    	\centering
    	\begin{subfigure}[b]{0.30\textwidth}
    		\includegraphics[width=\textwidth]{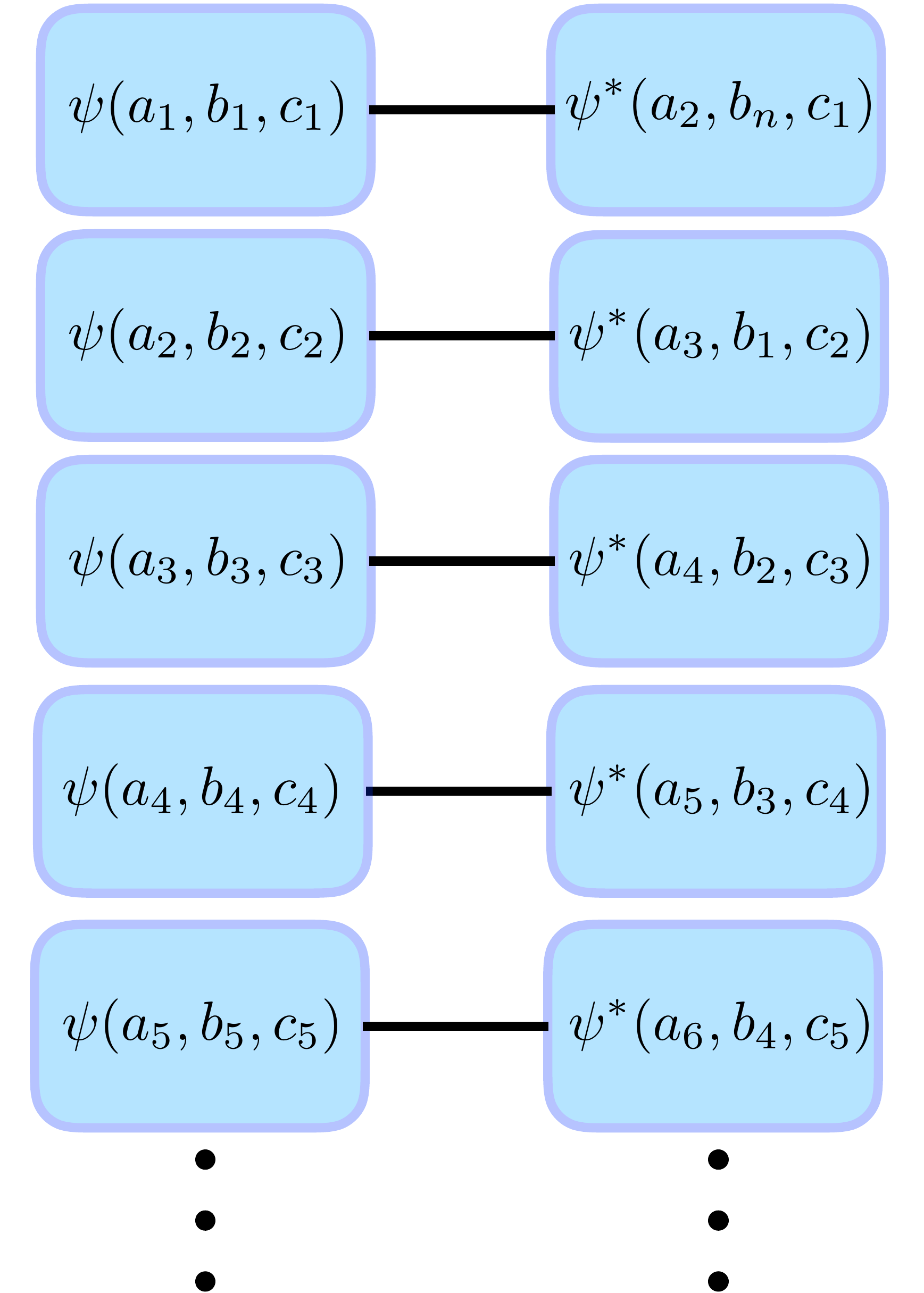}
    		\caption{$V_{AB}< V/2$    }
    		\label{fig:moment_transposed_less}
    	\end{subfigure}
    	\begin{subfigure}[b]{0.30\textwidth}
    		\includegraphics[width=\textwidth]{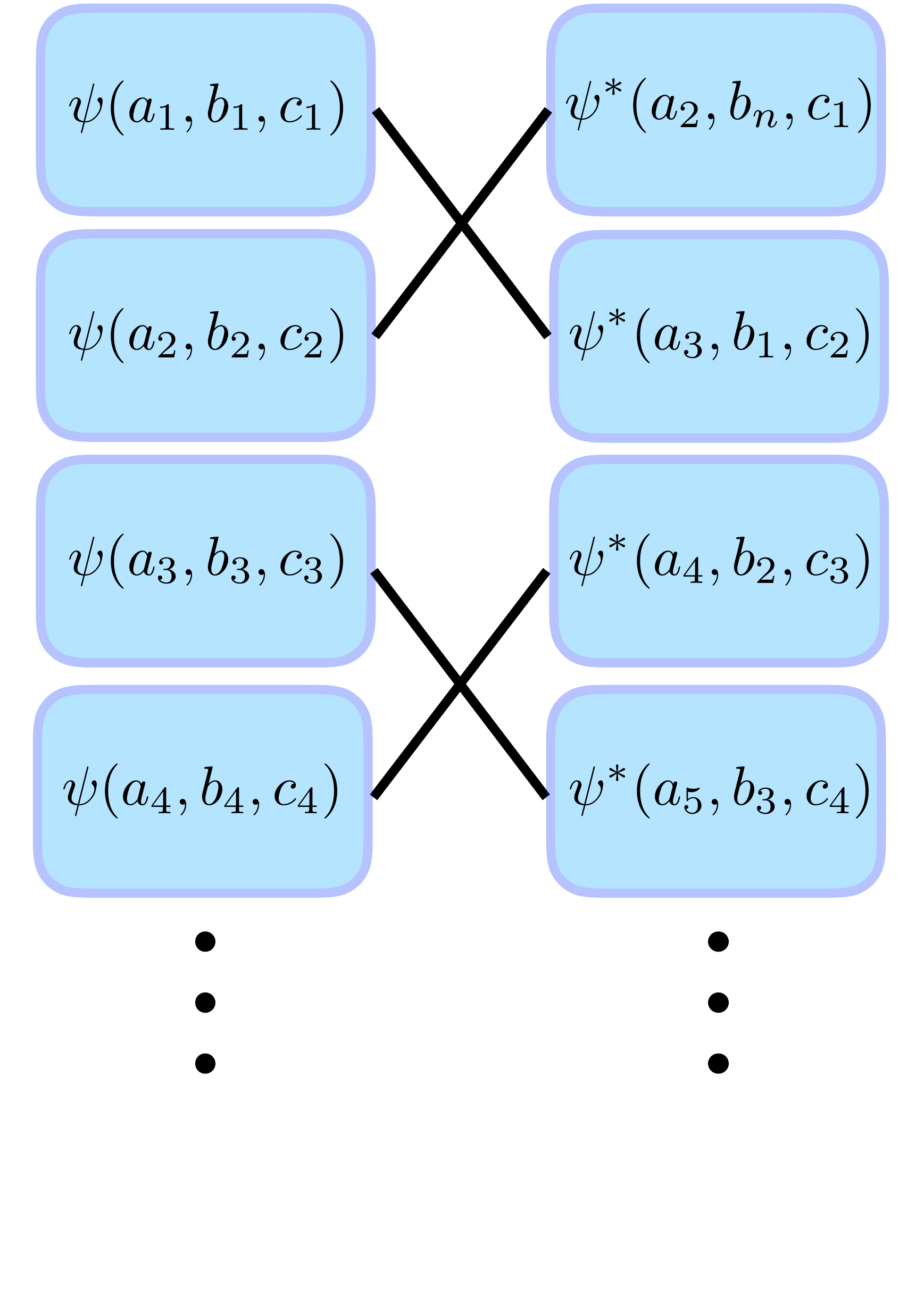}
    		\caption{$V_{AB}> V/2$; even $n$    }  \label{fig:moment_transposed_even}
    	\end{subfigure}
    	\begin{subfigure}[b]{0.30\textwidth}
    		\includegraphics[width=\textwidth]{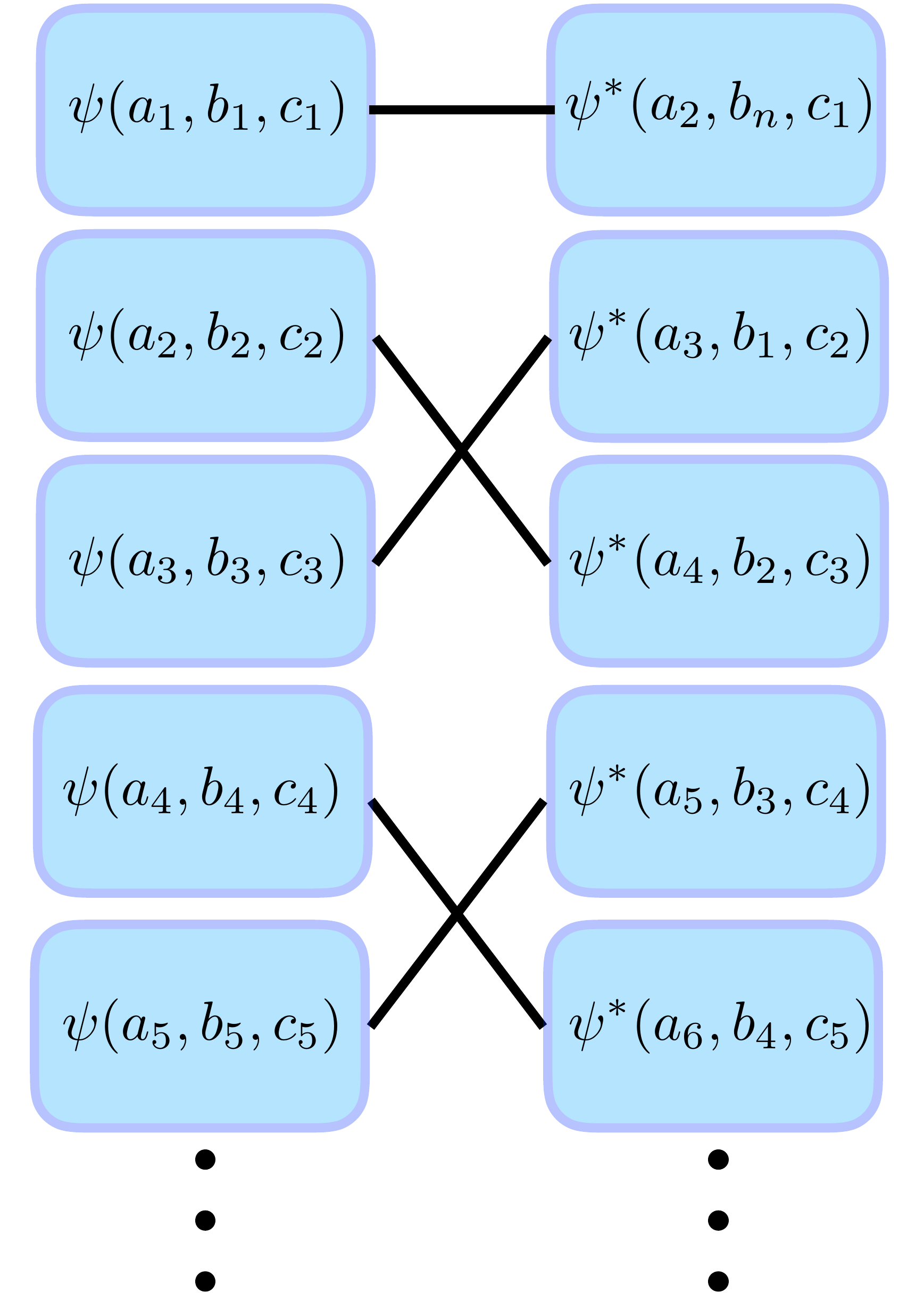}
    		\caption{$V_{AB}> V/2$; odd $n$    }  
    		\label{fig:moment_transposed_odd}
    	\end{subfigure}
    	\caption{ Dominating terms in $ \overline{ \tr\left[ \left(\rho^{T_B}_{AB}  \right)^{n}  \right] }   $       } 
    \end{figure}
    
    Now we calculate $ \tr\left[ \left(\rho^{T_B}_{AB}  \right)^{n}  \right]  =  \sum_{     \{a_i,b_i,c_i| i=1,\cdots, n  \}}   \prod_{i=1}^n  \left[       \psi(a_i,b_i,c_i)   \psi^*(a_{i+1},b_{i-1},c_i)\right]$. For $V_{AB}<\frac{1}{2}V$, the dominating term is given by Fig.\ref{fig:moment_transposed_less}, giving $d^{n}  \tr\left[ \left(\rho^{T_B}_{AB}  \right)^{n}  \right]= 2^{V_A+V_B+nV_C}$. On the other hand, the leading-order contraction pattern for $V_{AB}>\frac{1}{2}V$ depends on the parity of $n$. For even $n$, the pattern is given by Fig.\ref{fig:moment_transposed_even}. which results in a weight $2^{ (  \frac{n}{2}+1  ) (V_A+V_B) +\frac{n}{2} V_C }$. Note that one can vertically shift this contraction pattern by one replica to generate another contraction pattern of the same weight, contributing a factor of $2$ degeneracy. Thus $ d^{n} \tr\left[ \left(\rho^{T_B}_{AB}  \right)^{n}  \right]=2\cdot2^{ (  \frac{n}{2}+1  ) (V_A+V_B) +\frac{n}{2} V_C }$. 
    
    For odd $n$ at $V_{AB}>\frac{1}{2}V$, the leading-order contraction is given by Fig.\ref{fig:moment_transposed_odd}, giving the weight $2^{  \frac{n+1}{2}(V_A+V_B) +\frac{n+1}{2} V_C  }$. Note that the degeneracy is $n$ since there are $n$ possible choices for the horizontal contraction.  
    
    In sum, 
    \begin{equation}
    	\text{for even} ~n, \quad d^n  \tr\left[ \left(\rho^{T_B}_{AB}  \right)^{n}  \right]= \begin{cases}
    		2^{V_A+V_B +nV_C} \quad \quad \quad~~\quad  ~\text{for }~ V_{AB}<\frac{1}{2} V\\
    		2\cdot2^{ (  \frac{n}{2}+1  ) (V_A+V_B) +\frac{n}{2} V_C } \quad  \text{for }~ V_{AB}>\frac{1}{2} V
    	\end{cases}
    \end{equation}
    
    \begin{equation}
    	\text{for odd} ~n, \quad  d^n\tr\left[ \left(\rho^{T_B}_{AB}  \right)^{n}  \right]= \begin{cases}
    		2^{V_A+V_B +nV_C} \quad \quad\quad\quad\text{for }~ V_{AB}<\frac{1}{2} V\\
    		n 2^{  \frac{n+1}{2}(V_A+V_B) +\frac{n+1}{2} V_C  }\quad  \text{for }~ V_{AB}>\frac{1}{2} V
    	\end{cases}
    \end{equation}
    Combining the above equations, one finds:\\
    for even $n$
    \begin{equation}
    	\boxed{
    		R_n= \begin{cases}
    			0 \quad \text{for} ~ V_{AB}< V_C\\
    			\frac{1}{2}\left( V_{AB}-V_C \right)\log 2 - \frac{1}{n-2} \log 2  \quad \text{for} ~ V_{AB}> V_C\\
    	\end{cases}}
    \end{equation}
    for odd $n$: 
    \begin{equation}
    	\boxed{
    		R_n= \begin{cases}
    			0 \quad \text{for} ~ V_{AB}< V_C\\
    			\frac{1}{2}\left( V_{AB}-V_C \right)\log 2 - \frac{1}{n-1} \log n  \quad \text{for} ~ V_{AB}> V_C\\
    	\end{cases}}
    \end{equation}
    In other words, $R_n$ for $n>2$ has the same volume-law coefficient as entanglement negativity.

    \begin{subsection}{Proof that $ \abs{    \log \overline{\tr \rho_{AB}^n} - \overline{  \log \tr \rho_{AB}^n    }  }$ and $  \abs{     \log \left\{  \overline{\tr\left[ \left(\rho^{T_B}_{AB}  \right)^{n}  \right]   }  \right\} -  \overline{   \log \left\{  \tr\left[ \left(\rho^{T_B}_{AB}  \right)^{n}  \right] \right\}    }       }$ are exponentially small in the system volume}\label{appendix:prove_random_eq}
    	
    	The proof presented here is analogous to Ref.\cite{Lu_renyi_2019}, which we outline below. First we write

    	\begin{equation} \label{appendix:moment_proof}
    		\tr \rho_{AB}^n   = \overline{ \tr \rho_{AB}^n  } +  \left(  \tr \rho_{AB}^n  -  \overline{ \tr \rho_{AB}^n  } \right) = \overline{ \tr \rho_{AB}^n  }  \left( 1+x    \right), 
    	\end{equation}
    	where $x=  \frac{ \tr \rho_{AB}^n    }{   \overline{ \tr \rho_{AB}^n  }  }  -1  $. It follows that

    	\begin{equation}
    		\overline{  \log \tr \rho_{AB}^n    }    =    \log \overline{\tr \rho_{AB}^n}  + \overline{ \log (1+x)  },
    	\end{equation}
    	where the last term is the difference between two kinds of averages, and we calculate the variance of $x$ to show such difference is exponentially small. By definition $\overline{x}=0$, and the variance is
    	\begin{equation}
    		\overline{x^2  }   =   \overline{ \left(   \frac{  \tr \rho_{AB}^n  }{    \overline{  \tr \rho_{AB}^n  }     }  -1     \right)^2      }  =   \frac{ \overline{   \left(   \tr \rho_{AB}^n     \right)^2   }  }{  \left(  \overline{      \tr \rho_{AB}^n     }     \right)^2 }-1.
    	\end{equation}
    	
    	Given $
    	\tr \rho^{n}_{AB}=\sum_{     \{a_i,b_i,c_i| i=1,\cdots, n  \}}     \prod_{i=1}^n\left[ \psi(a_i,b_i,c_i)  \psi^*(a_{i+1},b_{i+1},c_i)  \right]$, we first consider the case for $V_{AB}/V <1/2$. Taking an ensemble average, Wick's theorem implies that the next-leading order term must be exponentially small in the system volume. Therefore, 
    	\begin{equation}
    		\overline{ \tr \rho_{AB}^n  }  = d^{-n} 2^{n(V_A+V_B) +V_C} \left[  1+  O(e^{-\alpha_1 V })    \right], 
    	\end{equation}
    	and 
    	\begin{equation}
    		\left(\overline{ \tr \rho_{AB}^n  } \right)^2 = d^{-2n}   \left(  2^{n(V_A+V_B) +V_C} \right)^2 \left[  1+  O(e^{-\alpha_2 V })    \right]. 
    	\end{equation}
    	On the other hand, $  \left(  \tr \rho_{AB}^n   \right)^2$ gives two copies
    	\begin{equation}
    		\left(  \tr \rho_{AB}^n   \right)^2  = \sum_{     \{a_i,b_i,c_i| i=1,\cdots, n  \}}     \prod_{i=1}^n\left[ \psi(a_i,b_i,c_i)  \psi^*(a_{i+1},b_{i+1},c_i)  \right]      \sum_{     \{a'_i,b'_i,c'_i| i=1,\cdots, n  \}}     \prod_{i=1}^n\left[ \psi(a'_i,b'_i,c'_i)  \psi^*(a'_{i+1},b'_{i+1},c'_i)  \right]. 
    	\end{equation}
    	Taking an average, the leading-order contraction of this $4n$-point function will be the leading-order contraction of the $2n$-point function from each copy, i.e. two copies decouple at the leading order. Consequently, 
    	
    	\begin{equation}
    		\overline{    \left(  \tr \rho_{AB}^n   \right)^2    } =     \left(  \overline{  \tr \rho_{AB}^n } \right)^2      \left[  1+ O(e^{-\alpha V})    \right], 
    	\end{equation}  
    	and
    	
    	\begin{equation}
    		\overline{x^2} =  \frac{ \overline{   \left(   \tr \rho_{AB}^n     \right)^2   }  }{  \left(  \overline{      \tr \rho_{AB}^n     }     \right)^2 }-1 = O(e^{-\alpha V}),
    	\end{equation}
    	where $\alpha$ is a positive $O(1)$ constant. This implies there is no fluctuation in $x$ as $V\to \infty$. Therefore, $\overline{  \log \tr \rho_{AB}^n    }   -   \log \overline{\tr \rho_{AB}^n}  = \overline{ \log (1+x)  }$ is exponentially small. Using the same approach, it is straightforward to perform a similar calculation for the partially transposed moment to prove $ \abs{     \log \left\{  \overline{\tr\left[ \left(\rho^{T_B}_{AB}  \right)^{n}  \right]   }  \right\} -  \overline{   \log \left\{  \tr\left[ \left(\rho^{T_B}_{AB}  \right)^{n}  \right] \right\}    }       }$ is exponentially small in the system volume as well.

    \end{subsection}

  \section{Additional numerical data of subsystem negativity}

   \subsection{Finite temperature eigenstates in a non-integrable spin chain}\label{appendix:finite_T_nega}
\begin{figure}
   	\centering
   	\begin{subfigure}[b]{0.3\textwidth}
   		\includegraphics[width=\textwidth]{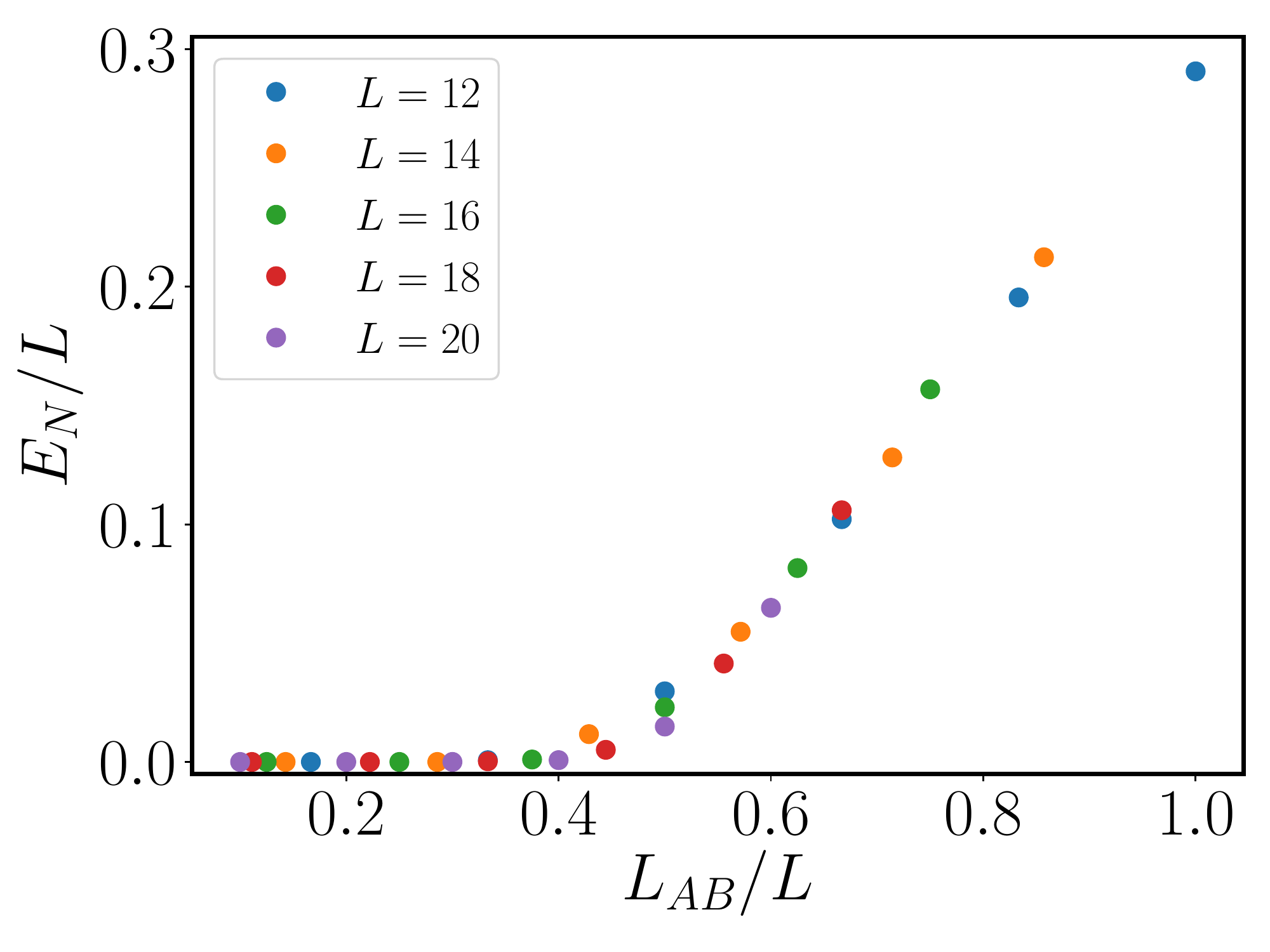}
   		\label{fig:}
   	\end{subfigure}
   	\begin{subfigure}[b]{0.3\textwidth}
   		\includegraphics[width=\textwidth]{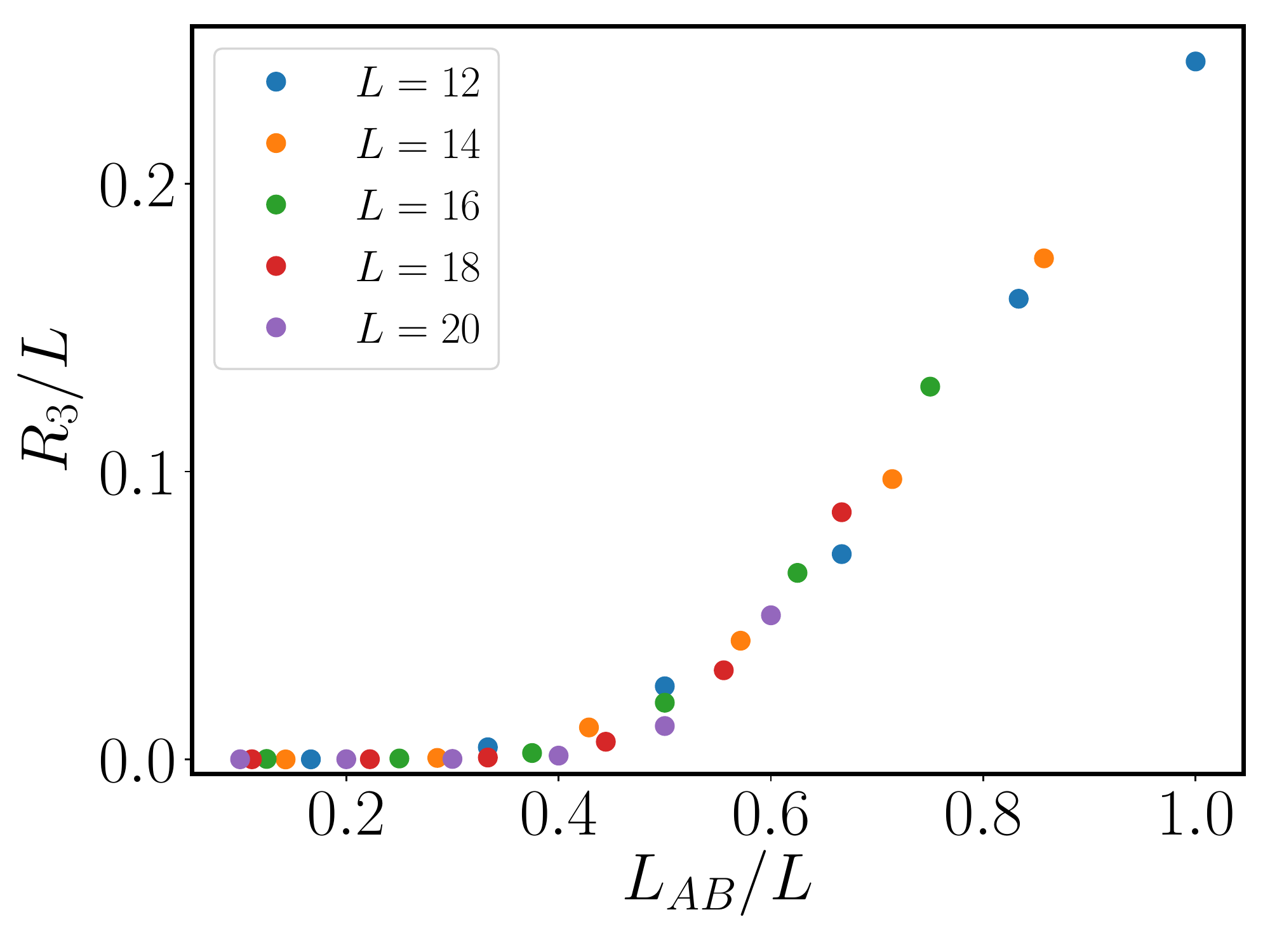}
   		\label{fig:}
   	\end{subfigure}
   	\begin{subfigure}[b]{0.3\textwidth}
   		\includegraphics[width=\textwidth]{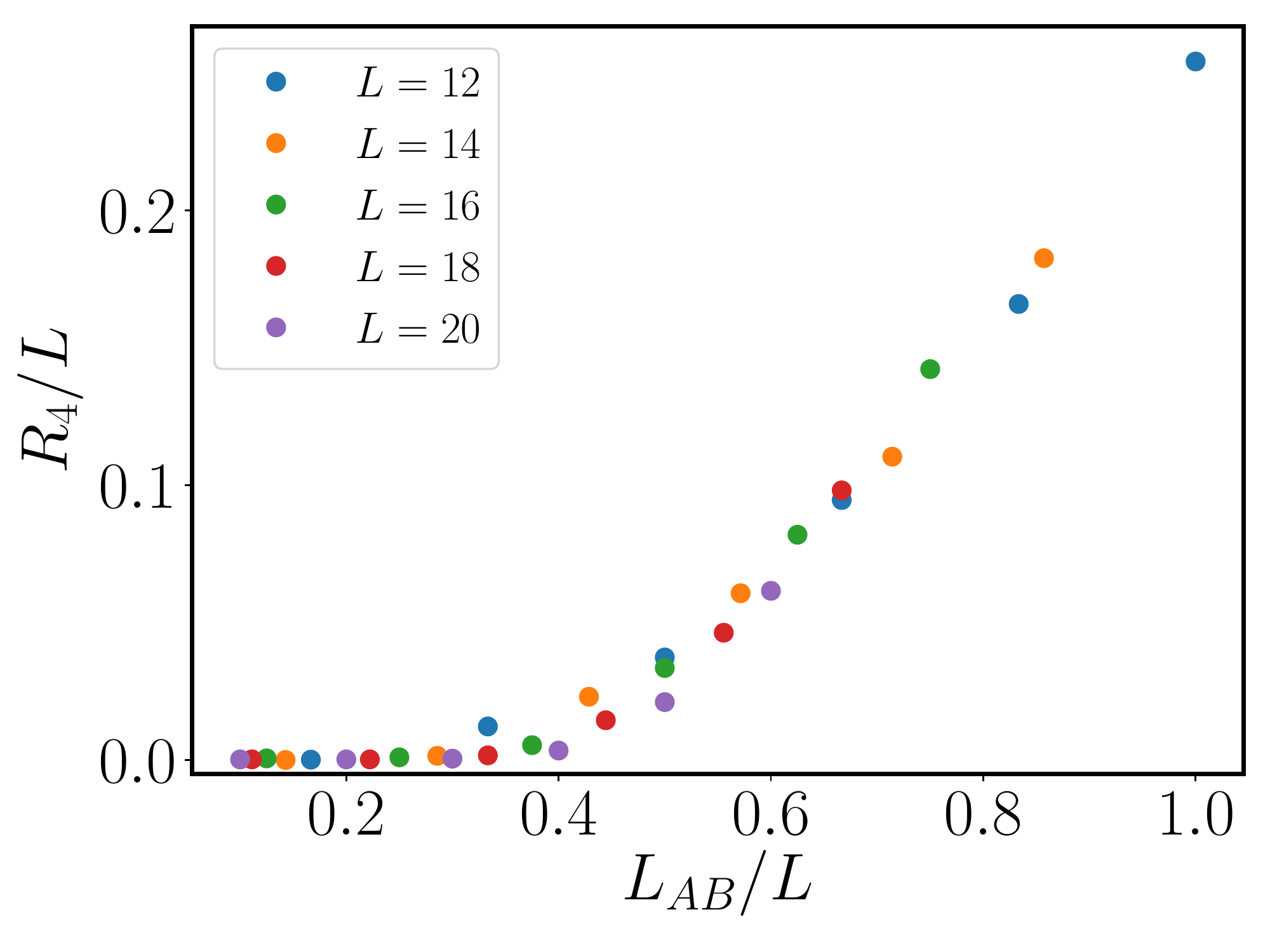}
   		\label{fig:}
   	\end{subfigure}
   	\caption{Subsystem (Renyi) negativity of a single energy eigenstate at inverse temperature $\beta=0.2$ in the non-integrable spin chain.}
   	\label{fig:finite_temp_chaotic_eigen}
   \end{figure}

  Here we report numerical data on negativity between $A$ and $B$ of finite temperature eigenstates in the non-integrable spin chain (Eq.\ref{main_eq:heisenberg} with $J_2=0.8$) in Fig.\ref{fig:finite_temp_chaotic_eigen}.
   
 \subsection{Eigenstates in a U(1) symmetric integrable spin chain}\label{appendix:U(1)_nega}
\begin{figure}
    	\centering
    	\begin{subfigure}[b]{0.3\textwidth}
    		\includegraphics[width=\textwidth]{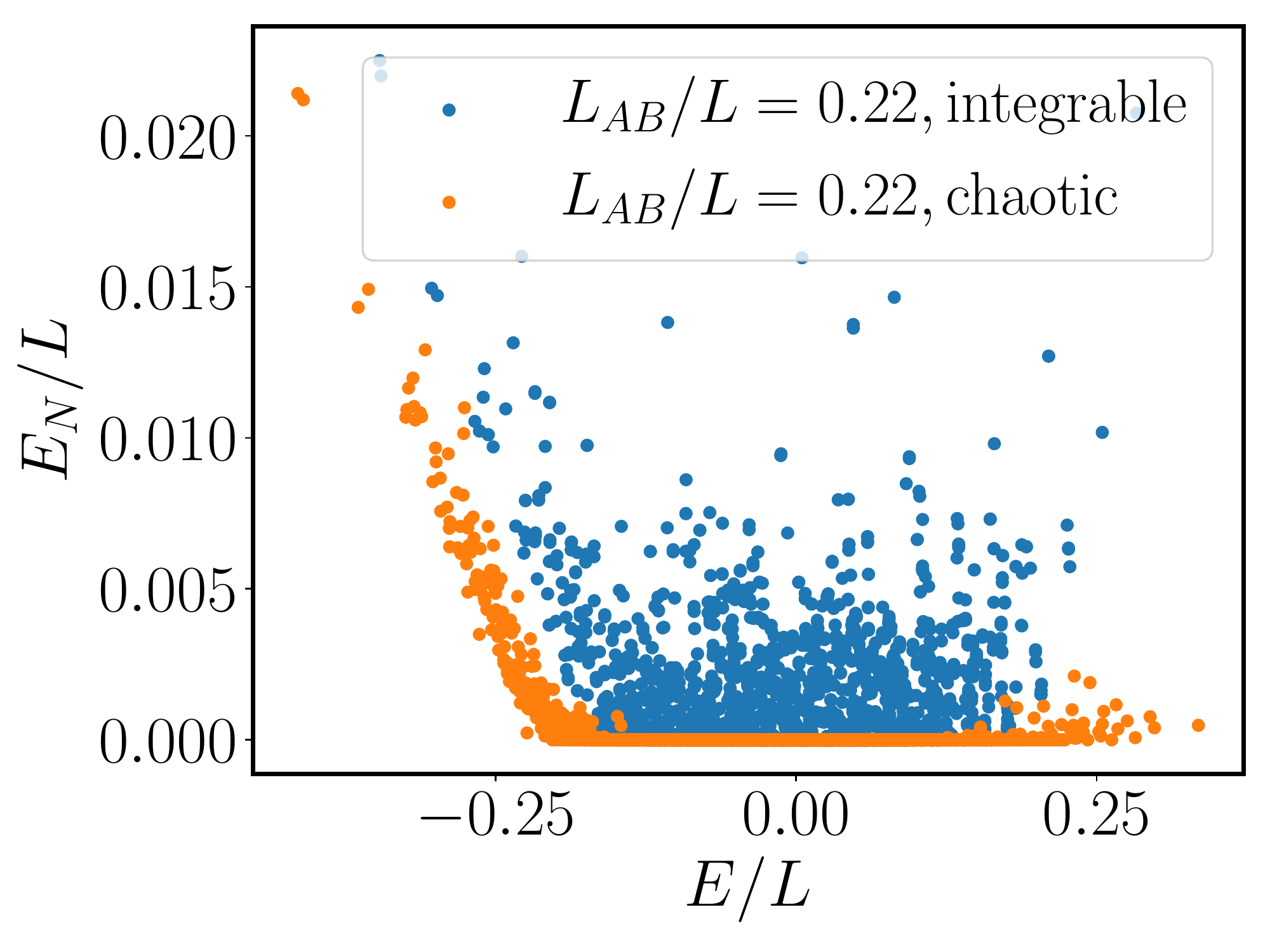}
    		\label{fig:}
    	\end{subfigure}
    	\begin{subfigure}[b]{0.3\textwidth}
    		\includegraphics[width=\textwidth]{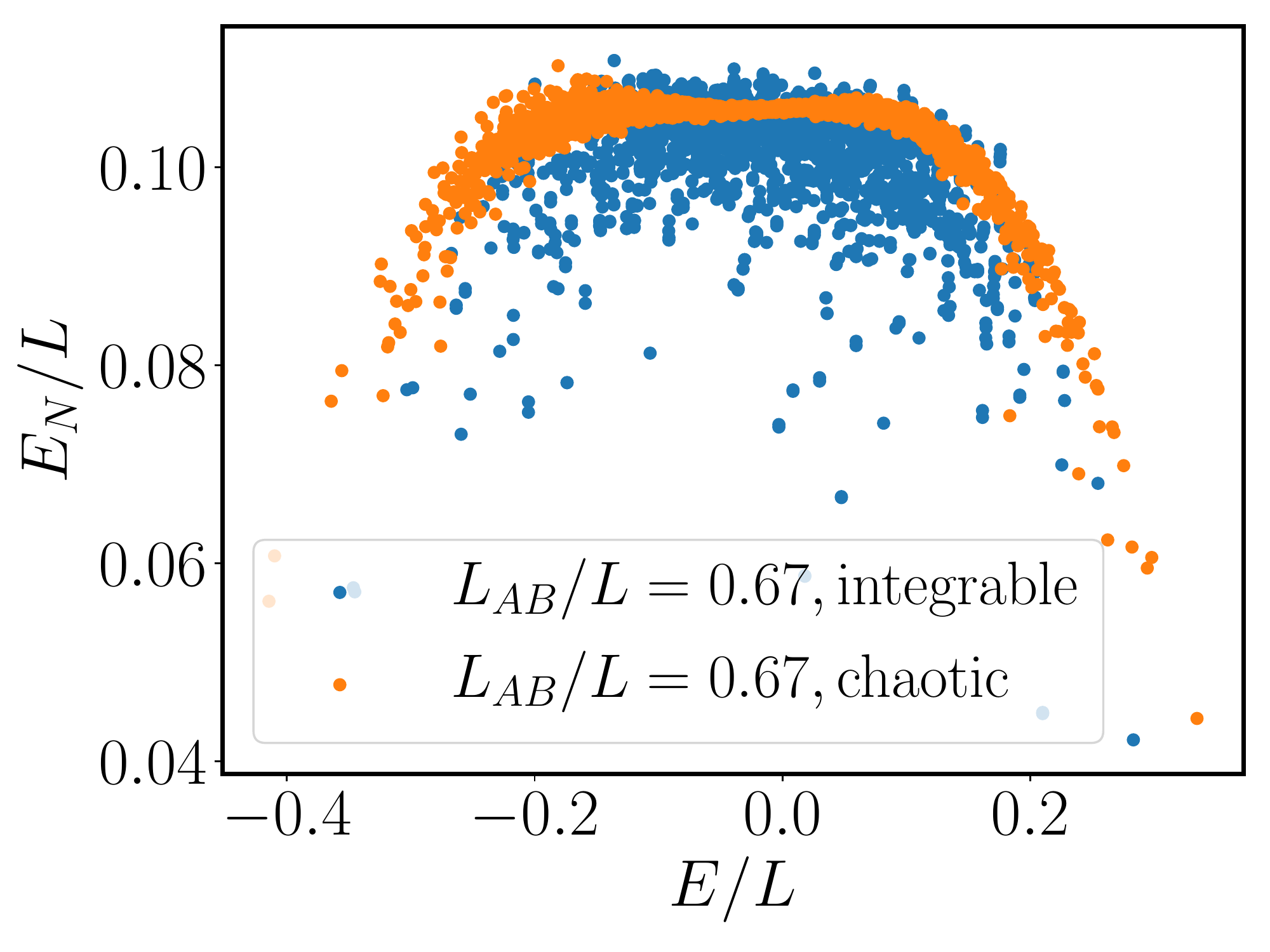}
    		\label{fig:}
    	\end{subfigure}
    	
    	\begin{subfigure}[b]{0.6\textwidth}
    		\includegraphics[width=\textwidth]{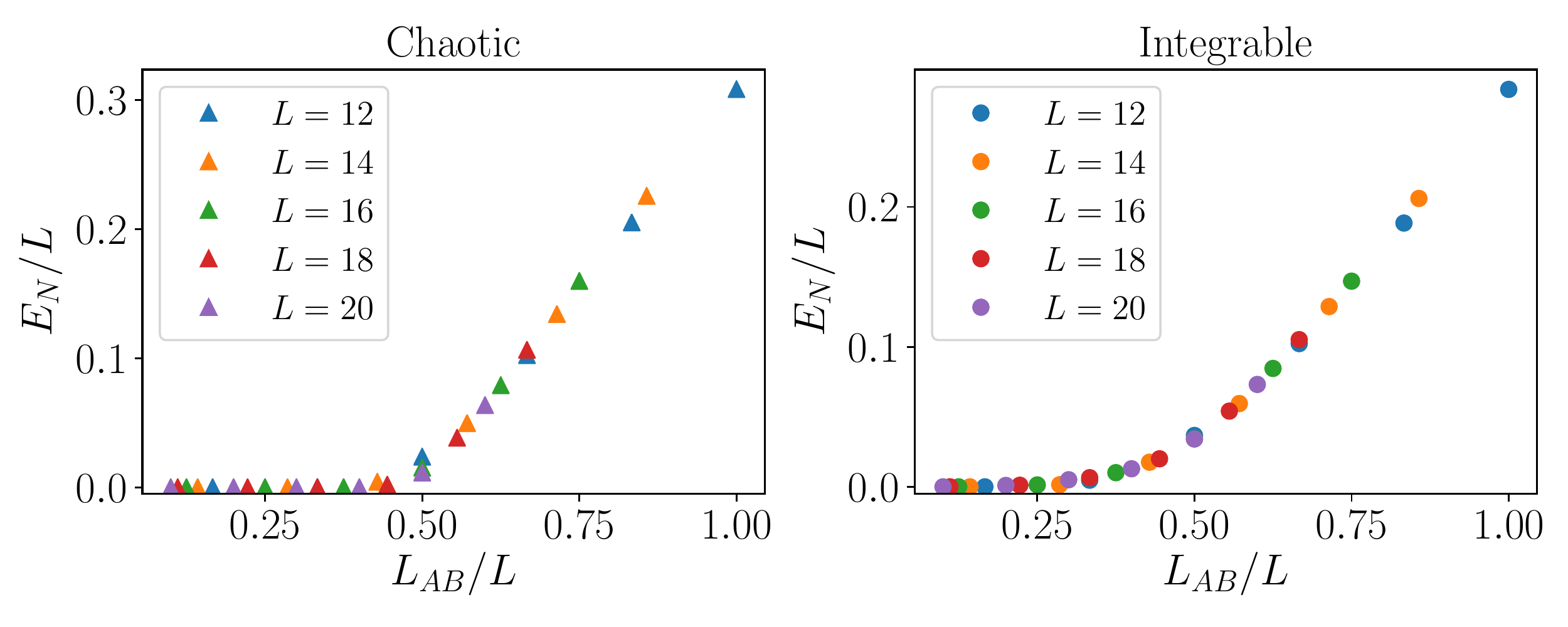}
    		\label{fig:}
    	\end{subfigure}
    	\caption{Comparison of subsystem negativity between a $U(1)$ symmetric integrable spin chain ($\Delta=0.4, J_2=0$) and a chaotic (non-integrable) spin chain ($\Delta=1, J_2=0.8$). Top panel: subsystem negativity of all energy eigenstates. Bottom panel: long-time averaged subsystem negativity in a global quench from the N\'{e}el state.}   
    	\label{fig:U(1) symmetric}
    \end{figure}

    In the main text, we numerically show that the finite-energy density eigenstates of the Heisenberg chain (Eq.\ref{main_eq:heisenberg} with $J_2=0$) exhibit volume-law subsystem negativity. Here we consider an integrable XXZ chain by introducing anisotropy in the Heisenberg chain to break the SU(2) symmetry down to U(1), and provide numerical evidence that subsystem negativity of eigenstates and long-time states in a global quench remains volume-law. Specifically we consider $H= \sum_{i=1}^L S_{i}^{x}S_{i+1}^{x}  +S_{i}^{y} S_{i+1}^{y}   + \Delta S_{i}^{z}  S_{i+1}^{z}  +J_2 S^z_iS^z_{i+2}$, and investigate an integrable point ($\Delta=0.4, J_2=0$) compared with a chaotic Hamiltonian ($\Delta=1, J_2=0.8$). See Fig.\ref{fig:U(1) symmetric} for results.

    \begin{section}{Proof of area-law subsystem negativity for $\frac{V_{AB}}{V}<  \frac{1}{2}$ in chaotic eigenstates}\label{appendix:area_law_proof}
    	
Given a chaotic eigenstate $\ket{\psi}$ with energy $E$ for a local Hamiltonian $H= H_{R}+ H_{\bar{R}}  +   H_{R\bar{R}}  $, subsystem ETH\cite{dymarsky2016subsystem} suggests that for $V_R< V_{\bar{R}}$, the reduced density matrix in $R$ takes the form 
    	\begin{equation}
    		\rho_{R}   =  \frac{1}{\mathcal{N}} \sum_i  e^{ S_{ \bar{R} } (E-E_i^R)   } \ket{i}\bra{i},
    	\end{equation}
    	where $\ket{i}$ is an eigenstate of $H_R$, and $e^{ S_{ \bar{R} } (E-E_i^R)   }  $ is the density of state of $H_{\bar{R}}$ at energy $E-E_i^R$. This equation indicates that the probability weight in $\ket{i}$ is proportional to the number of states in $\bar{R}$ consistent with the energy conservation. Using the expression of $\rho_R$, we will bound the negativity between two complementary subsystems $A,B$ in $R$. To proceed, we expand $ S_{ \bar{R} } (E-E_i^R) $: 
    	\begin{equation}
    		S_{ \bar{R} }(E-E_i^R)  =  \sum_{n=0}^{\infty}  \frac{  \left(-E^R_i  \right)^n   }{n!} \frac{\partial^n S_{\bar{R}}(E) }{  \partial E^n }. 
    	\end{equation}
    	Since microcanonical entropy is extensive, i.e. $S_{\bar{R}}(E) =V_{\bar{R}}  s_{th}( E/V_{\bar{R}}  )      $, one finds
    	\begin{equation}
    		\frac{\partial^n S_{\bar{R}}(E) }{  \partial E^n }   =  \frac{1}{V^{n-1}_{\bar{R}}}  \frac{  \partial^n  s_{th}(E/V_{\bar{R}}   )    }{   \partial \left(  E/ V_{\bar{R}}  \right)^n  }   =   \frac{1}{V^{n-1}_{\bar{R}}}  s^{(n)}_{th}(u),
    	\end{equation}
    	where $u\equiv E/V_{\bar{R}}$.
    	Therefore,
    	
    	\begin{equation}
    		\rho_R  =\frac{1}{Z}  e^{  M              } , \quad \text{where} \quad 	M=\sum_{n=1}^{\infty    }    \frac{ s_{th}^{(n)} \left(-H_{R} \right)^n  }{n!  V_{\bar{R}}^{n-1} }.
    	\end{equation}

 To proceed, we now recall a result from Ref.\cite{sherman2016}: given a thermal state $\rho\sim e^{ -\beta (H_A+H_B +H_{AB}       )  }$, negativity between $A$ and $B$ is bounded by $ E_N  \leq   \beta \left(  JK +   \norm{    H_{AB}    }  \right)$, where $\norm{ \cdots  }$ denotes the operator norm, i.e. the largest singular value of an operator. $K$ is the number of terms when expanding $H_{AB} =   \sum_{\alpha } H^A_{\alpha}  H_{\alpha}^B$, and $J$ is the upper bound of the interaction strength $\norm{H_{\alpha}^{A}}  \norm{H_{\alpha}^{B}} \leq  J  $. Using the triangular inequality, one also has $\norm{H_{AB}} \leq J K $ 
 so \begin{equation}
    		E_N  \leq   2  \beta JK,
    	\end{equation}
    	which results an area law for a local Hamiltonian. To apply this equation, we write $H_R= H_A+ H_B+H_{AB}$, where $H_A(H_B)$ contains the terms acting on region $A(B)$, and $H_{AB}$ denotes the interaction between $A$ and $B$. It follows that the $n$-th order term in $M$ reads 
    	\begin{equation}
    		\frac{  (-H_R)^n    }{  V_{\bar{R}}^{n-1}   } =  \frac{  (-H_A-H_B-H_{AB})^n    }{  V_{\bar{R}}^{n-1}   }. 
    	\end{equation}
    	Taking thermodynamic limit while fixing the subsystem volume fraction, the above quantity can be reduced to
    	\begin{equation}
    		\frac{  (-H_R)^n    }{  V_{\bar{R}}^{n-1}   } = \frac{1}{ V_{\bar{R}}^{n-1}  } \left[  (-H_{AB}) \left(  -H_A-H_B \right)^{n-1} +   \left(  -H_A-H_B \right)     (-H_{AB}) \left(  -H_A-H_B \right)^{n-2}   +\cdots\left(  -H_A-H_B \right)^{n-1}   (-H_{AB})      \right].
    	\end{equation}
    	Hence the $n$-th order term in $M$ contributes to the upper bound by
    	\begin{equation} 
    		\frac{  \abs{ s_{th}^{(n)} }  }{  n!   }   n\left(  \frac{  N_{AB}  }{V_{\bar{R}}}   \right)^{n-1} J^n \abs{ \partial V_{AB}},
    	\end{equation}
where $J$ is the upper bound of interaction strength in $H_R=H_A+H_B+H_{AB}$, $N_{AB}$ is the number of terms in $H_A+H_B$, and $\abs{  \partial V_{AB} }$ is the number of terms in $H_{AB}$. Finally one finds the upper bound of negativity:

    	\begin{equation}
    		E_N   \leq  2J\sum_{n=1}^{\infty}   \frac{   \left(J N_{AB}/V_{\bar{R}}  \right)^{n-1}      }{(n-1)!  }     \abs{ \frac{ \partial^{n} s_{th}(u) }{  \partial u^{n}  } }    \abs{  \partial V_{AB}} =  2J    \sum_{n=0}^{\infty}   \frac{   \left(J N_{AB}/V_{\bar{R}}  \right)^{n}      }{(n)!  }     \abs{ \frac{ \partial^{n+1} s_{th}(u) }{  \partial u^{n+1}  } }\abs{  \partial V_{AB}}.   
    	\end{equation}
    	Define the function $g(u, JN_{AB}/V_{\bar{R}})=  \sum_{n=0}^{\infty}   \frac{   \left(J N_{AB}/V_{\bar{R}}  \right)^{n}      }{(n)!  }     \abs{ \frac{ \partial^{n+1} s_{th}(u) }{  \partial u^{n+1}  } }    $, one finds  
    	
    	\begin{equation}
    		E_N   \leq   2J g(E/V_{\bar{R}}, JN_{AB}/V_{\bar{R}})\abs{\partial V_{AB}}.
    	\end{equation}
    	This completes the proof of area law in negativity. Note that $g(E/V_{\bar{R}}, JN_{AB}/V_{\bar{R}}  )$ is a function obtained by taking absolute value for each term in the Taylor series of $s'_{th}(E/V_{\bar{R}}+ JN_{AB}/V_{\bar{R}}   )    $ about $E/V_{\bar{R}}$. Since functions $g$ and $s'_{th}$, when expressed as power series of $JN_{AB}/V_{\bar{R}}$, have the same coefficient up to a minus sign, they share the same interval of convergence using a ratio test. Therefore, assuming $s(u)$ is an analytic function in $(u_{\text{min}},u_{\text{max}})$, where $u_{\text{min/max  }}$ are the lowest/highest energy density of the Hamiltonian, $s'_{th}(   E/V_{\bar{R}}+ JN_{AB}/V_{\bar{R}}   ) $ and $g(E/V_{\bar{R}}, JN_{AB}/V_{\bar{R}}  )$ will be convergent as long as $E/V_{\bar{R}}+ JN_{AB}/V_{\bar{R}} $ is in $(u_{\text{min}},u_{\text{max}})$. 
    	
    \end{section}

    \section{Subsystem Renyi negativity of ergodic tripartite states}\label{appendix:tripartite}
 Here we calculate the third Renyi negativity $R_3$ between two subsystems using the ergodic tripartite state ansatz: $\ket{ E  } = \sum_{a,b,c}^{'} \psi(a,b,c ) \ket{a}\otimes \ket{b}\otimes  \ket{c}$. The prime symbol in the summation imposes the energy conservation: $f_a u_{a}  +f_b u_{b}  +f_c u_{c} = u$, where $f_{\alpha}$, $u_{\alpha_i}$ denote the subsystem fraction and the energy density of the subsystem $\alpha\in\{ A,B,C \}$, and $u$ denotes the energy density of $\ket{E}$. To calculate the third Renyi negativity $R_3$:
    
    \begin{equation}
    	R_3= \frac{1}{1-3}\log \left\{  \frac{ \tr\left[ \left(\rho^{T_B}_{AB}  \right)^{3}  \right] }{  \tr \rho^{3}_{AB}} \right\}, 
    \end{equation}
    one introduces three replicas and compute the moments 
    \begin{equation}
    	\tr \rho^{3}_{AB}=\sum_{    \{a_i,b_i,c_i| i=1,\cdots, n  \}}^{'}  \psi( a_1,b_1,c_1   )  \psi( a_2,b_2,c_2   )   \psi( a_3,b_3,c_3   )      \psi^*( a_2,b_2,c_1   )    \psi^*( a_3,b_3,c_2   )    \psi^*( a_1,b_1,c_3   ), 
    \end{equation}
    and 
    \begin{equation}
    	\tr\left[ \left(\rho^{T_B}_{AB}  \right)^{3}  \right]  =  \sum_{     \{a_i,b_i,c_i| i=1,\cdots, n  \}}^{'}     \psi( a_1,b_1,c_1   )  \psi( a_2,b_2,c_2   )   \psi( a_3,b_3,c_3   )       \psi^*( a_2,b_3,c_1   )      \psi^*( a_3,b_1,c_2   )  \psi^*( a_1,b_2,c_3   ) ,
    \end{equation}
    where the energy conservation needs to be imposed on each individual replicas :$f_a u_{a_i}  +f_b u_{b_i}  +f_c u_{c_i}  = u$ for $i=1,2,3$. Taking the average for the moments gives $3! =6$ possible Wick's contraction patterns, and for example, a contraction denoted by 321 for $\tr \rho^{3}_{AB}$ gives 
    \begin{equation}
    	\contraction{  \psi( a_1,b_1,c_1   )  \psi( a_2,b_2,c_2   )       }{  \psi }{   ( a_3,b_3,c_3   )                }{ \psi}
    	\bcontraction{\psi( a_1,b_1,c_1   ) }{\psi}{( a_2,b_2,c_2   )   \psi( a_3,b_3,c_3   )    \psi^*( a_2,b_2,c_1   )   }{\psi} 
    	\contraction[2ex]{}{\psi}{    ( a_1,b_1,c_1   )  \psi( a_2,b_2,c_2   )   \psi( a_3,b_3,c_3   )    \psi^*( a_2,b_2,c_1   )    \psi^*( a_3,b_3,c_2   )      }{\psi}
    	\psi( a_1,b_1,c_1   )  \psi( a_2,b_2,c_2   )   \psi( a_3,b_3,c_3   )    \psi^*( a_2,b_2,c_1   )    \psi^*( a_3,b_3,c_2   )    \psi^*( a_1,b_1,c_3   ), 
    \end{equation} 
    i.e. the first $\psi$ contracts with the third $\psi^*$,  the second $\psi$ contracts with the second $\psi^*$, the third $\psi$ contracts with the first $\psi^*$. Using the contraction rule, 
    \begin{equation}
    	\contraction{}{\psi_{\alpha} }{}{\psi^{*}_\beta}
    	\psi_{\alpha} \psi^{*}_\beta= \frac{1}{d} \delta(\alpha=\beta),
    \end{equation}
    where $d$ denotes the total Hilbert space dimension, the contraction pattern 321 gives a term in $d^3\tr \rho_{AB}^3$:
    \begin{equation}
    	X_{321}=  \sum_{     \{a_i,b_i,c_i| i=1,\cdots, n  \}}^{'} \delta(  c_1=c_2  ) \delta(a_2=a_3) \delta(b_2=b_3)  =  \sum^{'}_{u_{a_1 },u_{a_2 },u_{b_1 },u_{b_2},u_{c_1 } ,u_{c_3}    }    e^{  V\left[ f_a \left(   s(u_{a_1}  )  +   s(u_{a_2}  )  \right)     +  f_b \left(   s(u_{b_1}  )  +   s(u_{b_2}  )  \right)   + f_c \left(   s(u_{c_1}+ s(u_{c_3}  )   \right)     \right]   },
    \end{equation}
    where $s(e)$ is the entropy density at the energy density $e$, and the energy density for each subsystem is still subject to the energy constraint. In the thermodynamic limit $V\to\infty$, one finds 
    \begin{equation}
    	X_{321} = e^{  V\left[ f_a \left(   s(u^*_{a_1}  )  +   s(u^*_{a_2}  )  \right)     +  f_b \left(   s(u^*_{b_1}  )  +   s(u^*_{b_2}  )  \right)   + f_c \left(   s(u^*_{c_1}+ s(u^*_{c_3}  )   \right)     \right]   },
    \end{equation}
    where $^*$ is used to denote the saddle point of the energy density. Thus, 
    \begin{equation}
    	d^3\tr \rho_{AB}^3 = \max\{  X_{123}, X_{231}, X_{312}, X_{213}, X_{321}, X_{132}   \}.    
    \end{equation}
    
    One can perform a similar analysis for $ d^3\tr\left[ \left(\rho^{T_B}_{AB}  \right)^{3}  \right]$: 
    \begin{equation}
    	d^3\tr\left[ \left(\rho^{T_B}_{AB}  \right)^{3}  \right] = \max\{  Y_{123}, Y_{231}, Y_{312}, Y_{213}, Y_{321}, Y_{132}   \}.    
    \end{equation}
    Therefore, for a given energy density of the ergodic tripartite state and given subsystem volume fractions $f_a,f_b,f_c$, comparing the saddle point value of each contraction patterns gives the volume-law coefficient of $R_3$. While we only concern the volume-law coefficient, we note that saddle point values from different contraction patterns can coincide, which induces an extra $O(1)$ constant.

\begin{figure}
    	\centering
    	\begin{subfigure}[b]{0.3\textwidth}
    		\includegraphics[width=\textwidth]{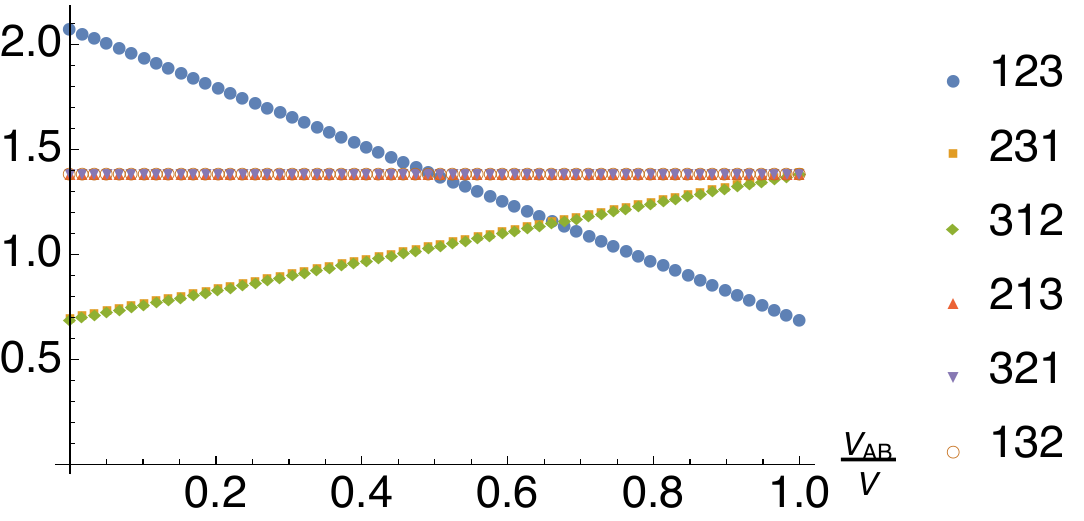}
    		\caption{$\beta=0$ }
    		\label{fig:pt_saddle_inf_beta}
    	\end{subfigure}
    	\begin{subfigure}[b]{0.3\textwidth}
    		\includegraphics[width=\textwidth]{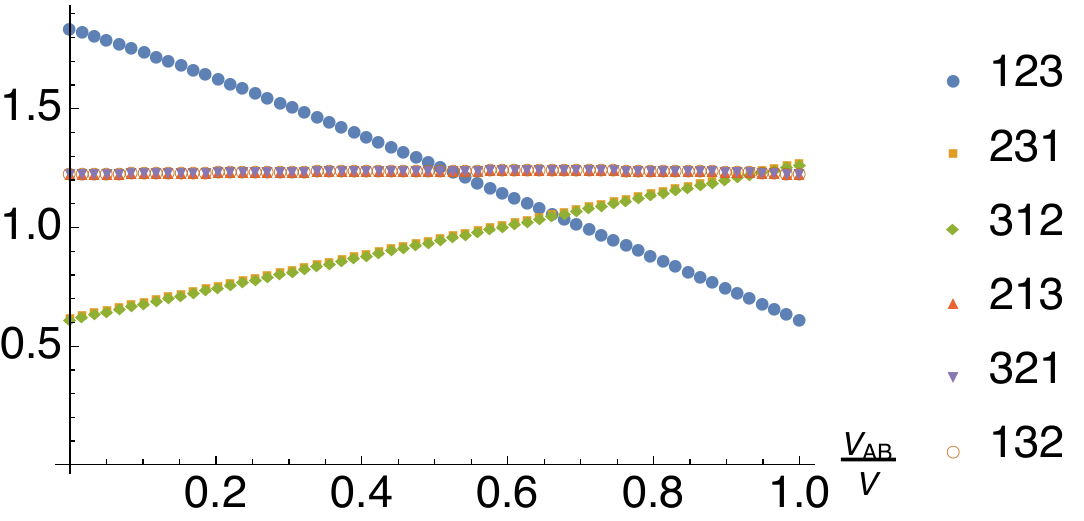}
    		\caption{$\beta=0.4$ }
    		\label{fig:pt_saddle_finite_beta}
    	\end{subfigure}
    	\caption{Saddle point value $\frac{1}{V}  \log Y $ for all possible contraction patterns.}
    	\label{}
    \end{figure}

\begin{figure}
    	\centering
    	\begin{subfigure}[b]{0.3\textwidth}
    		\includegraphics[width=\textwidth]{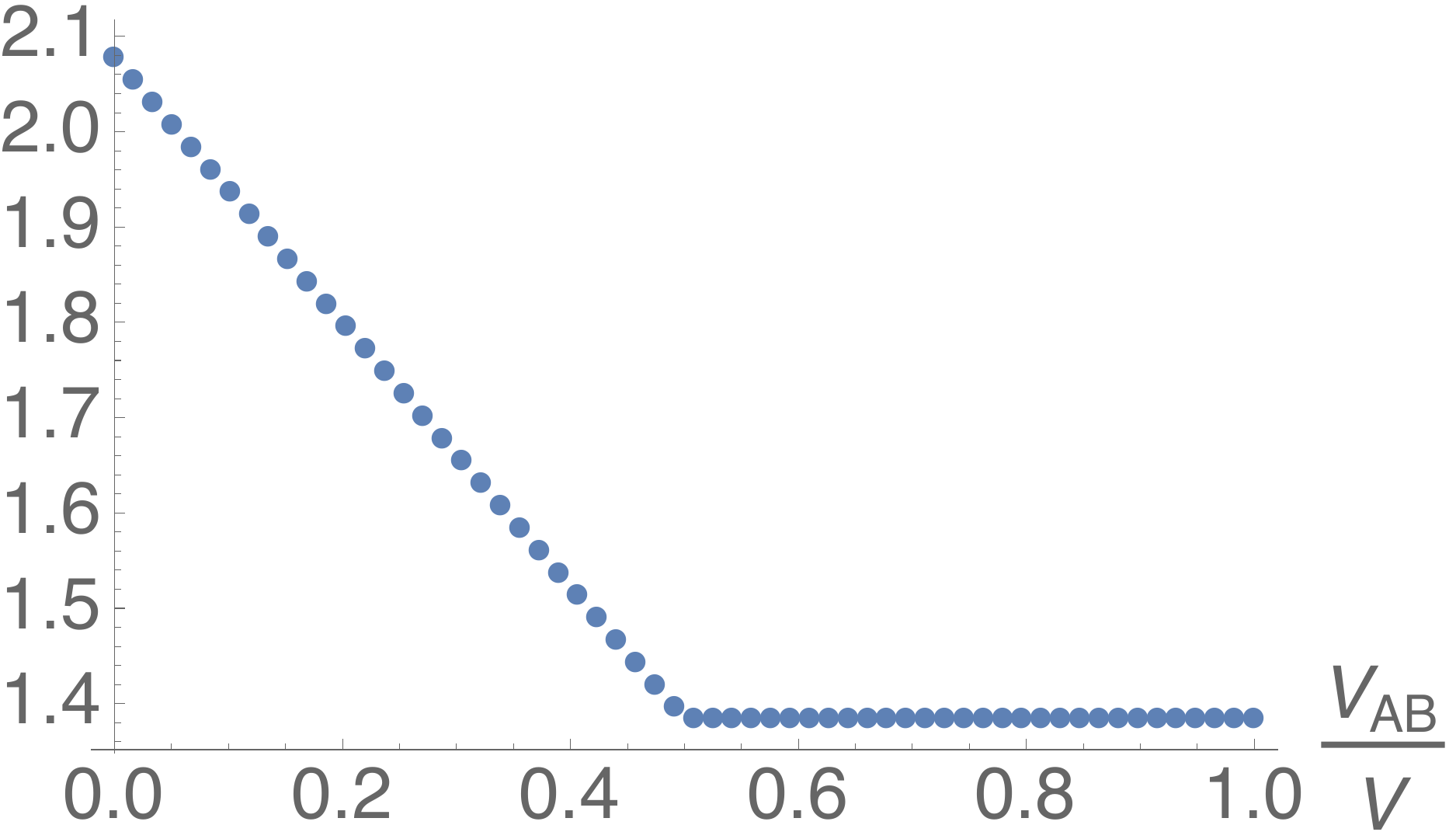}
    		\caption{$\beta=0$ }
    		\label{fig:pt_max_saddle_zero_beta}
    	\end{subfigure}
    	\begin{subfigure}[b]{0.3\textwidth}
    		\includegraphics[width=\textwidth]{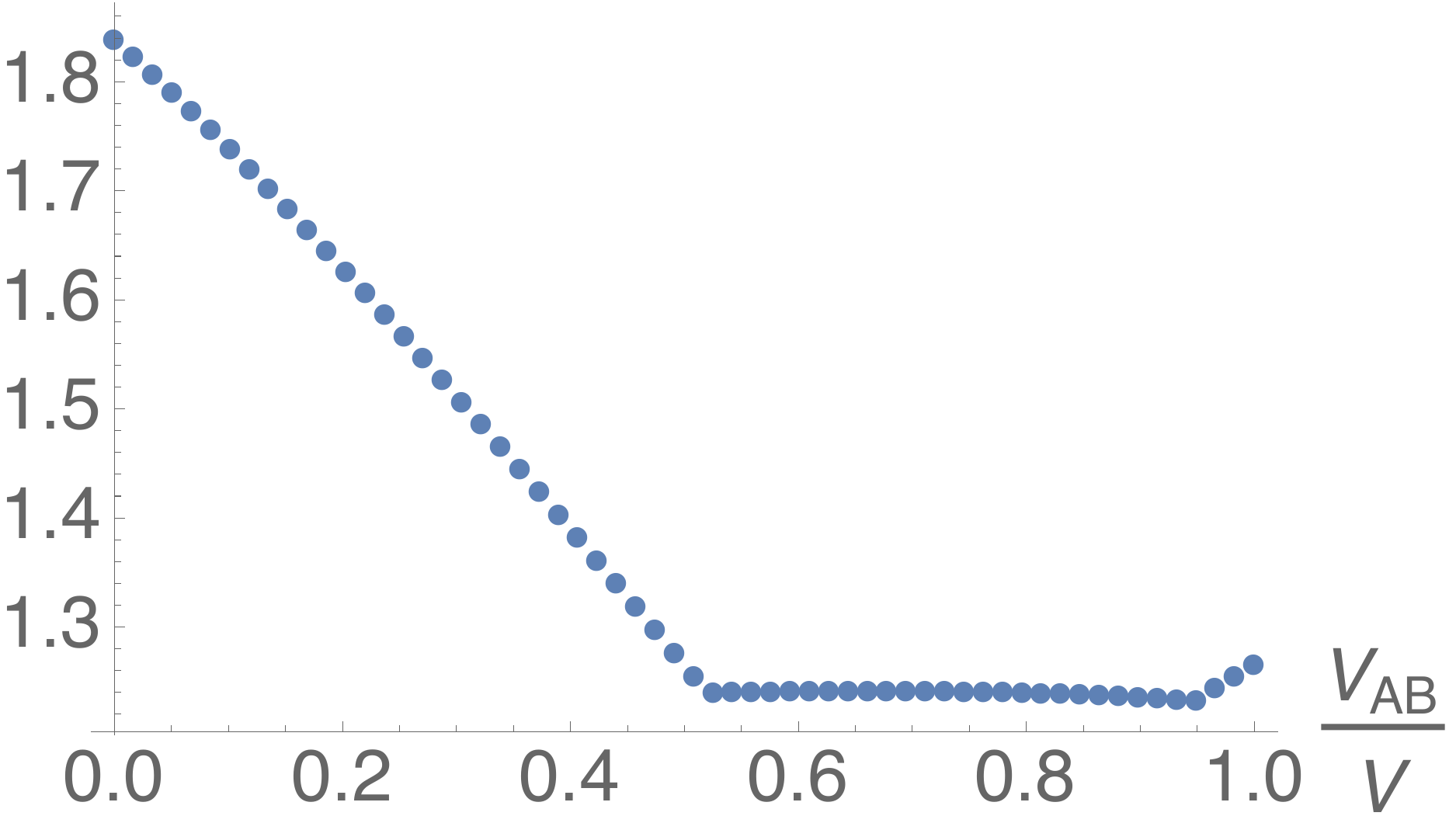}
    		\caption{$\beta=0.4$ }
    		\label{fig:pt_max_saddle_finite_beta}
    	\end{subfigure}
    	\caption{Maximum of saddle point value $\frac{1}{V}  \log Y $ among all possible contraction patterns.}
    	\label{}
    \end{figure}

\begin{figure}
    	\centering
    	\begin{subfigure}[b]{0.3\textwidth}
    		\includegraphics[width=\textwidth]{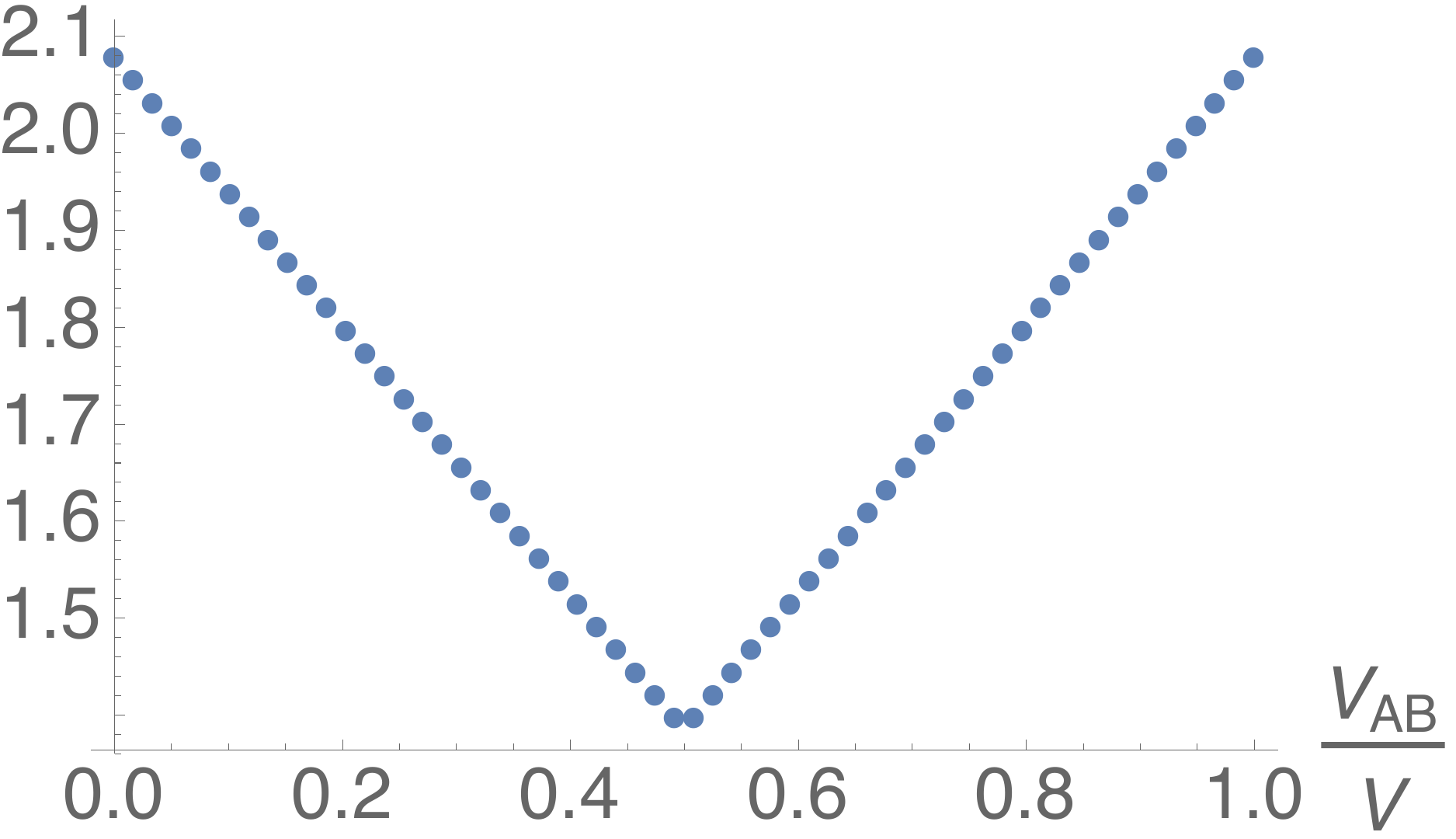}
    		\caption{$\beta=0$ }
    		\label{fig:max_saddle_zero_beta}
    	\end{subfigure}
    	\begin{subfigure}[b]{0.3\textwidth}
    		\includegraphics[width=\textwidth]{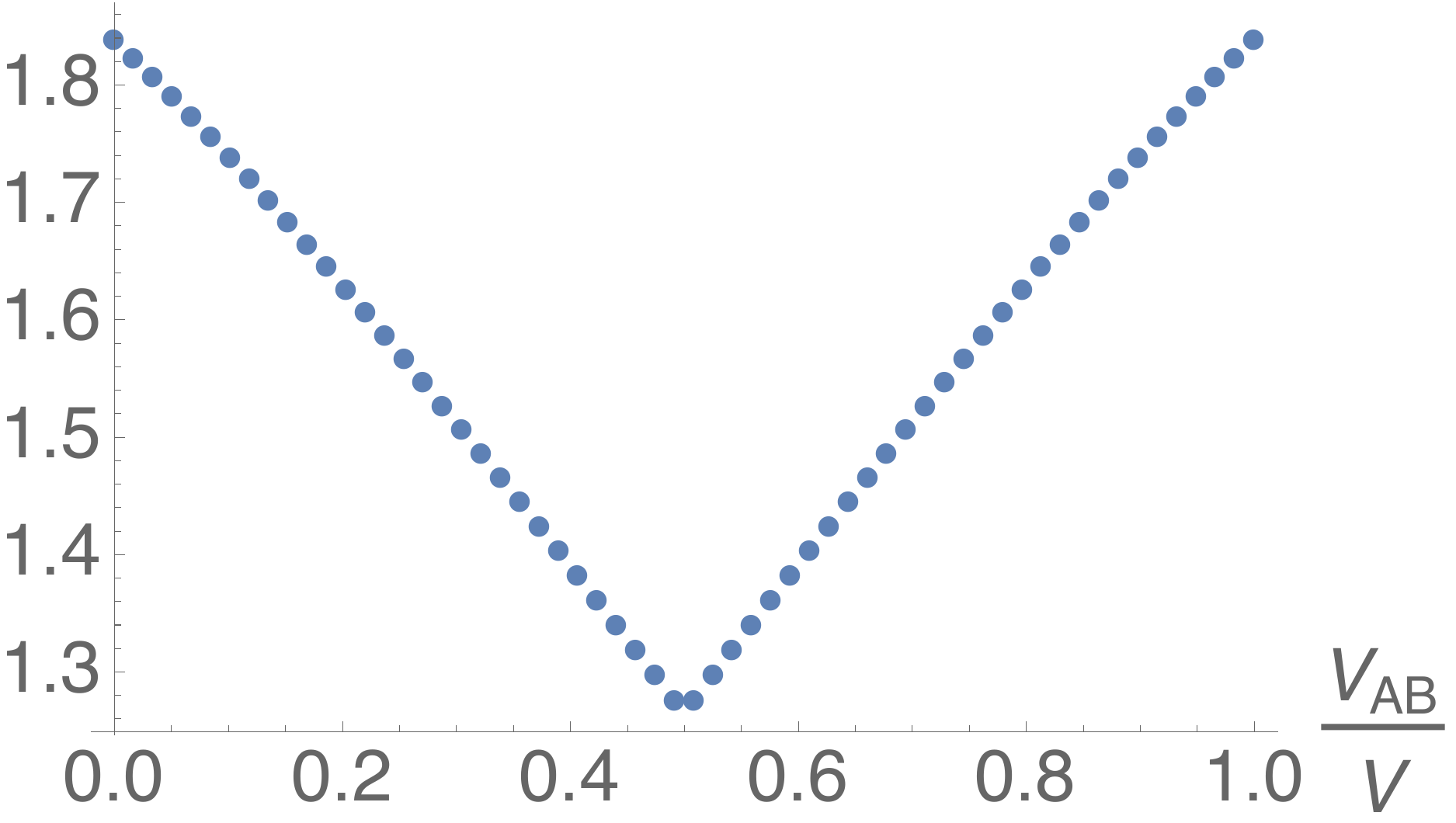}
    		\caption{$\beta=0.4$}
    		\label{fig:max_saddle_finite_beta}
    	\end{subfigure}   
    	\caption{Maximum of saddle point value $\frac{1}{V}  \log X $ among all possible contraction patterns.}
    	\label{}
    	
    	\label{fig:}
    \end{figure}

    Here we compute $R_3$ assuming that the many-body density of states $D(u) \sim e^{ V s(u)   }$ is a Gaussian, i.e. the entropy function is quadratic $s(u) = \log 2- \frac{1}{2}u^2$. For $\tr\left[ \left(\rho^{T_B}_{AB}  \right)^{3}  \right]$, we compare the saddle point values from 6 possible contraction patterns as a function of $f=V_{AB}/V$. While at $\beta=0$ (Fig.\ref{fig:pt_saddle_inf_beta}), the exchange of the saddle point values occurs at $f=1/2$, for $\beta \neq 0$, saddle point values exchange at two different subsystem fractions: one is slightly above $f=1/2$, and the other one is slightly below $f=1$ (Fig.\ref{fig:pt_saddle_finite_beta}). Choosing the maximal contraction patterns gives Fig.\ref{fig:pt_max_saddle_zero_beta},\ref{fig:pt_max_saddle_finite_beta}, indicating that $ \tr\left[ \left(\rho^{T_B}_{AB}  \right)^{3}  \right]$ has two singularities for $\beta \neq 0$ but only one singularity for $\beta=0$. On the other hand, $\tr{\rho_{AB}^3}$ has only one singularity at $f=1/2$ independent of the temperature as shown in Fig.\ref{fig:max_saddle_zero_beta},\ref{fig:max_saddle_finite_beta}. $R_3$ as a function of $f$ at different inverse temperatures is shown in the main text (Fig.\ref{fig:main_text_R_3_ansatz}). $R_3$ at $\beta=0$ exactly matches the result from the random pure state, where the singularity at $f=1/2$ corresponds to the area-law to volume-law transition. The same feature carries over to the finite temperature ergodic tripartite states ($\beta \neq 0 $). However at $\beta \neq 0$, there are two extra singularities in $R_3$ inheriting from the singularities of $\tr\left[ \left(\rho^{T_B}_{AB}  \right)^{3}  \right]$. It would be interesting to investigate in the future to see whether this is a unique feature of Renyi negativity which is not shared by negativity.  

    \section{Entanglement negativity in free fermion systems}\label{appendix:fermion}
  
    \subsection{Negativity of energy eigenstate $\ket{\psi}$}
       Considering a one dimensional lattice of $L$ sites with periodic boundary condition, we study the free fermion Hamiltonian with translation symmetry and $U(1)$ charge conservation: $H= - \sum_{x_1,x_2=1}^L t(x_1-x_2)  c_{x_1}^{\dagger}c_{x_2} + h.c.$, where the hopping amplitude $t(x_1-x_2)=t^*(x_2-x_1)$, and the operators $c_i$, $c_i^{\dagger}$ satisfy the fermionic algebra. Such Hamiltonian can be diagonalized as $H= \sum_k \epsilon_k d_k^{\dagger} d_k$ by Fourier transforming the operators $c_x=\frac{1}{\sqrt{L}}  \sum_k e^{ikx} d_k$. Divide the system into three parts labelled by $A$ (the sites from $x=1$ to $x=L_A$), $B$ (the sites from $x=L_A+1$ to $x=L_A+L_B$), and $C$ (the sites from $x=L_A+L_B+1$ to $x=L$), we study the negativity between $A$ and $B$ for energy eigenstates. A free fermion eigenstate $\ket{\psi}$ is fully characterized by its correlation matrix $C_{0,xy}=  \expval{c_x^{\dagger}c_y }$, where the expectation value is with respect to an eigenstate $\ket{\psi}$. Note that $C_{0}$ is a Hermitian matrix, and its $k$-th eigenvalue specifies the occupation number on $k$-th single particle modes for $\ket{\psi}$, which can only be 0 or 1.

    To calculate negativity between $A$ and $B$, we consider the reduced density matrix in $AB$: $\rho_{AB}=\tr_C\ket{\psi}\bra{\psi}$, which is a Gaussian state characterized by the correlation matrix $C$ restricted in the region $AB$ of size $L_{AB}=L_A+L_B$:
    
    \begin{equation}
    	C_{x_1x_2}= \expval{c_{x_1}^{\dagger}c_{x_2} } =\frac{1}{L} \sum_k e^{ik(x_2-x_1)} \expval{d_k^{\dagger} d_k }.
    \end{equation}
    The spatial coordinates $x_1,x_2$ are restricted in $AB$:   $x_1,x_2\in \{1,2,\cdots, L_{AB} \} $. Note that eigenvalues of the matrix $C$ are bounded between $0$ and $1$ since $C$ is a sub-block from the correlation matrix $C_{0}$ (See Appendix \ref{appendix:bound_sub_matrix} for proof). Below we apply the correlation matrix method to calculate the negativity between $A$ and $B$\cite{Shapourian2018}. Let $\rho_{AB}^{T_B}$ be the partial transposed density matrix, one defines the normalized composite density matrix (remains a Gaussian) $\widetilde{\rho} =\rho_{AB}^{T_B} \left(  \rho_{AB}^{T_B}  \right)^{\dagger} /\widetilde{Z}$ , where $\widetilde{Z} = \tr \left[   \rho_{AB}^{T_B} \left(  \rho_{AB}^{T_B}  \right)^{\dagger}   \right]= \tr \rho_{AB}^2$. The negativity reads\cite{Shapourian2018} 
    \begin{equation}\label{appendix:fermion_main}
    	\begin{split}
    		E_N= \log \left(   \tr \sqrt{\rho_{AB}^{T_B} \left(  \rho_{AB}^{T_B}  \right)^{\dagger}   }   \right) =  \log \left(   \tr   \widetilde{\rho }^{\frac{1}{2}}  \right)+\frac{1}{2} \log \left(  \tr \rho^2_{AB}  \right),
    	\end{split}
    \end{equation}
    where the above two terms can be individually calculated using the correlation matrix method:
    
    \begin{equation}
    	\begin{split}
    		&\log \left(   \tr   \widetilde{\rho }^{\frac{1}{2}}  \right)      =     \tr{ \log\left[ \widetilde{C}^{\frac{1}{2}} + \left( 1-\widetilde{C} \right)^{\frac{1}{2}}   \right] } \\
    		&    \frac{1}{2} \log \left(  \tr \rho^2_{AB}  \right)     =      \frac{1}{2}  \tr{  \log \left[  C^2+\left(  1-C \right)^2    \right]     }.
    	\end{split}
    \end{equation}
    $\widetilde{C}$ and $C$ are the correlation matrices of $\widetilde{  \rho  }$ and $\rho_{AB}$ respectively. To obtain $\widetilde{C}$, we first define $\Gamma= \mathbb{I} -2C$
    \begin{equation}
    	\Gamma=\begin{pmatrix}
    		\Gamma^{AA} &\Gamma^{AB} \\ 
    		\Gamma^{BA} & \Gamma^{BB} 
    	\end{pmatrix},
    \end{equation}
    and the transformed matrices
    \begin{equation}
    	\Gamma_{\pm}  =\begin{pmatrix}
    		-\Gamma^{AA} &  \pm i \Gamma^{AB}\\
    		\pm  i \Gamma^{BA} & \Gamma^{BB}
    	\end{pmatrix},
    \end{equation}
    then the correlation matrix is $\widetilde{C}=\frac{1}{2} \left(    1- \widetilde{\Gamma} \right)$, where $\widetilde{\Gamma} $ is

    \begin{equation}
    	\widetilde{ \Gamma  }  =   \left( \mathbb{I}+\Gamma_{+} \Gamma_{-}  \right)^{-1} \left( \Gamma_+  +  \Gamma_-  \right).
    \end{equation}
    Below we apply this formalism to calculate the negativity averaged over all eigenstates for free fermions. To proceed, we find it more convenient to work with $\Gamma$ and $\widetilde{ \Gamma  }$, which gives

    \begin{equation}
    	\begin{split}
    		&\log \left(   \tr   \widetilde{\rho }^{\frac{1}{2}}  \right)   =     \tr{ \log\left[    \left( \frac{1}{2}(1-\widetilde{ \Gamma  })  \right)^{\frac{1}{2}} + \left(  \frac{1}{2}   (  1+\widetilde{ \Gamma  }  ) \right)^{\frac{1}{2}}   \right] } \\
    		&   \frac{1}{2} \log \left(  \tr \rho^2_{AB}  \right)  =  \frac{1}{2}   \tr{ \log\left[    \left( \frac{1}{2}(1-\Gamma)  \right)^2 + \left(  \frac{1}{2}   (  1+ \Gamma   ) \right)^2 \right] }.
    	\end{split}
    \end{equation}

    \subsection{Volume-law coefficient of negativity in $\frac{L_{AB}}{L} \ll  1$ limit.}
    Here we calculate the volume law coefficient of negativity in $\frac{L_{AB}}{L} \ll 1$ limit averaged over all eigenstates. The central idea is to perform the expansion about $\widetilde{\Gamma}=0$ in powers of $\widetilde{\Gamma}$ 
    
    \begin{equation}\label{appendix:expansion}
    	\log \left(   \tr   \widetilde{\rho }^{\frac{1}{2}}  \right) =   \frac{\log 2}{2} L_{AB} - \sum_{n=1}^{\infty}  a_n  \tr \widetilde{ \Gamma  }^{2n}   =    \frac{\log 2}{2} L_{AB}  -\frac{  \tr \widetilde{ \Gamma  }^2}{8}   -     \frac{3 \tr\widetilde{ \Gamma  }^4}{64} - \frac{5  \tr \widetilde{ \Gamma  }^6}{192}+O\left(  \tr\widetilde{ \Gamma  }^8\right).
    \end{equation}
    Note this is a convergent series since the eigenvalues of $\widetilde{ \Gamma  }$ is bounded between $-1$ and $1$. First we calculate $\tr \widetilde{ \Gamma  }^2$:
    \begin{equation}
    	\tr \widetilde{ \Gamma  }^2 = \tr \left[    \left( \mathbb{I}+\Gamma_{+} \Gamma_{-}  \right)^{-1} \left( \Gamma_+  +  \Gamma_-    \right) \right]^2.
    \end{equation} 
    Since the eigenvalues of $\Gamma_+ \Gamma_- $ are bounded between $0$ and $1$ (see Appendix.\ref{appendix:bound_pm} for proof), we can expand the matrix $ \left( \mathbb{I}+   \Gamma_{+} \Gamma_{-}  \right)^{-1}=      \sum_{m=0}^{\infty} \left( -\Gamma_+  \Gamma_-   \right)^m     =  \mathbb{I} -   \Gamma_{+} \Gamma_{-} + \left(   \Gamma_{+} \Gamma_{-}  \right)^2+\cdots $, and hence
    \begin{equation}\label{eq:series}
    	\tr \widetilde{ \Gamma  }^2 = \sum_{m_1,m_2=0}^{\infty }   \tr \left[    \left( -\Gamma_+  \Gamma_-   \right)^{m_1} \left(    \Gamma_{+}  +\Gamma_{-} \right)     \left( -\Gamma_+  \Gamma_-   \right)^{m_2} \left(    \Gamma_{+}  +\Gamma_{-} \right)        \right].
    \end{equation}
    In the thermodynamic limit $L\to \infty$ with $L_{AB}/L$ fixed, taking the average over all eigenstates, one can show $\overline{ \tr \widetilde{ \Gamma  }^2}$ is in the form $L_{AB}\left[b_1 \left( L_{AB}/L \right)+ b_2 \left( L_{AB}/L  \right)^2+ \cdots \right]$. When $\frac{L_{AB}}{L} \ll 1 $, we only need to consider the leading order $L_{AB}^2/L$, which corresponds to $m_1=m_2=0$ term in the series:
    \begin{equation}
    	\overline{ \tr \widetilde{ \Gamma  }^2  } =    \overline{  \tr \left(  \Gamma_{+ }+\Gamma_{- }  \right)^2    } =4 \overline{      \tr \left( \Gamma^{AA} \right)^2} +   4  \overline{ \tr \left( \Gamma^{BB} \right)^2}.
    \end{equation}
    Since
    
    \begin{equation}
    	\tr \left(  \Gamma^{AA}  \right)^2  =  \sum_{x_1,x_2 \in A}   \Gamma^{AA}_{x_1x_2} \Gamma^{AA}_{x_2x_1} 
    	=\frac{1}{L^2} \sum_{x_1,x_2\in A}  \sum_{k_1,k_2} e^{-ik_1(x_1-x_2)} e^{-ik_2(x_2-x_1)} n_{k_1}n_{k_2},
    \end{equation}
    where $n_k\equiv 1-2\expval{d_k^{\dagger} d_k }\in \{ \pm 1 \}$. Taking average over all eigenstates gives $\overline{  n_{k_1}n_{k_2}  }  =\delta_{k_1k_2}$. Therefore, one finds
    
    \begin{equation}
    	\overline{  \tr \left(  \Gamma^{AA}  \right)^2  } = \frac{L_A^2}{L},
    \end{equation}
    and similarly 
    \begin{equation}
    	\overline{  \tr \left(  \Gamma^{BB}  \right)^2  } = \frac{L_B^2}{L},
    \end{equation}
    Setting $L_A=L_B= \frac{1}{2} L_{AB}$ gives 
    \begin{equation}\label{eq:}
    	\overline{\tr \widetilde{ \Gamma  }^2   }=   2\left(  \frac{L_{AB}}{L}   \right)  L_{AB}.
    \end{equation}
    Thus, 
    \begin{equation}\label{eq:exact_first_term}
    	\overline{  \log \left(   \tr   \widetilde{\rho }^{\frac{1}{2}}  \right)   }  = \left[ \frac{\log 2 }{2} - \frac{1}{4} \frac{L_{AB}}{L}  +O\left( \left(  \frac{L_{AB}}{L} \right)^2   \right) \right]  L_{AB}.
    \end{equation}
    Similarly, one can expand $\frac{1}{2} \log \left(  \tr \rho^2_{AB}  \right)= \frac{1}{2}   \tr{ \log\left[    \left( \frac{1}{2}(1-\Gamma)  \right)^2 + \left(  \frac{1}{2}   (  1+ \Gamma   ) \right)^2 \right] } $ about $\Gamma=0$:
    \begin{equation}
    	\frac{1}{2} \log \left(  \tr \rho^2_{AB}  \right)  = -\frac{\log 2 }{2}  L_{AB} - \sum_{n=1}^{\infty }   \frac{ (-1)^n \tr \Gamma^{2n}     }{2n}.
    \end{equation}
    Taking average over all eigenstates, to the leading order in $\frac{L_{AB}}{L}$ as $L\to \infty $, one finds 
    \begin{equation}\label{eq:exact_second_term}
    	\overline{  \frac{1}{2} \log \left(  \tr \rho^2_{AB}  \right) }  =   -\frac{\log 2 }{2}  L_{AB}  + \frac{1}{2} \overline{ \tr \Gamma^2 }+\cdots  =  \left[  -\frac{\log 2 }{2}   + \frac{1}{2}\frac{L_{AB}}{L}   +O\left( \left(  \frac{L_{AB}}{L} \right)^2   \right) \right]L_{AB},
    \end{equation}
    where we have performed the similar calculation to obtain $\overline{\tr \Gamma^2} = \frac{L_{AB}^2}{L}$.
    Combining Eq.\ref{appendix:fermion_main}, Eq.\ref{eq:exact_first_term}, and Eq.\ref{eq:exact_second_term}, we find as $L\to \infty$ with $\frac{L_{AB}}{L}$ fixed, the negativity averaged over all eigenstates follows a volume law scaling, where the volume law coefficient $\alpha$ is a power series of $ \frac{L_{AB}}{L}$: 
    
    \begin{equation}
    	\boxed{
    		\overline{E_N}  =  \alpha L_{AB}      =  \left[\frac{1}{4} \frac{L_{AB}}{L} + \sum_{n=2}^{\infty} \alpha_n \left( \frac{L_{AB}}{L}    \right)^n     \right]  L_{AB}}.
    \end{equation}

    \subsection{Some useful mathematical results}

    \subsubsection{Bounds on the eigenvalues of ~$C$}\label{appendix:bound_sub_matrix} 
    Given a $L \cross L$ Hermitian matrix $C_0$ with all eigenvalues being 0 or 1, consider the $l\cross l$ sub-block matrix $C$ obtained by restricting the row and column index $i=1,2,\cdots, l$ in $C_0$, all eigenvalues $\lambda_i$ of $C$ satisfy $0\leq \lambda_i  \leq 1$. \\
    \noindent \underline{\textit{Proof}}:\\

    First, we show $\lambda_i\geq 0$ (i.e. $C$ is positive semi-definite). To see this, consider $\bra{v}C\ket{v}$, where $\ket{v}$ is a normalized vector with $l$ components: $\ket{v}=  (v_1,v_2,\cdots ,v_l)^T$, one can embed $\ket{v}$ in a larger vector space of $L$ dimension so that $\ket{v} \to \ket{v_0}=  (v_1,v_2,\cdots ,v_l,0,\cdots,0)^T$, which implies $\bra{v}C\ket{v} = \bra{v_0}C_0 \ket{v_0}$. The fact that $C_0$ is a positive semi-definite matrix means $\bra{v_0}C_0 \ket{v_0}\geq 0$ for any $\ket{v_0}$. Thus $\bra{v}C\ket{v} \geq 0 $ for all $\ket{v}$, and $C$ is  positive semi-definite. 
    
    Second, consider a normalized vector $\ket{v}$, we show that the norm $\norm{C\ket{v}}\leq 1$, which implies eigenvalues of $C$ satisfies $\abs{\lambda_i} \leq 1 $. To see this, we again consider $\ket{v}=  (v_1,v_2,\cdots ,v_l)^T \to \ket{v_0}=  (v_1,v_2,\cdots ,v_l,0,\cdots,0)^T$. Then $\norm{ C \ket{v}  }=\norm{PC_0 \ket{v_0}   }$, where $P$ is the projector from the $L$ dimensional vector space back to the $l$ dimensional vector space. Since projection can not increase the norm of a vector, one finds $\norm{ C \ket{v}  }= \norm{PC_0 \ket{v_0}    }\leq \norm{C_0 \ket{v_0}   }$. Because the eigenvalues of $C_0$ are less than or equal to one, it follows that $\norm{  C\ket{v}  }\leq 1$, implying the eigenvalues of $C$ satisfy $0\leq \abs{\lambda_i} \leq 1$.

    Combining the above two results proves $0\leq \lambda_i\leq 1$.

    \subsubsection{Bounds on the eigenvalues of~ $\Gamma_{+}\Gamma_{-}$ }\label{appendix:bound_pm}
    Here we prove all eigenvalues $\lambda_i$ of $\Gamma_{+}\Gamma_{-}$ satisfy $0\leq \lambda_i\leq 1$.\\ 
    \noindent \underline{\textit{Proof}}:\\
    Given the Hermitian matrix $\Gamma=\begin{pmatrix} A &B \\ B^{\dagger} & C \end{pmatrix} $ in the block matrix form with $A=A^{\dagger}$ and $C=C^{\dagger}$, $\Gamma_{+ }$ is defined as     $ \begin{pmatrix}  -A & iB \\ iB^{\dagger} & C   \end{pmatrix}$ and $\Gamma_{- } = \Gamma_{+ }^{\dagger}$. We notice that $\Gamma_{+}$ can be written as $\Gamma_{+}=S \Gamma S$ using a unitary matrix $S=\begin{pmatrix}   i \mathbb{I} & 0\\ 0 &\mathbb{I}
    \end{pmatrix}$, and thus $\Gamma_{+}\Gamma_{-} =S\Gamma SS^{\dagger} \Gamma  S^{\dagger}= S\Gamma^2 S^{\dagger}$. Since $S$ is unitary, $\Gamma_{+ }\Gamma_{-}$ and $\Gamma^2$ have exactly the same spectrum. Note that the spectrum of $\Gamma$ is bounded between -1 and 1 due to $\Gamma= \mathbb{I}-2C$, where $C$ is the correlation matrix with eigenvalues bounded between 0 and 1. Hence the eigenspectrum of $\Gamma_{+ }\Gamma_{- }$ and $\Gamma^2$ is bounded between 0 and 1.

 \end{document}